\documentclass[12pt]{article}
\usepackage[margin=0.8in]{geometry}
\usepackage[utf8]{inputenc}

\usepackage{amsmath,amssymb,amsfonts,amsthm}

\usepackage{breqn}

\usepackage{bold-extra}

\usepackage{graphicx}
\usepackage{xcolor}
\usepackage{url}
\usepackage{hyperref}
\hypersetup{
  colorlinks   = true, 
  urlcolor     = blue, 
  linkcolor    = black, 
  citecolor   = black 
}
\usepackage[numbers,sort&compress]{natbib}
\let\OLDthebibliography\thebibliography
\renewcommand\thebibliography[1]{
  \OLDthebibliography{#1}
  \setlength{\parskip}{4pt}
  \setlength{\itemsep}{0pt plus 0.3ex}
}

\allowdisplaybreaks
\usepackage{tabularx}
\usepackage{bbm}
\usepackage{listings}
\usepackage{longtable}
\usepackage{array}

\usepackage{titlesec}
\titleformat*{\subsection}{\bfseries\boldmath}
\titleformat*{\subsubsection}{\bfseries\boldmath}

\usepackage{multicol}
\usepackage{enumitem}

\newcommand\refeq[1]{Eq.~(\ref{#1})}

\newcommand\refse[1]{Sect.~\ref{#1}}

\newcommand\citere[1]{Ref.~\cite{#1}}
\newcommand\citeres[1]{Refs.~\cite{#1}}
\newcommand\refap[1]{App.~\ref{#1}}
\def\reffi#1{\mbox{Fig.~\ref{#1}}}

\usepackage{soul}

\newcommand{\gev}{\ \mathrm{GeV}}
\newcommand{\tev}{\ \mathrm{TeV}}

\graphicspath{{Plots/}}

\begin{document}

\thispagestyle{empty}

\def\thefootnote{\fnsymbol{footnote}}

\begin{flushright}
  DESY 21-125 ~~
\end{flushright}

\vspace*{1cm}

\begin{center}

  {\Large Reconciling Higgs physics and
    pseudo-Nambu-Goldstone \\[0.4em] dark matter
    in the S2HDM using a genetic algorithm}

  \vspace{1cm}

\setcounter{footnote}{3} 

  Thomas Biekötter\footnote{thomas.biekoetter@desy.de},
  Mar\'ia Olalla
    Olea-Romacho\footnote{maria.olalla.olea.romacho@desy.de}\\[0.4em]
  {\textit{
   Deutsches Elektronen-Synchrotron DESY,
   Notkestra{\ss}e 85, 22607 Hamburg, Germany\\[0.8em]
 }}

\vspace*{1cm}

\begin{abstract}
We investigate a possible realization of
pseudo-Nambu-Goldstone (pNG) dark matter
in the framework of a singlet-extended 2 Higgs
doublet model (S2HDM). pNG dark matter gained
attraction due to the fact that direct-detection
constraints can be avoided naturally because of the
momentum-suppressed scattering cross sections,
whereas the relic abundance of dark matter can
nevertheless be accounted for via the usual thermal
freeze-out mechanism. We confront the S2HDM with
a multitude of theoretical and experimental
constraints, paying special attention to the
theoretical limitations on the scalar potential,
such as vacuum stability and perturbativity.
In addition, we discuss
the complementarity between constraints
related to the dark matter sector, on the
one hand, and to the Higgs sector, on the other hand.
In our numerical discussion we explore the
Higgs funnel region with dark matter
masses around $60 \gev$ using a genetic algorithm.
We demonstrate that the S2HDM can easily
account for the measured relic abundance while
being in agreement with all relevant constraints.
We also discuss whether the so-called
galactic center excess and the
antiprotron excess can be
accommodated, 
possibly in
combination with a Higgs
boson at about $96\gev$ that can be the origin
of the LEP- and the CMS-excess observed at this mass
in the $b \bar b$-quark and
the diphoton final state, respectively.
\end{abstract}

\end{center}

\renewcommand{\thefootnote}{\arabic{footnote}}
\setcounter{footnote}{0} 

\newpage

\tableofcontents

\section{Introduction}
\label{secintro}

The discovery of a Higgs boson 
with a mass of approximately $125\gev$
at the LHC
by both the ATLAS~\cite{ATLAS:2012yve}
and the CMS~\cite{CMS:2012qbp}
collaborations is a milestone for the
understanding of the laws of nature.
So far, the experimental observation
related to the discovered
particle, denoted by $h_{125}$ in the
following, agree with the
interpretation of a fundamental scalar
particle that behaves according to the
prediction of the Standard Model
(SM)~\cite{Khachatryan:2016vau,
Aad:2019mbh,Sirunyan:2018koj}.
As a result, any model has to incorporate
a particle state that resembles a SM-like
Higgs boson within the current
experimental uncertainties.
In contrast to the measurement
of the Higgs-boson mass, the measurements
of the couplings of $h_{125}$ are much less
precise, with uncertainties at the level of
about ten to twenty
percent~\cite{Aad:2019mbh,Sirunyan:2018koj}.
This leaves room for interpretations
of the discovered Higgs boson in models
beyond the SM (BSM), in which the theoretical
predictions of the properties of $h_{125}$
only agree with the SM interpretation at
the level of the current experimental
uncertainties. In such models, phenomena
can be accommodated
that cannot be explained in the SM,
and the precise measurements
of the properties of $h_{125}$ will be crucial
in order to shed light on the
question which of the
proposed BSM scenarios could be realized
in nature.

Another experimental milestone
consists of
various indications for the
existence of dark matter (DM)
through a conjoint of data gathered from, among others,
rotation curves of spatial
\mbox{galaxies~\cite{Zwicky:1933gu,Rubin:1970zza}},
gravitational lensing~\cite{Massey:2010hh},
and the Bullet cluster merger~\cite{Clowe:2006eq}. 
The Planck collaboration \cite{Planck:2018vyg},
using the precise map of the cosmic
microwave background (CMB), reports
the most precise measurement of today's
DM relic abundance,
${(\Omega h^2)_{\rm Planck} = (0.119 \pm
0.003)}$. Hence, the DM sector
constitutes about $26\%$ of
the energy-matter content of the Universe.
Even though there are many indirect
indications for the existence of DM via
gravitational effects, so far there has
not been any direct discovery of a DM
particle that could give rise to more
information about the properties of DM.
The elusive nature of DM has opened up 
an interesting landscape of BSM theories
that can provide one or more DM candidates.
One of the most studied
scenarios in such SM extensions is
the weakly-interacting particle (WIMP),
a particle with weak couplings to the
SM particles and a mass around the 
electroweak (EW) scale whose
existence could
potentially be probed also
at present or future colliders.
In view of the fact that the
DM particle(s) might not be charged under
the SM gauge groups, in which
case they are also not
coupled directly to the quarks and leptons,
the possibility 
of 
coupling the DM to the SM only via
the Higgs sector, often called
Higgs portal~\cite{Patt:2006fw,Barbieri:2006dq},
is an interesting scenario.
Many extended Higgs sectors
provide a (pseudo)scalar DM candidate fitting
the WIMP paradigm. However, they
are stringently constrained 
by DM direct-detection
experiments~\cite{Schumann:2019eaa}.
A possibility to evade the
constraints from direct-detection experiments
is given by the fact that the scattering
cross sections between the DM and the
nuclei can be momentum-suppressed. A particle
that naturally has this feature is
the so-called
pseudo-Nambu-Goldstone boson (pNG)
DM~\cite{Barger:2008jx,Barger:2010yn,
Barducci:2016fue,Gross:2017dan,
Balkin:2018tma,Huitu:2018gbc,Karamitros:2019ewv}.
As a result, BSM models that predict
the existence of a stable pNG in order to
account for the DM relic abundance have recently gained
a lot of attention~\cite{Azevedo:2018exj,
Ishiwata:2018sdi,Arina:2019tib,
Chiang:2019oms,Cline:2019okt,
Ahmed:2020hiw,Abe:2020dut,Glaus:2020ihj,
Zhang:2021alu,Haisch:2021ugv}.

The most economic way to introduce pNG DM
is to extend the SM by a complex singlet
field $\phi_S$
and demanding that the Lagrangian respects a
softly broken global U(1) symmetry under
which the singlet field is
charged~\cite{Barger:2008jx}.
The pNG DM is then given by the imaginary
part of $\phi_S$, where the global U(1)
symmetry prevents the particle from decaying,
and the soft U(1)-breaking
term gives rise to the mass of the pNG state.
The SM has short-comings beyond the
fact that 
it does not provide an explanation for
the observation of DM,
such that it is also compelling to investigate
pNG DM in models with Higgs sectors that
compared to the SM contain additional
fields together with $\Phi_S$.
One of such possibilities is that
pNG DM can be
incorporated into
2~Higgs~doublet~models (2HDM)
(see \citere{Branco:2011iw} for a review),
in which, 
in contrast to the pNG DM
model with only one Higgs
doublet~\cite{Kannike:2019wsn},
a first-order EW phase transition
can be realized~\cite{Zhang:2021alu,Biekotter:2021ysx}.
Such a transition
provides a departure from thermal
equilibrium, required to 
explain the matter-antimatter asymmetry
via the mechanism of EW
baryogenesis~\cite{Kuzmin:1985mm, Cline:1996mga}.
A cosmological first-order phase transition can
also source the formation of a
stochastic gravitational-wave
background that could be observable
at future space-based gravitational-wave
interferometers, such as LISA~\cite{Zhang:2021alu}.
Moreover,
we emphasize that in
the S2HDM the
presence of the second Higgs doublet gives
rise to two additional neutral and two
charged scalars.
For DM masses comparable to the masses of these
additional states, new annihilation
processes can become important
for the prediction of the DM relic abundance.
As a result, the predicted relic abundance
for a certain DM mass can differ
substantially between the S2HDM and
the simpler model
with only one Higgs doublet,
and new parameter regions can
become physically viable, where
the corresponding parameter space
of the model with only one Higgs
double predicts a too large
relic abundance~\cite{Jiang:2019soj}.

In addition to the
above mentioned phenomenological
reasons to consider a model with
a second Higgs doublet field,
one should also note that
supersymmetric extensions of the SM,
in which the hierarchy problem can be
addressed~\cite{Veltman:1980mj,
Dimopoulos:1981zb,Witten:1981nf},
require the existence of at least two
Higgs doublet fields $\phi_{1,2}$
in order to account for the masses of
all quarks and leptons.
Also the most commonly studied
solutions to the strong CP-problem
incorporating the so-called QCD axion
require the presence of two doublet
fields~\cite{Kim:1986ax}.
Moreover, new axially coupled U(1) interactions,
resulting in extra gauge bosons weakly
coupled to standard model particles
and which behaves very much as an
axion-like particle,
provide a possible bridge to a
new dark sector and also
demand an additional electroweak
doublet~\cite{Fayet:2020bmb}.
Other models
to solve the hierarchy problem rely
on a unification of the gauge
interactions and the
fact that $h_{125}$ arises as a
(composite)
pNG, but where also additional
(potentially stable) pNG can be
present~\cite{Frigerio:2012uc}. Such models
could resemble at low energy
a model with two Higgs doublets~\cite{Mrazek:2011iu,
DeCurtis:2016tsm,DeCurtis:2018zvh,Vieu:2018nfq}.

In this paper we study a singlet
extension of the 2HDM (S2HDM), where the
real part of the singlet field $\phi_S$
gives rise to a third Higgs boson, and
the imaginary part of $\phi_S$ gives
rise to the pNG DM, as discussed above,
while the terms of the
scalar potential incorporating the
doublet fields $\phi_{1,2}$ are identical
to the 2HDM with softly broken
$\mathbb{Z}_2$ symmetry.
In total, the physical scalar
particle spectrum consists of three
CP-even Higgs bosons $h_{1,2,3}$ that
are mixed with each other, two CP-odd
states $A$ and $\chi$ that do not mix
and where $\chi$ is the stable
DM candidate, and finally the charged
Higgs bosons $H^\pm$.
We focus on the 
S2HDM with type~II Yukawa structure
and the parameter space 
that gives rise to the so called
\textit{Higgs funnel} scenario,
i.e.\ resonant DM annihilation
via $s$-channel diagrams
mediated by the SM-like Higgs boson
$h_{125}$.
The possibility of relatively
light pNG DM
in a doublet extension of the SM
suggest an interesting interplay
of collider phenomenology and
astrophysics that can be 
constrained by various 
experimental requirements coming from flavour physics,
electroweak precision observables,
searches for additional scalars
and measurements of the properties
of $h_{125}$. 
Additionally, there are experimental
constraints related to the presence
of the DM candidate. In particular,
the limitation imposed by the fact that a 
too large relic abundance after thermal
freeze-out would overclose the universe
and indirect-detection limits coming
from the observation of dwarf spheroidal
galaxies (dSph) by the Fermi-LAT space
telescope~\cite{Fermi-LAT:2016uux}
play an important role.
We also take into account
several theoretical constraints that must be
 imposed in order to ensure
the validity of the
perturbative treatment of the theory
and the stability of the 
EW vacuum.
While the S2HDM was already studied
in regards to the DM phenomenology
and its cosmological
history in \citeres{Vieu:2018nfq,
Jiang:2019soj,Zhang:2021alu}, a
careful treatment of the S2HDM
taking into account the large number of
constraints has not been carried out yet.
In our analysis we will demonstrate that
the combined consideration of the
experimental and theoretical constraints
is crucial in order to make reliable
predictions for the phenomenology of
the S2HDM.

In this context, we explore benchmark
scenarios featuring 
pNG DM in the mass range
${40 \gev \leq m_\chi \leq 80 \gev}$.
This region is particularly relevant
from the experimental point of view,
since it belongs to 
parameter space of the S2HDM
where it is possible to accommodate
a sizable fraction of the DM relic
abundance, and, whenever this is the case,
the presence of the DM candidate
is currently, and even more so
in the new future, probed by DM
indirect detection experiments.
As a result,
important limitations on the parameter
space of the S2HDM will arise.
In this regard, it is interesting
to note that the corresponding
parameter space is
also suitable to realize
the excess of gamma rays from the galactic
center observed by the Fermi Large Area
Telescope (LAT)~\cite{Fermi-LAT:2017opo,
Fermi-LAT:2015sau}.
It has been argued that these observations
could be originated by DM annihilations
in the galactic center~\cite{Hooper:2010mq, Hooper:2011ti,
Hooper:2013rwa, Daylan:2014rsa,
Calore:2014xka, Zhou:2014lva, Abazajian:2014fta}
where a large concentration of DM
is expected to
reside~\cite{Navarro:1996gj, Kaplinghat:2013xca}.
However, they are also consistent
with an unresolved population of thousands 
of millisecond pulsars in the Galactic
bulge~\cite{Bartels:2015aea, Lee:2015fea, Calore:2015bsx}.
At the same time, the Alpha Magnetic
Spectrometer (AMS) \cite{AMS:2016oqu},
onboard the International Space Station, 
reported an excess over the expected
flux of cosmic ray antiprotons 
consistent with DM annihilating
into $b$-quark pairs with a similar 
range of DM masses~\cite{Cirelli:2014lwa,
Cuoco:2016eej, Cui:2016ppb, Cholis:2019ejx,
Lin:2019ljc, Carena:2019pwq}.
We will address the question whether the
DM candidate of the S2HDM can account for
the two cosmic-ray excesses, potentially
in combination with a Higgs boson at
roughly $96\gev$ that could give rise to
the so-called LEP excess in the $b \bar b$
final state and the CMS excess in the diphoton
final state, which was already investigated
in a singlet-extension of the SM
featuring pNG DM
in \citere{Cline:2019okt}. Therein, it was found that
the CMS excess requires new charged particles
in order to account for a sufficiently
strong signal. In contrast, here
we will follow the results of
\citere{Biekotter:2019kde} obtained
in the Next-to 2HDM (N2HDM)
and in models with
supersymmetry~\cite{Cao:2016uwt,Biekotter:2017xmf,
Domingo:2018uim,Hollik:2018yek,
Choi:2019yrv,Biekotter:2019gtq}.
It was shown that the presence of a second
Higgs doublet in addition to
a singlet scalar field allows for
an explanation of both collider
excesses without having
to rely on new charged states that
appear in the loops of the loop-induced
coupling of the possible Higgs
boson at $96\gev$ in order to enhance
its diphoton rate.

This paper is organised as follows. In 
\refse{secs2hdm}, the model is presented. In \refse{secconsts}, we describe the relevant
experimental and theoretical constraints
that we apply to its parameter space.
In \refse{secnumanal}, we describe
the genetic algorithm that was used
to scan the parameter space and
to determine the parameter points 
that pass the various theoretical
and experimental requirements. In 
\refse{secnumanal1}, we explore the 
Higgs funnel region after imposing
the previously described constraints
and disregarding the 
explanation of the excesses at LEP and CMS,
whereas in \refse{secnumanal2} we
additionally demand that the collider excesses
are accommodated. Finally, we
conclude in \refse{conclu}.

\section{The S2HDM}
\label{secs2hdm}
The scalar sector of the S2HDM consists of
two SU(2) doublets and a complex gauge
singlet field, which can be expressed as
\begin{equation}
\phi_1 =
\begin{pmatrix}
\phi_1^+ \\
\left( \rho_1 + \mathrm{i} \sigma_1 \right)
    / \sqrt{2}
\end{pmatrix} \ , \quad
\phi_2 =
\begin{pmatrix}
\phi_2^+ \\
\left( \rho_2 + \mathrm{i} \sigma_2 \right)
    / \sqrt{2}
\end{pmatrix} \ , \quad
\phi_S =
\left( \rho_S + \mathrm{i} \chi \right)
    / \sqrt{2} \ ,
\end{equation}
where the pseudoscalar component $\chi$
gives rise to
the DM candidate of the model.
Assuming the absence of
explicit CP violation, the
scalar potential of the S2HDM
is given by
\begin{align}
V &=
\mu_{11}^2 \left( \phi_1^\dagger \phi_1 \right)
+ \mu_{22}^2 \left( \phi_2^\dagger \phi_2 \right)
- \mu_{12}^2 \left( \left( \phi_1^\dagger \phi_2 \right)
    + \left( \phi_2^\dagger \phi_1 \right) \right)
+ \frac{1}{2} \mu_S^2 \left| \phi_S \right|^2
- \frac{1}{4} \mu_\chi^2 \left( \phi_S^2
    + \left(\phi_S^*\right)^2 \right) \notag \\
&+ \frac{1}{2} \lambda_1 \left( \phi_1^\dagger \phi_1 \right)^2
+ \frac{1}{2} \lambda_2 \left( \phi_2^\dagger \phi_2 \right)^2
+ \lambda_3 \left( \phi_1^\dagger \phi_1 \right)
    \left( \phi_2^\dagger \phi_2 \right)
+ \lambda_4 \left( \phi_1^\dagger \phi_2 \right)
    \left( \phi_2^\dagger \phi_1 \right) \notag \\
&+ \frac{1}{2} \lambda_5 \left(
    \left( \phi_1^\dagger \phi_2 \right)^2
        + \left( \phi_2^\dagger \phi_1 \right)^2
\right)
+ \frac{1}{2} \lambda_6 \left( |\phi_S|^2 \right)^2
+ \lambda_7 \left( \phi_1^\dagger \phi_1 \right) | \phi_S |^2
+ \lambda_8 \left( \phi_2^\dagger \phi_2 \right) | \phi_S |^2
\ .
\label{eqscalpot}
\end{align}
Here the terms that exclusively involve
the doublet fields are
identical to the scalar potential of the
2HDM, where a $\mathbb{Z}_2$ symmetry
defined by the transformations
$\phi_1 \to \phi_1$, $\phi_2 \to - \phi_2$
and $\phi_S \to \phi_S$ is only
softly broken by the terms proportional
to $\mu_{12}$.
After the
generalization of the $\mathbb{Z}_2$
to the Yukawa sector,
which is unchanged in the S2HDM
compared to the 2HDM,
the $\mathbb{Z}_2$ symmetry gives
rise to the absence of flavour-changing
neutral currents at tree level.
Depending on the assigned charges
of the fermions, this results
in the usual four Yukawa types
(see \citere{Branco:2011iw} for details). We will
focus on the so-called type~II, familiar
from supersymmtric BSM scenarios, in which
$\phi_2$ is only coupled to up-type quarks,
and $\phi_1$ is only coupled to
down-type quarks and the charged leptons.
The remaining terms of the scalar
potential $V$ involve
the singlet field $\phi_S$
and respect a global
U(1) symmetry, except for the
term proportional to $\mu_\chi^2$.
This term softly breaks the U(1)
symmetry, thus
providing a non-zero mass for the
pNG DM.

Without loss of generality,
the field configuration of the vacuum
can be expressed as
\begin{equation}
\langle \phi_1 \rangle =
\begin{pmatrix}
0 \\
v_1 / \sqrt{2}
\end{pmatrix} \ , \quad
\langle \phi_2 \rangle =
\begin{pmatrix}
v_{\mathrm{C}} / \sqrt{2} \\
\left( v_2 + \mathrm{i} v_{\mathrm{CP}}
    \right) / \sqrt{2}
\end{pmatrix} \ , \quad
\langle \phi_S \rangle =
\left( v_S + \mathrm{i} v_{\mathrm{DM}}
\right) / \sqrt{2} \ ,
\end{equation}
where we made use of the fact that
redundant degrees of freedom related
to the gauge symmetries
can be removed
via the
gauge transformations.
We will focus on
the case in which the EW symmetry is
broken by non-zero values of $v_1$ and $v_2$, and
an accidental $\mathbb{Z}_2$
symmetry, under which $\rho_S$ changes the sign,
is broken
by $v_S > 0$. The
charge-breaking
vev $v_{\mathrm{C}}$, the CP-breaking vev
$v_{\mathrm{CP}}$, and $v_{\mathrm{DM}}$
are considered to be vanishing,
noting that a non-zero value
of $v_{\rm DM}$ would give rise to decays
of the DM candidate $\chi$.
In our numerical analysis, we verify
for each parameter point whether there
exists a global minimum of the potential
with $v_1,v_2,v_S > 0$ and $v_{\mathrm{C}},
v_{\mathrm{CP}}, v_{\mathrm{DM}} = 0$.
Otherwise, we
remove such a point,
since it potentially features an
EW vacuum that is short-lived compared
to the age of the universe
(see \refse{secconstraintstheo} for details).
In order to make a connection to the
SM and the 2HDM we define the parameters
$v^2 = v_1^2 + v_2^2 \sim (246\gev)^2$
and $\tan\beta = v_2 / v_1$.

Assuming the EW vacuum configuration as described above,
the CP-even fields $\rho_{1,2,S}$ mix, giving rise
to the mass eigenstates $h_{1,2,3}$, where throughout
this paper the mass hierarchy ${m_{h_1} < m_{h_2}
< m_{h_3}}$ will be assumed.
The mixing in the CP-even sector can be written
in terms of an orthogonal transformation given by
a matrix $R$, such that
\begin{equation}
\begin{pmatrix}
h_1 \\ h_2 \\ h_3
\end{pmatrix} =
R \cdot
\begin{pmatrix}
\rho_1 \\ \rho_2 \\ \rho_S
\end{pmatrix} \ , \text{ with }
\label{mixingmatrix}
R=
\begin{pmatrix}
c_{\alpha_1}c_{\alpha_2} &
  s_{\alpha_1}c_{\alpha_2} &
    s_{\alpha_2} \\
-(c_{\alpha_1}s_{\alpha_2}s_{\alpha_3}+s_{\alpha_1}c_{\alpha_3}) &
  c_{\alpha_1}c_{\alpha_3}-s_{\alpha_1}s_{\alpha_2}s_{\alpha_3}  &
    c_{\alpha_2}s_{\alpha_3} \\
-c_{\alpha_1}s_{\alpha_2}c_{\alpha_3}+s_{\alpha_1}s_{\alpha_3} &
-(c_{\alpha_1}s_{\alpha_3}+s_{\alpha_1}s_{\alpha_2}c_{\alpha_3}) &
c_{\alpha_2}c_{\alpha_3}
\end{pmatrix}~,
\end{equation}
where $- \pi / 2 \leq \alpha_1, \alpha_2, \alpha_3
\leq \pi / 2$ are the three mixing angles, and
we use the short-hand notation $s_x = \sin x$, $c_x = \cos x$.
The charged scalar sector remains unchanged compared
to the 2HDM. It contains two physical charged
Higgs bosons $H^\pm$ with mass $m_{H^\pm}$
and the charged Goldstone bosons
related to the gauge symmetries.
The pseudoscalar components
$\sigma_1$, $\sigma_2$ and $\chi$
form a neutral Goldstone boson
and two physical states $A$ and $\chi$
with masses $m_A$ and $m_\chi$, respectively. 
Here it is important to note that
the remnant $\mathbb{Z}_2$ symmetry
that is present when $v_{\rm DM} = 0$,
preventing the DM candidate $\chi$ from
decaying, also forbids
the mixing between $\chi$ and $A$.
Hence, the pseudoscalar $A$ has effectively
the same couplings to the fermions as the
one of the 2HDM.

Given the definitions of the parameters
as defined above, it is possible to replace
most of the parameters of the scalar potential
shown in \refeq{eqscalpot} by more physically meaningful
parameters. In our numerical analysis, we will
sample the parameter space of the
type~II S2HDM in
terms of the parameters
\begin{equation}
m_{h_{1,2,3}} \ , \quad
m_A \ , \quad
m_{H^\pm} \ , \quad
m_\chi \ , \quad
\alpha_{1,2,3} \ , \quad
\tan\beta \ , \quad
M = \sqrt{\mu_{12}^2 / \left(
    s_\beta c_\beta \right) } \ , \quad
v_S \ .
\label{eqparas}
\end{equation}
The relations between the parameters shown
in \refeq{eqparas} and the Lagrangian parameters
of the potential in \refeq{eqscalpot}
are given in \refap{secparas}.

\section{Constraints}
\label{secconsts}

In this section we briefly discuss
the constraints on the parameter space
of the S2HDM that we applied in our
analyses. In each case, we also illustrate
the impact of the constraint in order
to give an impression on their relevance
for our numerical discussion.
Some of the constraints, such as the ones
arising from demanding a stable EW vacuum,
are similar
to the corresponding constraints known
from the Next-to 2HDM (N2HDM)~\cite{Muhlleitner:2016mzt}.
However, there are also
important differences which we will point out below.

\subsection{Theoretical constraints}
\label{secconstraintstheo}
The parameters that appear in the scalar
potential of the S2HDM are subject to
important theoretical constraints.
These constraints
ensure the stability of the EW vacuum for
a given parameter point, and they exclude
parameter values which give rise to
parameter points that could not be treated
perturbatively.

\subsubsection*{Boundedness-from-below}
We apply bounded-from-below (BfB)
conditions on the tree-level scalar
potential,
which determine whether the potential
is bounded from below along all field directions.
Due to the fact that the quartic
part of the potential $V$ is unchanged compared
to the N2HDM, we can apply the same conditions that
were found for the N2HDM~\cite{Klimenko:1984qx,
Muhlleitner:2016mzt}. We exclude
all parameter
points from our analyses
that do not feature
a scalar potential that is BfB. It
was shown in \citere{Staub:2017ktc}
that large loop corrections
can transform a bounded tree-level 2HDM potential into
an unbounded one, potentially
destabilizing
the EW vacuum. This effect is expected to
be also present in the S2HDM, such that
our tree-level analysis of the boundedness
could permit potentially unphysical
parameter points. However,
the possibility of loop corrections changing
the boundedness of the potential
was shown to be present only
in regions of the parameter space with splittings
between $m_A$, $m_H$ and $m_{H^\pm}$
larger than $\sim 250\gev$, where
consequently
large quartic couplings
are present~\cite{Staub:2017ktc},
which then give rise to the corrections.
In our analysis, we demand an upper limit
of $200 \gev$
on the splitting of the heavy Higgs-boson masses 
compared to the mass scale $M$
defined in \refeq{eqparas} (see
also the discussion below), such that we
expect that the boundedness
of the potential, and therefore the stability
of the EW vacuum,
are not to be severely
affected by the loop corrections.

\subsubsection*{EW vacuum stability}
In the next step, we demanded that the EW minimum
as described in \refse{secs2hdm} is the global minimum
of $V$,
such that no vacuum decay into other unphysical
minima is possible, and
the stability of the EW vacuum is guaranteed.
To test whether the EW minimum is the global
one, we first determined all extrema
of $V$ by solving the stationary
conditions $\partial V / \partial (v_1, v_2,
v_{\mathrm{C}},v_{\mathrm{CP}},v_S,v_{\mathrm{DM}})$,
where we used the code
\texttt{Hom4PS-2}~\cite{Lee:hom4ps2} to solve the
system of polynomial equations. For each
extrema we calculated the value of $V$ in this point
of field space.
One can conclude that, if for any of the extrema the value
of $V$ is smaller than for the field values of the
EW vacuum,
the EW minimum is not the global minimum
of $V$. In this case,
the EW vacuum is potentially short-lived
compared to the age of the universe, such that the
corresponding parameter point might be unphysical,
and we rejected it from the analyses.\footnote{A
(zero temperature)
calculation of the lifetime of an unstable EW vacuum
shows that in some cases the EW vacuum can be considered
to be sufficiently long-lived, even though there are
deeper minima present, such that a parameter
point with a non-global EW minimum could
still be viable (see \citere{Ferreira:2019iqb} for an
N2HDM analysis). However, in such cases it is
still unclear whether the universe would have adopted
the (meta-stable) EW vacuum at some point within
the thermal history of the universe, or would
have rather transitioned into a deeper
unphysical minimum. The analysis of
the thermal history of the scalar potential of
the S2HDM is beyond the scope of this paper
(see \citere{Biekotter:2021ysx}
for an N2HDM analysis), such
that we demand the most conservative constraint, i.e.\
excluding all parameter points for which the EW minimum
is not the global minimum of the potential.}

\subsubsection*{Perturbative unitarity}
In order to verify whether a perturbative treatment
of the model is valid for a given
parameter point,
we applied the so-called tree-level perturbative
unitarity constraints. We derive these constraints
by calculating the scalar $2 \times 2$ scattering
matrix in the high-energy limit, in which only
the quartic contact interactions are relevant.
The precise form of the conditions is given in
\refap{apppert} for completeness. They set upper limits on the
absolute values of the parameters $\lambda_i$ and
combinations thereof, such that they are particularly
relevant when there are large mass splittings
between the heavy doublet states $A$, $H^\pm$
and one of the scalars $h_i$. Due to the fact that
compared to the N2HDM the only additional degree
of freedom is the CP-odd component of the singlet
field $\phi_S$, the perturbativity conditions are
in most parts very similar to the N2HDM
conditions~\cite{Muhlleitner:2016mzt}.
However, an important difference is that
an additional condition on the singlet self-coupling
of the form $| \lambda_6 | < 8 \pi$ appears.
In addition, the constraints related to scattering
amplitudes involving the singlet field components
and the field components
of the doublet fields
(see \refeq{eqroots})
are modified with respect to the N2HDM.

\subsubsection*{Energy scale dependence of the theoretical constraints}
Both the perturbative unitarity constraints and
the BfB conditions are in many
analyses of the 2HDM or its extensions
applied exclusively at a certain energy
scale.
However, it is known that the model parameters
obtain an intrinsic energy scale dependence
due to radiative corrections, which is governed
by their evolution
under the running group equations (RGE).
It is therefore possible that even though at the initial
scale, here assumed to be $\mu = v$
such that the input scale of
the parameters corresponds to the
EW scale, a parameter
point passes the theoretical constraints, the point
becomes unphysical at larger energy scales $\mu > v$.
Due to the fact that the pertubativity conditions
allow for values of $|\lambda_ i| > 1$,
the energy range $v \leq \mu \leq \mu_v$ in which
the theoretical constraints are fulfilled can be
small, with values of $\mu_v$ even smaller than
the energy scales that are probed at the LHC.
In consequence, we apply the
previously described theoretical constraints
taking into account the energy-scale dependence
of the parameters, utilizing the two-loop $\beta$-functions
of the S2HDM and demanding that the theoretical
constraints are respected up to a certain
energy scale $\mu_v$.
The $\beta$-functions for the S2HDM
were obtained with the help
of the public code \texttt{SARAH}~\cite{Staub:2013tta,
Schienbein:2018fsw}, solving the general expressions
known in the literature~\cite{Machacek:1983fi,
Machacek:1984zw,Machacek:1983tz}.
We also calculated the $\beta$-functions with the code
\texttt{PyR@TE 3}~\cite{Sartore:2020gou} to be able to cross
check the expressions and found exact agreement.
We discarded a parameter point
when the scale $\mu_v$ at which the scalar potential
becomes unbounded or at which the perturbative
unitarity constraints are violated is
below $1\tev$, which was also chosen as the upper
limit on the Higgs-boson masses in the
numerical discussion (see \refse{secnumanal}).

\subsection{Experimental constraints}
\label{secexpconstr}
The S2HDM offers a rich phenomenology that
can be probed experimentally by various means.
The corresponding experimental (null)-results
give rise to numerous constraints that have
to be taken into account. We start by
discussing the constraints related to the
Higgs sector of the model. Subsequently,
we describe the manner in which the constraints
from measurements from DM experiments were
taken into account.

\subsubsection*{Searches for additional scalars
and properties of
\texorpdfstring{$h_{125}$}{h125}}
Regarding the Higgs phenomenology of the
model, we used the public code
\texttt{HiggsBounds v.5.9.0}~\cite{Bechtle:2008jh,
Bechtle:2011sb,Bechtle:2013gu,Bechtle:2013wla,
Bechtle:2015pma,Bechtle:2020pkv}
to test the parameter points against
a large number
of cross-sections limits from
direct searches for Higgs bosons at LEP,
the Tevatron and the LHC. For each Higgs
boson, \texttt{HiggsBounds} selects the
potentially most sensitive experimental
search based on the expected limits.
For the selected searches, the code then
compares the predicted cross sections
against the observed upper limits on
the $95\%$ confidence level and excludes
a parameter point whenever the theoretical
prediction lies above the experimental
limit for one of the Higgs bosons.

Regarding the discovered Higgs boson
at $125\gev$, we use the public code
\texttt{HiggsSignals v.2.6.1}~\cite{Bechtle:2013xfa,
Stal:2013hwa,Bechtle:2014ewa,Bechtle:2020uwn}
to verify whether an  
S2HDM parameter point
features a particle $h_i$ that resembles
the properties of the discovered particle
$h_{125}$
within the experimental uncertainties.
\texttt{HiggsSignals} performs a $\chi^2$-analysis
confronting the predicted signal rates against
the experimentally measured signal rates.
In our more general parameter scan discussed
in \refse{secnumanal1}, we applied as constraint that
the resulting $\chi^2$ value (called $\chi^2_{125}$
in the following) fulfills $\chi^2_{125} \leq
\chi^2_{\mathrm{SM},125}  + 5.99$, where
$\chi^2_{\mathrm{SM},125} = 84.41$ is the fit
result assuming a SM Higgs boson at $125\gev$,
and where the allowed penalty of $5.99$ corresponds
to a $95\%$ confidence interval for
two-dimensional parameter
distributions.\footnote{See
\citere{Bechtle:2020uwn} for
details on the interpretations of the $\chi^2$
analysis of \texttt{HiggsSignals}.}
In \refse{secnumanal2}, in which we aim for accommodating
the collider excesses observed at about $96\gev$,
we combine the value of $\chi^2_{125}$ obtained
from \texttt{HiggsSignals} with a value $\chi^2_{96}$ that
quantifies the fit to the excesses. The precise
criterion applied will be given in \refse{secnumanal2}.

\texttt{HiggsBounds} and \texttt{HiggsSignals}
require as input effective coupling coefficients,
which are defined as the couplings of
the physical scalars
normalized to the coupling of a SM
Higgs boson of the same mass.
With the help of these coupling
coefficients, the codes compute the
relevant cross sections for the
scalars by rescaling the predictions
for a hypothetical SM Higgs boson.
In the S2HDM, the coupling coefficients
can be expressed in terms of $\tan\beta$ and
(for the states $h_i$) in terms of the mixing
angles $\alpha_i$. The precise expressions are
identical to the N2HDM expressions and can
be found in \citere{Muhlleitner:2016mzt}.
Moreover, the user has to provide
the branching ratios of the Higgs bosons.
We calculated these in two steps. 
First, we
used the public \texttt{Fortran} code
\texttt{N2HDECAY}~\cite{Djouadi:1997yw,
Butterworth:2010ym,Djouadi:2018xqq,
Muhlleitner:2016mzt,Engeln:2018mbg}
implemented in the \texttt{anyhdecay}
\texttt{C++} library to calculate the decay widths
of $h_i$, $A$ and $H^\pm$ for decays into
SM particles and for cascade decays with
one or two Higgs bosons in the final
state.\footnote{The \texttt{anyhdecay}
library can be downloaded at
\url{https://gitlab.com/jonaswittbrodt/anyhdecay}.}
In a second step we calculated the decay
widths for the invisible decay into a
pair of $\chi$ as described below
(see \refeq{eqgaminv}).
We finally divided each partial decay widths
by the total widths to obtain the branching
ratios for each possible decay mode.

In addition to the global constraints on
the measured signal rates of $h_{125}$,
the S2HDM can also be probed via possible
decays of $h_{125}$ into a pair of
DM particles $\chi$ with a mass
of $m_\chi < 125 / 2 \gev$.
At leading order, the partial decay width
of the invisible decay is given by
\begin{equation}
\Gamma_{\rm inv} \left( h_i \to \chi \chi \right) =
\frac{1}{32 \pi m_{h_i}}
\sqrt{1 - \frac{4 m_{\chi}^2}{m_{h_i}^2}}
\left( R_{i1} \lambda_7 v_1 +
R_{i2} \lambda_8 v_2 +
R_{i3} \lambda_6 v_S \right)^2 \ ,
\; m_{h_i} > 2 m_\chi^2 \ .
\label{eqgaminv}
\end{equation}
The most recent upper limit on
the branching ratio of
the invisible decay
$\mathrm{BR}_{\rm inv}$
of $h_{125}$ has recently been reported by ATLAS
and is given by $\mathrm{BR}_{\rm inv} < 0.11$
at the $95\%$ confidence level~\cite{ATLAS:2020kdi}.
We applied this limit as additional
constraint complementary to the \texttt{HiggsSignals}
analysis.
However, as will be demonstrated in \refse{secnumanal1},
in most cases parameter points
with sizable values of the corresponding
branching ratios $\mathrm{BR}_{\rm inv}$ are already
excluded by the global constraints on the
measured signal rates of $h_{125}$, since the
additional decay mode $h_{125} \to \chi \chi$
suppresses the ordinary decays of $h_{125}$
into SM final states.

\subsubsection*{Electroweak precision observables}
Further constraints originating from the presence
of the BSM Higgs bosons that have to be taken into
account are the ones related to EW precision
observables (EWPO). Since the S2HDM extends
the SM particle content exclusively
by scalar states, one can
to a very good approximation apply the formalism
of the oblique parameters $S$, $T$ and
$U$~\cite{Peskin:1990zt,Peskin:1991sw}.
These parameters are theoretically defined by
loop corrections to the gauge-boson self-energies,
in which the BSM particles appear in the loops,
giving rise to modifications of the values of
the oblique parameters compared to the SM.
In order to predict the oblique parameters,
we applied the general expressions at the
one-loop level from \citeres{Grimus:2007if,
Grimus:2008nb} to the S2HDM.
Experimentally, $S$, $T$ and $U$
are constrained via global fits to the EWPO, where
we utilize here the results (including their
uncertainties) found in \citere{Haller:2018nnx}.
In 2HDM-like extensions of the SM,
the most sensitive parameter
is the $T$ parameter, whereas the modifications of
the $U$ parameter in practically all cases are
orders of magnitude smaller than the experimental
sensitivity, and we explicitly checked this
to hold in the S2HDM.\footnote{We found that
at the one-loop level the
theoretical predictions for $S$, $T$ and $U$
in the S2HDM and the N2HDM (given the
same values of $m_{h_i}$, $m_A$ and
$m_{H^\pm}$) are identical,
because they do not depend on the additional
state $\chi$ of the S2HDM as long as $v_{\rm DM}
= 0$.}
We therefore performed a
two-dimensional $\chi^2$ test regarding $S$ and
$T$, written as $\chi^2_{ST}$ in
the following, and discarded
parameter points for which
the predicted values 
were not in agreement with
the experimental fit
result~\cite{Haller:2018nnx} at
the $95\%$ confidence level.
This gives rise to the requirement
$\chi^2_{ST} \leq 5.99$.
The $T$ parameter is sensitive to the
breaking of the custodial symmetry.
As a result, one finds strong exclusions
when there is a sizable mass splitting between
the states $A$, $H^\pm$ and (depending on the
doublet-admixture) one of the
\textit{heavy} CP-even state $h_2$ or $h_3$.

\subsubsection*{Flavour-physics observables}
Also the theoretical predictions for flavour-physics
observables are modified compared to
the SM via  contributions
from the additional Higgs bosons of the S2HDM.
In particular, the presence of the charged
Higgs bosons gives rise to robust constraints
in the parameter plane of $\tan\beta$
and $m_{H^\pm}$.
Since there are no public results
for the theoretical predictions
in singlet extensions of the 2HDM for some
of the most relevant flavour observables,
we simply applied hard cuts on the
ranges of $\tan\beta$ and $m_{H^\pm}$ in our
numerical analysis, where the cuts were determined
by assuming that the exclusion regions known
from the 2HDM are not severely
modified by the presence of the additional
field of the S2HDM, which we expect
to be the case due to the singlet nature
of this field.
Consequently, working in the type~II S2HDM,
we set lower limits of $\tan\beta > 1.5$
and of $m_{H^\pm} > 600\gev$ in order
to not be in conflict with constraints from
radiative and (semi-)leptonic $B$ meson decays
and from their mixing
frequencies~\cite{Haller:2018nnx}.\footnote{A
more recent result suggests a lower limit
of $m_{H^\pm} > 800\gev$ in the type~II 2HDM
from the measurement of the radiative
$B$ meson decay~\cite{Misiak:2020vlo}, whereas
\citere{Bernlochner:2020jlt}
claims that theoretical uncertainties
might have been underestimated in the
literature, potentially
giving rise to a weaker lower limit. We emphasize
that the conclusions drawn from our
numerical analysis do not depend on the precise
value of the lower limit chosen for $m_{H^\pm}$.}

\subsubsection*{Dark matter observables}
We now turn to the experimental constraints
that are related to the presence of the
dark matter candidate $\chi$.
The most important limitation arises from
the fact that a too large relic abundance
of $\chi$ after thermal freeze-out would
overclose the universe. The currently most
precise measurement of today's
DM relic abundance
$\Omega h^2$ is given by surveying
the cosmic microwave background by
the Planck satellite, leading to a
measurement of $(\Omega h^2)_{\rm Planck} = (0.119 \pm
0.003)$~\cite{Planck:2018vyg}.
We will use this value as an upper limit
on the relic abundance of $\chi$ in our
analysis, taking into consideration that
in case the relic abundance of $\chi$ is
smaller than $(\Omega h^2)_{\rm Planck}$
there is room for additional (particle or
astrophysical) contributions to the relic
abundance. We focus the analysis
on the \textit{Higgs funnel region}
with DM masses of $40 \leq m_\chi \leq 80$,
where there are good prospects
to be able to explain most (or
all) of the observed DM relic abundance
via the thermal freeze-out of~$\chi$~\cite{Barger:2008jx,
Cline:2019okt,Arina:2019tib,Jiang:2019soj,Zhang:2021alu}.
For the theoretical prediction of the relic
abundance, we wrote an S2HDM modelfile for
the \texttt{Mathematica} package
\texttt{FeynRules v.2}~\cite{Christensen:2008py,
Christensen:2009jx,Alloul:2013bka}, which we utilized
to obtain a \texttt{CalcHEP}~\cite{Belyaev:2012qa}
input for the public code
\texttt{MicrOMEGAs v.5}~\cite{Belanger:2018ccd} written
in \texttt{C} and \texttt{Fortran}.
With this input, \texttt{MicrOMEGAs} is capable
of calculating the relic abundance and the
freeze-out temperature, where for the computation
of the annihilation cross section all $2 \times 2$
processes and also processes with off-shell
vector bosons in the final state are
taken into account.

As already pointed out in \refse{secintro},
one of the attractive features of the S2HDM
is that due to the pNG nature of the DM particle
the cross sections for the scattering of $\chi$
on nuclei vanish at leading order
in the limit of vanishing
momentum transfer~\cite{Jiang:2019soj},
such that at this order
direct-detection experiments are not
sensitive to the presence of $\chi$.
In addition, it was shown
in models with a single Higgs doublet field
and a complex singlet field
that the loop contributions to the direct-detection
cross sections are small,
and the predicted direct-detection scattering cross
sections remain far below the current (and
near future)
sensitivity of direct-detection
experiments~\cite{Azevedo:2018exj,
Ishiwata:2018sdi,Glaus:2020ihj}.
In the type~II S2HDM
the masses of
the additional doublet particle states
$H(=h_2 \text{ or } h_3)$, $A$ and $H^\pm$
are required to be substantially larger than
the DM masses $m_{\chi}$
considered in our analysis
(see discussion above), such that we can safely
assume that the relevant loop corrections
to the DM-nuclei scattering cross
sections are captured
by the pNG DM model with only one
Higgs doublet. One should note also
that for light DM an additional suppression
of the scattering cross section given by
the factor $(m_\chi / m_{h_i})^4$ is present,
which reflects the pNG nature of $\chi$
and the fact that the loop corrections
vanish in the limit
$m_\chi \to 0$~\cite{Gross:2017dan}.
Consequently, in our scenario the
additional
loop corrections arising from the
presence of the second Higgs doublet
are even smaller than the corrections
known from the pNG DM
model with one Higgs doublet, and thus
there are no
relevant constraints from
direct-detection experiments
that have to be taken into account in our
analysis (see also discussions in
\citeres{Jiang:2019soj,Zhang:2021alu}).

On the other hand, constraints from DM
indirect-detection experiments are
important, in particular in the Higgs funnel
region investigated here, in which $\chi$ mainly
annihilates into $b$ quark pairs, typically via
$h_{125}$ in the $s$-channel.
The most stringent constraints on the annihilation
cross sections of DM come from
the observation of dwarf spheroidal galaxies (dSph)
by the Fermi-LAT space telescope \cite{Fermi-LAT:2016uux}. In order to account
for these constraints, we used \texttt{FeynRules}
to generate UFO~\cite{Degrande:2011ua}
model files for the S2HDM, which were then 
used as input for the public code
\texttt{MadDM v.3}~\cite{Backovic:2013dpa,
Ambrogi:2018jqj}. \texttt{MadDM} is a plugin
for \texttt{MadGraph5_atMC v.3.1.1}~\cite{Alwall:2014hca}
that can be used to
compute the relevant velocity-averaged
annihilation cross sections
$\langle \sigma v_{\rm rel} \rangle_{b \bar b}$,
and to subsequently compare the theoretical
predictions to the upper limits on
the velocity weighted cross section
for DM particles annihilating into $b\bar{b}$
final states
from the Fermi measurements of gamma rays from
dSph at the 95 \% CL.\footnote{We also computed
$\langle \sigma v_{\rm rel} \rangle$ for other
two body final states. However, for the range of
$m_\chi$ investigated here the $b$ quark final state
was always the dominant one. In addition,
we applied the so-called fast mode of \texttt{MadDM}
in order to reduce the duration of the calculation.
We checked for several parameter points of
our scans that
the difference between the values of
$\langle \sigma v_{\rm rel} \rangle$ in the
fast and the precise mode are very similar.}
The Fermi-LAT collaboration utilizes a
likelihood analysis to fit the spectral
and spatial features of dSphs to obtain upper
limits on the annihilation cross section 
as a function of the DM
mass~\cite{Fermi-LAT:2016uux}.
The analysis accounts
for point-like sources from the latest LAT
source catalog, models
the galactic and isotropic diffuse
emission, and incorporates
uncertainties in the determination
of astrophysical $J$-factors,
which depend on both the DM density profile
and the distance.
The observed limits are sensitive to 
the determination method of the 
$J$-factors. In \citere{Fermi-LAT:2016uux} an
evaluation of the uncertainties arising
from targets lacking measured
$J$-factors was performed.
Using only predicted J-factors
for the whole sample weakened the
observed limits by a factor of about
2 to 3,
depending on the choice of $J$-factor
uncertainty, with respect
to the limits obtained by using both
predicted and measured $J$-factors.
Considering these uncertainties will be 
important
for the discussion of the tension between
the constraints coming from dSph and
the gamma-ray excesses and anti-protons
measured from the galactic center,
as will be demonstrated in
\refse{secnumanal1}.

For the comparison between the predicted annihilation
cross section and the Fermi bounds from
dSph observations, we rescaled
(when not explicitly said otherwise)
the cross sections with a factor
\begin{equation}
\xi^2 = \left(
\frac{\Omega h^2}{( \Omega h^2 )_{\rm Planck}}
\right)^2 \ ,
\label{eqxiani}
\end{equation}
in order to account for the suppression of
today's annihilation cross section of $\chi$
due to the smaller number density when
the relic abundance of DM is not
made up completely out of $\chi$.\footnote{For
the calculation of $\xi$ we used the value
of $\Omega h^2$ as predicted by \texttt{MicrOMEGAs}.
In principle, also \texttt{MadDM} can calculate
the relic abundance. However, by default
\texttt{MadDM} does not take into account
the contributions to the annihilation cross
section with off-shell gauge bosons, which are
relevant in our analysis. Moreover, the calculation
of the relic abundance is much faster
using \texttt{MicrOMEGAs.}}
We also point out that the
velocity-averaged annihilation cross
sections 
can be considered here
to be velocity-independent
in the non-relativistic limit
to a very good approximation,
since in our scan range of
$m_\chi$ they are dominantly
generated via diagrams with $s$-channel exchange
of either $h_1$ or $h_2$~\cite{Bauer:2017qwy}.
Nevertheless, we calculated
$\langle \sigma v_{\rm rel} \rangle$
with different relative velocities
$v_{\rm rel}$ for
the comparison against the Fermi-LAT dSph
constraints, on the one hand,
and for the comparison
against the preferred regions regarding
the gamma-ray and the anti-proton
excesses, on the other hand.
In both cases we used the default
values of \texttt{MadDM}, which are $v_{\rm rel} =
2 \cdot 10^{-5}$ for the DM
in dSph and $v_{\rm rel} = 10^{-3}$
for DM in the center of the galaxy as relevant
for the galactic center excess. In agreement with our
expectation, the differences of the
annihilation cross sections for the
two values of $v_{\rm rel}$ stayed below
a few percent 
and are not
relevant for our discussion.

\section{Numerical analysis}
\label{secnumanal}
As was already discussed in
\refse{secintro}, we divide our numerical
analysis of the type~II S2HDM
into two parts. In the
first part discussed
in \refse{secnumanal1}, we will demonstrate
in a broad parameter scan how the
Higgs funnel region with
$40\gev \leq m_\chi \leq 80\gev$ is
affected by the various theoretical
and experimental constraints.
Here the DM particle $\chi$ is
the lightest BSM state, and $h_{125} = h_1$
is the lightest of the three CP-even
Higgs bosons $h_i$.
We will describe in detail the predictions
for the DM relic density and its interplay
with the Higgs sector of the model.
In addition, we investigate whether
the annihilation of $\chi$ in this
scenario could give rise to the
cosmic-rays anomalies from observations
of the spectra of cosmic rays
coming from the center of the galaxy.
We emphasize at this point that
due to the large mass gap between
the DM mass $m_\chi$ studied here and the masses of
the heavy scalar states $h_3$, $A$
and $H^\pm$ in the type~II S2HDM,
the predictions for the DM relic abundance 
and today's DM annihilation cross section
mainly depend on the couplings of $\chi$
to the SM-like Higgs boson and (when present)
the light singlet-like Higgs boson.
Accordingly, the properties of
the DM sector will be similar compared to
the predictions from the pNG DM model with
only one Higgs doublet, because additional
annihilation processes involving the heavier
states (see also the discussion in
\refse{secintro}) do not play a role.
However, differences between both models
can still arise due to the richer mixing
patterns of the states $h_i$ in the S2HDM,
where the mixing angles $\alpha_{1,2,3}$
enter the couplings of $h_{i}$ to $\chi$.

In the second part of our
analysis, discussed in
\refse{secnumanal2}, we focus on the parameter space
in which at the same time the collider
excesses at about $96\gev$ could be
accommodated. Consequently, here
the presence of a singlet-like
Higgs boson $h_{96} = h_1$ with $m_{h_1} = 96\gev$
is enforced as an additional constraint
on the parameter space. As a result,
the SM-like Higgs boson
$h_{125}$ is the second lightest Higgs boson $h_2$,
and its mixing with $h_{96}$ is subject
to the constraints from the LHC
measurements of the signal rates
of $h_{125}$.
Going beyond the discussion of the
collider phenomenology and the
excesses at $96\gev$, we will illustrate
in detail how the presence of $h_{96}$ has
also important consequences for the
DM phenomenology in the Higgs funnel,
in particular giving rise to
a second $s$-channel contribution to
the thermal freeze-out cross section
and today's annihilation cross section
relevant for DM indirect-detection experiments.

In both parameter scan presented in the
following, we sampled the multi-dimensional
parameter space of the model
utilizing a genetic algorithm.
In contrast to random
or uniform (grid)-scans of the model parameters,
a genetic algorithm has the advantage that
it focuses on the relevant
parameter region by minimizing a so-called
loss function, which has to be suitably defined
in each case. The definition of the loss functions
used in both parts of our analysis will be given in
\refse{secnumanal1} and \refse{secnumanal2}.
Apart from the loss function, the properties of
the genetic algorithm applied were
in large parts identical in both scans.
For the interested reader
we briefly describe the main design choices here,
where we made use of the public
\texttt{python} package
\texttt{DEAP}~\cite{DEAP_JMLR2012}
to perform the algorithm.

The algorithm starts by generating an
initial sample (also called population)
of $50\,000$ parameter points. Each parameter point
(also called individual) is defined by a list
of 14 numbers (also called attributes
or genes), where each number of this list
defines a value of one of the
model parameters within
a given parameter range.
The population is then subject to an evolution
including the three steps: selection, mating
and mutation. These three steps
are performed in a loop for a total number
of $N$ cycles (also called generations), such that
each cycle gives rise to a new population
of parameter points with (desirably)
better fitnesses.
The fitness of each individual is defined
by the corresponding value of the loss function:
the smaller the value of the loss function
given the parameter values of an individual,
the better is the fitness of the individual.

The first step of each cycle, i.e.\ selection,
determines which of the individuals of the
population are allowed to take part in
the following two steps, i.e.\ mating and
mutation. As a selection function we used
the so-called tournament selection with
size three. This function selects the individual
with the best fitness from three randomly
picked individuals of the population.
In total $50\,000$ individuals
are selected in this way
(where
each individual was allowed to be selected
more than once) 
and these then proceed to the mating stage.
Since the selection is based on the
fitness values,
individuals with better fitness have a higher
chance of producing new individuals
(called offspring).

For the mating process,
we divided the selected individuals into
two distinct groups, and then we performed
a uniform crossover of pairs of individuals
from each group. A uniform crossover creates
two child individuals from each pair
of parent individuals, where the child
individuals are
defined by swapping the attributes of the two
parent
individuals, in our case according to a probability
of 0.2. Hence, the two parent individuals
produce two offspring individuals which have on average
$20\%$ of the attributes from one parent
and $80\%$ of the attributes from the other parent.
In addition, we included a so-called mating
probability of 0.8, such that for $20\%$ of
the pairs of parent individuals no mating
was performed and the parent individuals
were just kept in the population
without changing their attributes.

Afterwards, the mutation stage
is performed, which modifies some of the
individuals of the offspring via a randomized
function, potentially giving rise to
new individuals with
good fitness values that belong to
so far unexplored parameter regions.
As a mutation function we applied the 
so-called float uniform mutator function
with a mutation probability of 0.2.
This function multiplies the attributes
of an individual
with a random number between
0.8 and 1.2
according to a probability of 0.1.
As a result, $20\%$ of the
individuals of the offspring are mutated,
and the mutations modify on average
$10\%$ of the attributes of such individual.

At the end of each cycle, we replace
the initial population with the offspring
and enter a new cycle, until either
an individual is found that corresponds
to a value of the loss function below
a certain threshold, or until the maximum
number of cycles is reached.
Since it is possible that the individual
in the parent population
with the best fitness would be lost
when the population is replaced,
we append this best-fit
individual
to the offspring population in order
to ensure that it always survives
the complete cycle.
Finally, when the algorithm has
completed, we save the parameter point
with the best fitness. Accordingly,
the above described
algorithm is performed as many times
as the number of desired parameter points
in the final sample.

For the two
scans discussed in \refse{secnumanal},
we compared the performance of the
genetic algorithm to the one of
a random scan over
the free parameters using a flat prior.
For a machine-independent estimate
of the performances of both algorithms,
we chose the number of evaluations of
the loss function $L$ (see \refeq{eqlossfunc})
that is required until
a parameter point featuring a value
of $L$ below a certain threshold is found.
We found for the first scan
discussed in \refse{secnumanal1}
that, on average, the genetic
algorithm succeeds in finding a value
of $L<90$ with roughly 60\% to 70\% fewer
evaluations of $L$
compared to the random scan, such
that the improvment is only moderate.
For the second scan discussed
in \refse{secnumanal2},
in which $L$
receives an additional term,
our computations indicate
that the genetic algorithms
outperforms the random scan drastically.
Here we found that using the
genetic algorithm the average number
of evaluations of $L$ in order to find
a parameter point with $L < 150$
was approximately 35 times smaller
than using a random scan.
Since in this scan the parameter points
with the desired features with regards
to the collider excesses (see
\refse{secnumanal2} for details) require values of
$L$ that are even smaller than $L = 150$,
we conclude that the
usage of the genetic algorithm was a vital
piece of our numerical analysis.
The reason for the fact that the genetic
algorithm performs so much better in the
second scan, whereas the improvement was only
moderate in the first scan, can be attributed
to the fact that the simultaneous minimization
of the values of $\chi_{125}$ (see \refse{secexpconstr})
and the value $\chi_{96}$ (defined in \refse{secnumanal2}),
which quantifies the fit to the collider excesses,
requires additional relations between the
mixing angles $\alpha_i$ and $\tan\beta$, which
the genetic algorithm is able to find
more quickly by
successively adjusting the parameters
of the points
with the lowest values of $L$ that have been
found in the previous generation.

\subsection{pNG DM in the Higgs funnel region}
\label{secnumanal1}
In order to explore the Higgs funnel
region, we scanned the parameter space of the
S2HDM within the parameter ranges
\begin{align}
1.5 \leq \tan\beta \leq 10 \ , \quad
m_{h_1} = 125.09 \gev \ , \quad
140 \gev \leq m_{h_{2,3}} \leq 1 \tev \ , \notag \\
40 \gev \leq m_\chi \leq 80 \gev \ , \quad
40 \gev \leq v_S \leq 1 \tev \ , \quad
-\pi / 2 \leq \alpha_{1,2,3} \leq \pi / 2 \ ,
\notag \\
400 \gev \leq M \leq 1 \tev \ , \quad
600 \gev \leq m_{H^\pm} \leq 1 \tev \ , \quad
m_A \leq 1 \tev \ , \notag \\
\Delta M_{\rm max}
= \max \left( |m_H - M|,
|m_A - M|, |m_{H^\pm} - M| \right)
< 200 \gev \ ,
\label{eqranges1} 
\end{align}
where in the last line
$m_H = m_{h_2}$ when $\Sigma_{h_2}
< \Sigma_{h_3}$ or $m_H = m_{h_3}$ when
$\Sigma_{h_2} > \Sigma_{h_3}$. Thus, this
condition on $\Delta M_{\rm max}$ ensures that
the masses of the heavy doublet-like states
$A$, $H^\pm$ and $H= h_2$ or $=h_3$ are
not further than $200\gev$ away from the
mass scale $M$. As explained in
\refse{secconstraintstheo},
and as will also be demonstrated
in the following, this condition excludes
parameter points that have a very small
energy range $v \leq \mu \leq \mu_v$ 
in which the parameter points
fulfill the theoretical constraints,
with potentially $\mu_v \ll 1\tev$.
The lower limits of $\tan\beta \geq 1.5$
and $m_{H^\pm} \geq 600 \gev$ exclude parameter
points that are potentially in conflict
with constraints from flavour-phyiscs
observables. The mass hierarchy of the
CP-even Higgs bosons $h_i$ is fixed such
that $h_{125} = h_1$ is the lightest one.
Their mixing angles $\alpha_i$ are scanned
over the theoretically possible range, where
it should be noted that their values are
strongly constrained by the measurements of
the signal rates of $h_{125}$, as will also
be demonstrated below. The vev of the singlet
field $v_S$ is allowed to take on values up
to $1 \tev$, which coincides with the upper
value chosen for the masses of the heavier
BSM states $H^\pm$, $A$ and $h_{2,3}$.
If we would have allowed for larger values
of $v_S$ and $M$, the heavy states could
acquire also larger masses and decouple
from the lighter states $h_1$ and $\chi$.
However, we wanted to focus on the parameter
space region in which the collider constraints
from direct searches at the LHC play a role,
such that we limited our scan to the case
in which all particle states could be
produced (and discovered)
at the LHC.

The scan points that we will present
were obtained in a two step procedure.
In the first step we applied the genetic
algorithm as described before in order
to find parameter points that minimize
the loss function
\begin{equation}
L =
\chi^2_{125} +
\max \left[0,(r_{\rm obs}^{\rm HB} - 1)
    \cdot 100 \right]
+ \left\{
\begin{matrix}
C \ , \ \text{when $\chi^2_{ST} > 5.99$ or theo.}
    \hspace*{3cm} \\
  \hspace*{2cm}  \text{constraint violated at }
    \mu = v \\
0 \ , \ \text{otherwise} \hspace*{5.9cm}
\end{matrix}
\right. \hspace*{-0.4cm} .
\label{eqlossfunc}
\end{equation}
Here $\chi^2_{125}$ is the result of
the \texttt{HiggsSignals} test, and
$r_{\rm obs}^{\rm HB}$ is provided from
the \texttt{HiggsBounds} test. $r_{\rm obs}^{\rm HB}$
is defined as the ratio of predicted
cross section for the most sensitive channel
divided by the experimentally observed
upper limit (see \refse{secexpconstr} for details).
As a result, parameter points featuring
a value of $r_{\rm obs}^{\rm HB} > 1$
should be rejected, and the second term in
the loss function quantifies the penalty
of this requirement. The factor 100
is included in order to enhance the importance
of this exclusion in terms of the
loss function compared to the values
of $\chi^2_{125}$, thus making sure that
all parameter points with low values of
the loss function have
$r_{\rm obs}^{\rm HB} < 1$ and are
consequently not excluded by direct searches.
Finally, the third term
is a huge constant $C$ that is added when a
parameter point does not fulfill the
theoretical constraints at the initial
energy scale $\mu = v$, or when the
constraints from the EWPO are not
fulfilled. With this
definition of the loss function, the
genetic algorithm finds parameter points
that pass the theoretical constraints,
the constraints from the collider
experiments and the EWPO.

In a second step, all the
parameter points found with the genetic
algorithm were subject to the remaining
constraints: according to the discussion
in \refse{secconstraintstheo},
we applied the theoretical
constraints for scales $\mu > v$ and
verified whether they are fulfilled
up to at least $\mu = 1\tev$. In addition,
we verified that, regarding the SM-like
Higgs boson, we have ${\Delta\chi^2_{125}
 = \chi^2_{125} - \chi^2_{\mathrm{SM},125}
\leq} 5.99$
and $\mathrm{BR}_{\rm inv} < 0.11$,
and, regarding the DM candidate,
that the predicted relic abundance is
is not larger than the Planck value,
i.e.\ $\Omega h^2 \leq (\Omega h^2)_{\rm Planck}$.
We also ensured that the constraints from the
indirect-detection experiments from
the observation of dSph are respected.
The DM observables were not taken into
account already in the definition of the
loss function, because the computation of
the relevant theoretical predictions were
the most time-consuming part of the
analysis, such that it was much more
efficient to perform these computations
only for the parameter points that otherwise
passed all the other theoretical and
experimental constraints.

As was already mentioned before, the main
purpose of this analysis is to illustrate
the combined impact of the various constraints
on the model parameters. In particular, we
will point out which of the constraints
give rise to limitations on which subset
of parameters, and whether the constraints cover
similar or clearly distinct regions of
the S2HDM parameters. In the following,
we start the discussion with the theoretical
constraints that were applied according to
the discussion in \refse{secconstraintstheo}.
In the next step,
we examine the impact of the collider constraints
by taking into account both the constraints
from direct searches and from the constraints
on the properties of $h_{125}$
(see \refse{secexpconstr}). Finally, we consider the
physics related to the DM candidate $\chi$,
and how its properties are interconnected
to the Higgs sector.

\begin{figure}[t]
\centering
\includegraphics[width=0.48\textwidth]{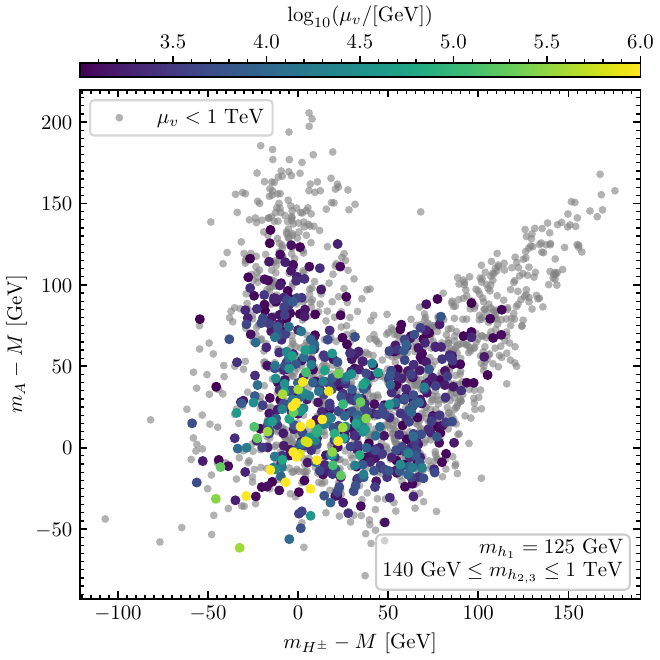}~
\includegraphics[width=0.48\textwidth]{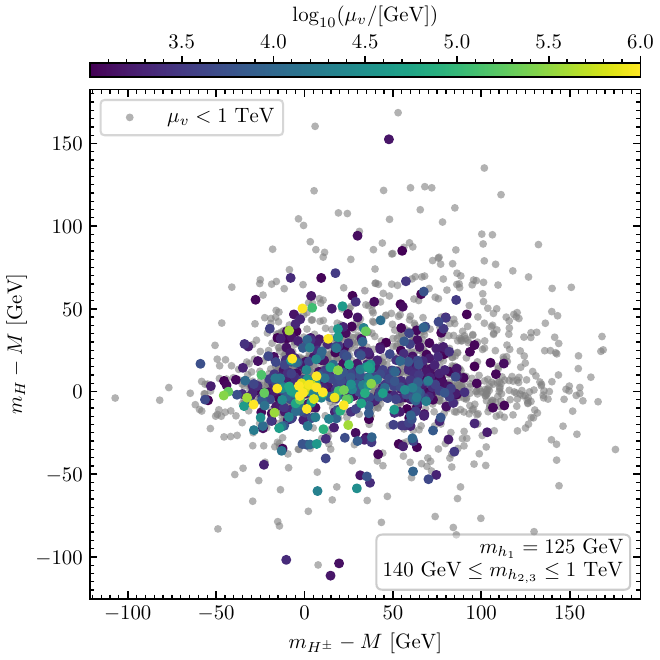}\\[0.4em]
\includegraphics[width=0.48\textwidth]{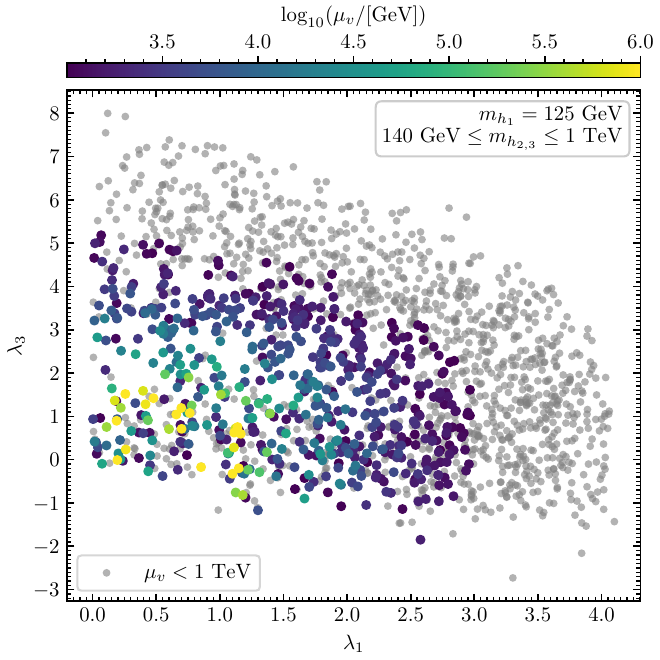}~
\includegraphics[width=0.48\textwidth]{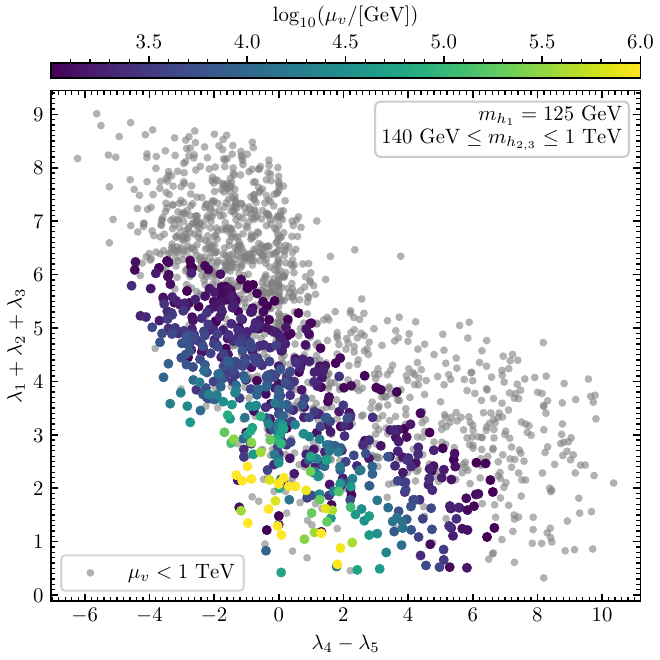}
\caption{\small
Top row:
$m_A - M$ (left) and $m_H -M$
(right) in dependence of
$m_{H^\pm} - M$, where $m_H = m_{h_2}$
or $m_H = m_{h_2}$
depending on whether $\Sigma_{h_2} < \Sigma_{h_3}$
or vice versa. Bottom row: $\lambda_3$ in dependence of
$\lambda_1$ (left) and $\lambda_1 + \lambda_2
+ \lambda_3$ in
dependence of $\lambda_4 - \lambda_5$ (right).
The colour coding indicates the
value of $\mu_v$. Also shown in grey
are discarded parameter points
with $\mu_v < 1 \tev$.}
\label{figtheo}
\end{figure}

In order to analyze the impact of the
theoretical constraints, we show in
\reffi{figtheo} the parameter points with
the colour coding indicating the energy
scale $\mu_v$ until which the theoretical
constraints are respected. We remind the
reader that all parameter points fulfill
the theoretical constraints at the
initial scale $\mu = v$. All points for
which $\mu_v < 1 \tev$ are shown in grey.
We performed the RGE running up to
$\mu = 100 \tev$, such that points that
have $\mu_v = 100 \tev$ (yellow points)
are potentially valid up to much higher
energy scales. In the upper left plot
we show the parameter points in the
plane $m_{H^\pm} - M$ and $m_A - M$.
One can see that only points for which
these differences are below roughly
$50\gev$ are valid at energy scales much beyond
$1\tev$. On the other hand, parameter
points with values of $|m_{H^\pm} - M|
\gtrsim 120\gev$ and/or $|m_A - M| \gtrsim
150\gev$ are always in contradiction
with one of the theoretical constraints
already at scales $\mu_v < 1\tev$.
The same observation can be made in
the upper right plot, in which $|m_H - M|$
is depicted on the vertical axis.
Points that feature values of $\mu_v$
much larger than about $1\tev$ are concentrated
at values of $|m_H - M| \lesssim 50 \gev$,
whereas points with larger values
of $|m_H - M|$ are
almost always only well behaved within
a small range of energies.

We find that the relevant constraint
that give rise to the low values of
$\mu_v$ are in most cases
the tree-level perturbative
unitarity constraints. These constraints
effectively provide upper limits on
the absolute values of the quartic
scalar couplings $\lambda_i$ and
combinations thereof (see also
\refap{apppert}).
It is therefore easy to understand why
they are more severe in region of parameter
space with relatively large splittings
between the masses of
the heavy BSM states and the mass
scale $M$, since such
splittings are induced by large absolute
values of $\lambda_{1,2,3,4,5}$ (see also
\refap{secparas}). Moreover, for obvious
reasons also the
energy scale dependence of the quartic
couplings is stronger
when their absolute
values are larger. As a result, points with
large mass splittings, which potentially
were already on the edge of being excluded
via the tree-level perturbative unitarity
constraints at the initial energy scale,
quickly break one of these constraints
once the RGE evolution is considered.
This is also reflected in the plots
in the lower row of \reffi{figtheo}, in which
we show the points in the planes
$\lambda_1$-$\lambda_3$ on the left
and $(\lambda_4 - \lambda_5)$-$(\lambda_1
+ \lambda_2 + \lambda_3)$ on the right.
In the left plot one can see that
verifying the theoretical constraints
exclusively at the initial scale $\mu = v$
gives rise to
parameter points with values of
$\lambda_1 \lesssim 4$ and
$-3 \lesssim \lambda_3 \lesssim 8$,
whereas demanding that the constraints
are respected within a range of energy of
at least $v \leq \mu \leq 1\tev$,
the allowed ranges shrink
to $\lambda_1 \lesssim 3$ and
$-2 \lesssim \lambda_3 \lesssim 5$.\footnote{$\lambda_1$
has to be positive according to the BfB conditions
on the tree-level scalar potential.}
A similar observation can be made in the
right plot, in which the allowed values
change from $\lambda_1 + \lambda_2 + \lambda_3 \lesssim 9$
and $-6 \lesssim \lambda_4 - \lambda_5 \lesssim 10$
to $\lambda_1 + \lambda_2 + \lambda_3 \lesssim 6$
and $-4 \lesssim \lambda_4 - \lambda_5 \lesssim 7$
once the RGE running and the additional
constraint $\mu_v > 1\tev$ is taken into
account.
Note that the limits on the values
of the couplings that we found are much
below the naive perturbativity criterion
$|\lambda_i| < 4 \pi$, which is often applied
in the analysis of extended Higgs sectors
in order to exclude non-perturbative
parameter regions.

Consequently, we conclude that
regarding the collider phenomenology
the main impact of our choice
to demand the theoretical constraints to
be respected at least until $\mu = 1\tev$
is that the masses of the heavy states are
closely related to the overall mass scale
$M$, which, however,
does not significantly
constrain the values of
$\lambda_{6,7,8}$, since they
do not depend directly on $M$
(see \refap{secparas}).
Thus, the only exception to the
constraints on the mass splittings
arises when there is
a Higgs boson $h_2$ or $h_3$
with almost $100\%$ singlet component present,
in which case its mass would be dominantly related
to the value of $v_S$ instead of $M$,
and the mass could differ
substantially from $m_A$, $m_{H^\pm}$
and $m_H$, as will also be further discussed
below.
Thus, our approach of including the theoretical
constraints 
drives the model predictions
towards the decoupling limit of
the S2HDM, where the masses
of the heavy states $m_{A}$,
$m_{H^{\pm}}$ and $m_{H}$ are
approximately determined by
the scale $M$ of the soft-breaking
of the discrete $\mathbb{Z}_2$.
Considering the theoretical
constraints described above
has in some aspects the same effect
as applying the
constraints from the EWPO, which are also
sensitive to large mass splittings between
the scalar states~\cite{Haller:2018nnx}.
This fact on its own is not very surprising
since also the EWPO observables arise from
the radiative corrections.
More interesting, however, is that
while it is
sufficient to have either $m_H \sim m_{H^\pm}$
or $m_A \sim m_{H^\pm}$ in order to be
in agreement with the constraints
from EWPO (at one-loop level), the inclusion of
the RGE running and the requirement
$\mu_v > 1 \tev$ gives rise to the fact
that both conditions should be approximately
fulfilled, i.e.\ $m_H \sim m_A \sim m_{H^\pm}$.

The low values of $\mu_v$ that
we found for values of
$|\lambda_i| \gtrsim 1$ are
relevant also for
cosmological aspects of the S2HDM,
where we stress again that one of the
main motivations of the model
is the possibility of 
accommodating a first-order
EW phase transition. In order to
achieve such a transition,
it is required (just as in the 2HDM)
to consider parameter space regions
where large loop corrections to
the scalar potential are 
present, since
at tree level 
the scalar potential does not allow
for an EW phase transition of
first order.
The required loop corrections have their origin in
values of one or more
${|\lambda_{1,2,3,4,5}| > 1}$~\cite{Biekotter:2021ysx}.
As a result, our analysis indicates
that for a perturbative study
of the parameter regions of the S2HDM
relevant for
possible first-order EW phase transitions,
it is of crucial importance to take
into account constraints
in relation to the perturbative
unitarity and the RGE running
of the quartic couplings.\footnote{For
instance, both type~II
benchmark scenarios in Tab.\ I of
\citere{Zhang:2021alu},
where first-order
phase transitions are discussed
in the context of the
S2HDM, would
be excluded in our analysis
due to the large absolute
values of $\lambda_4 \sim 5$
and $\lambda_5 \sim -7$
(see also \refeq{eqkiller}).}
On the other hand, if one restricts
an analysis of the S2HDM to regions
of the parameter space in which
the couplings $\lambda_i$ have absolute
values substantially below one, then
the model can be valid to energy scales
much beyond the TeV scale. In this case,
however, the S2HDM cannot accommodate
a first-order EW phase transition
and its related phenomenology,
and also the heavy BSM states are
largely decoupled from the EW scale
(as  discussed above).

\begin{figure}
\centering
\includegraphics[width=0.48\textwidth]{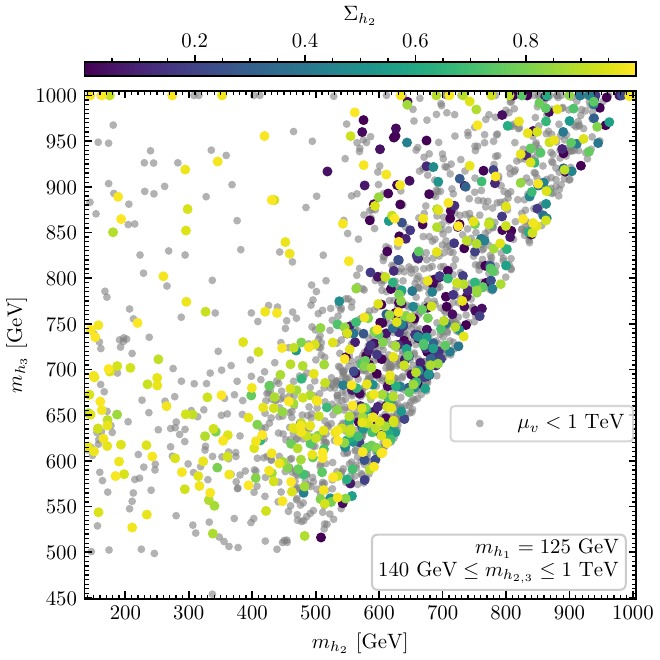}~
\includegraphics[width=0.48\textwidth]{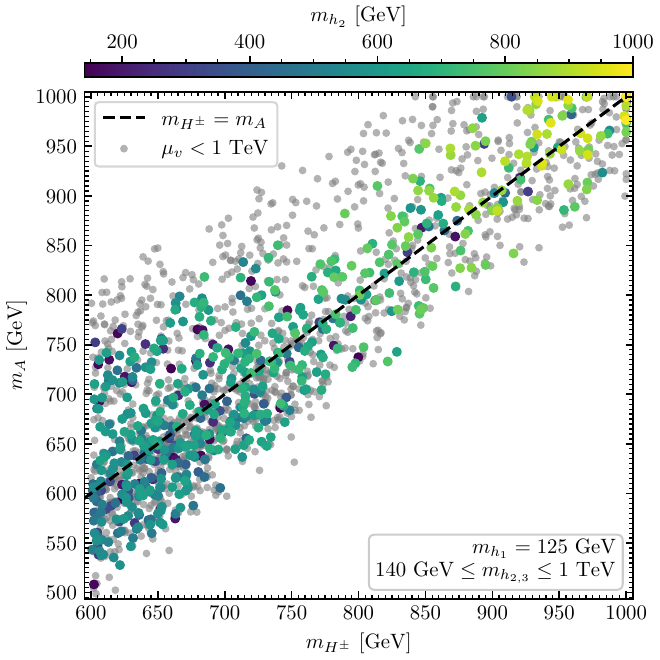}
\caption{\small Spectrum of the heavy Higgs bosons
$h_{2,3}$, $A$ and $H^\pm$
for the parameter points fulfilling the
theoretical and experimental constraints. Left:
$m_{h_3}$ in dependence of $m_{h_2}$, with the colours
indicating $\Sigma_{h_2}$. Right: $m_A$ in dependence
of $m_{H^\pm}$, with the colours indicating $m_{h_2}$.
Also shown in grey are excluded parameter
points with $\mu_v < 1 \tev$.
The dashed line indicates where
$m_{H^\pm} = m_A$.}
\label{figspec}
\end{figure}

To shed more light on the spectrum
of the Higgs bosons, we show in
\reffi{figspec} the mass $m_{h_3}$
in dependence of $m_{h_2}$ on the left
and $m_A$ in dependence on $m_{H^\pm}$
on the right. In the left plot one
can see that it is possible that
$h_2$ is substantially lighter than
$h_3$ when it has a large singlet component
of $\Sigma_{h_2} > 90\%$, as indicated
by the colours of the points. On the
other hand, when $h_2$ and $h_3$ are
sizably mixed, the masses of
both states have to
be relatively close to $M$ in order
to comply with the theoretical constrains.
The same observation can be made in
the right plot regarding the masses
of $A$ and $H^{\pm}$. Note here that all
the points with $m_A - m_{H^\pm} \gtrsim
150 \gev$ are grey, indicating that they
feature values of $\mu_v < 1\tev$.
In this plot the colour coding indicates
the values of $m_{h_2}$, and a correlation
can be seen between the mass of $h_2$
and the masses of $A$ and $H^\pm$. The
heavier the latter states, the larger
also tend to be the values of $m_{h_2}$.
Since by definition $m_{h_2} < m_{h_3}$,
one can conclude that in most parameter
points all six states $h_{2,3}$, $A$
and $H^\pm$ are relatively close in
mass, with the only exception
being a very
singlet-like state $h_2$ with
$m_{h_2} \ll M$, as mentioned
earlier already.
Hence, our analysis shows a trend
towards the decoupling limit of the S2HDM,
in which at low energies
the model could become practically
indistinguishable from the SM.
In this case, the only possibility
to observe a BSM effect would arise from
the DM phenomenology or a possible
invisible branching ratio of $h_{125}$
if the decay $h_{125} \to \chi \chi$
is kinematically allowed
(see the discussion below).
The presence of an invisible branching
ratio of $h_{125}$ could also allow for
a distinction between the S2HDM and
the 2HDM, whereas the S2HDM in the
decoupling limit could be
practically indistinguishable from
the pNG model with one Higgs doublet,
in which case only a discovery of one of the
additional particles of the S2HDM
at a collider could shed light
on the model realized in nature.

Going beyond the theoretical limitations,
the spectrum of the Higgs bosons is also
severely constrained by direct searches
at colliders, where due to the
fact that we focus here on the
mass ordering with $h_1 = h_{125}$
only the LHC results play
a role in our discussion.\footnote{
See also
\citeres{vonBuddenbrock:2016rmr,
Baum:2018zhf} for investigations
of the collider phenomenology of a
2HDM extended with a complex singlet
scalar, in which no additionally
U(1) symmetry
is imposed on the singlet.}
Without
going into the details of each of the
relevant search channels, we list here
the searches that were selected
by \texttt{HiggsBounds} and which led to
exclusions of parameter points in the
scenario under investigation:
\begin{itemize}
\item[-] ATLAS~\cite{ATLAS:2018oht}:
    $g g \to A \to (h_2) Z \to
    (b \bar b) l^+ l^-$
    at $\sqrt{s} = 13 \tev$
\item[-] ATLAS/CMS~\cite{ATLAS:2018sbw,
CMS:2015hra}:
    $p p \to h_2 , h_3 \to V V$
    at $\sqrt{s} = 13\tev/7+8\tev$
\item[-] ATLAS/CMS~\cite{ATLAS:2017tlw,
CMS:2018amk}:
    $p p \to h_2 \to Z Z$
    at $\sqrt{s} = 13\tev$
    and including width effects
\item[-] ATLAS~\cite{ATLAS:2020zms}:
    $p p \to h_2, h_3, A \to \tau^+ \tau^-$
    at $\sqrt{s} = 13\tev$
\item[-] ATLAS~\cite{ATLAS:2018rnh}:
    $p p \to h_2 , h_3 \to h_1 h_1 \to
    b \bar b b \bar b$
    at $\sqrt{s} = 13\tev$
\item[-] CMS~\cite{CMS:2019qcx}:
    $g g \to A \to (h_1) Z \to (b \bar b)
    l^+ l^-$ at $\sqrt{s} = 13\tev$
    assuming $h_1 = h_{125}$
\item[-] ATLAS~\cite{ATLAS:2021upq}:
    $p p \to (H^\pm) t b \to (t b) t b$
    at $\sqrt{s} = 13\tev$
\item[-] ATLAS~\cite{ATLAS:2019qdc}:
    $p p \to h_2 , h_3 \to h_1 h_1$
    at $\sqrt{s} = 13\tev$
    assuming $h_1 = h_{125}$
\item[-] CMS~\cite{CMS:2019pzc}:
    $g g \to h_2 , h_3 \to t \bar t$
    at $\sqrt{s} = 13\tev$
    and including width effects
\end{itemize}
In general, the most promising searches
at the lower end of the $\tan\beta$ range
are the searches for the charged Higgs bosons
or the searches for
the neutral states $h_2$, $h_3$ and
$A$ dominantly produced in the gluon
fusion channel, where depending on their
masses they then mostly decay into
pairs of $t$ quarks, pairs of
vector bosons or into
a lighter Higgs boson and a $Z$ boson.
For the upper end of the $\tan\beta$ range,
the most promising channel is the resonant search
for new Higgs bosons in the invariant
mass spectrum of two $\tau$ leptons.
Here it should be noted that
the resulting exclusions in the S2HDM can
be substantially different in comparison
to the 2HDM, because
$h_3$ and $A$ can have sizable branching ratios
for the decays into final states containing a
potentially much lighter singlet-like state
$h_2$, in which case the branching ratios
in regards to the decays of $h_3$ and $A$
into a pair of $\tau$-leptons are suppressed.
As a result, for a fixed value of $\tan\beta$
both states can be lighter in the S2HDM compared
to the 2HDM without being in conflict with
the searches for heavy Higgs bosons
decaying into two $\tau$
leptons~\cite{ATLAS:2021upq}.
Finally, for parameter points in the
intermediate $\tan\beta$ range with
$3 \lesssim \tan\beta \lesssim 6$,
the bosonic decays of the neutral states
are most relevant, such that the searches
with two vector bosons in the final state or Higgs
cascade decays can probe parts of the
parameter space of the S2HDM.

\begin{figure}
\centering
\includegraphics[width=0.48\textwidth]{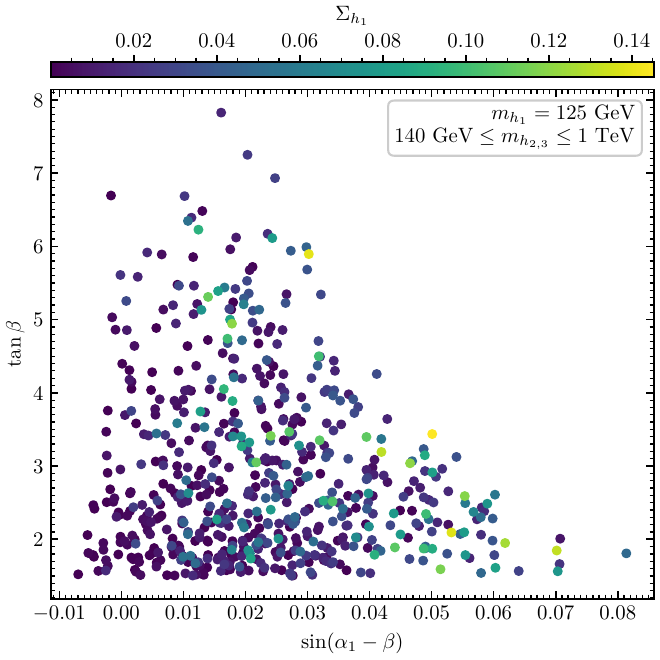}~
\includegraphics[width=0.48\textwidth]{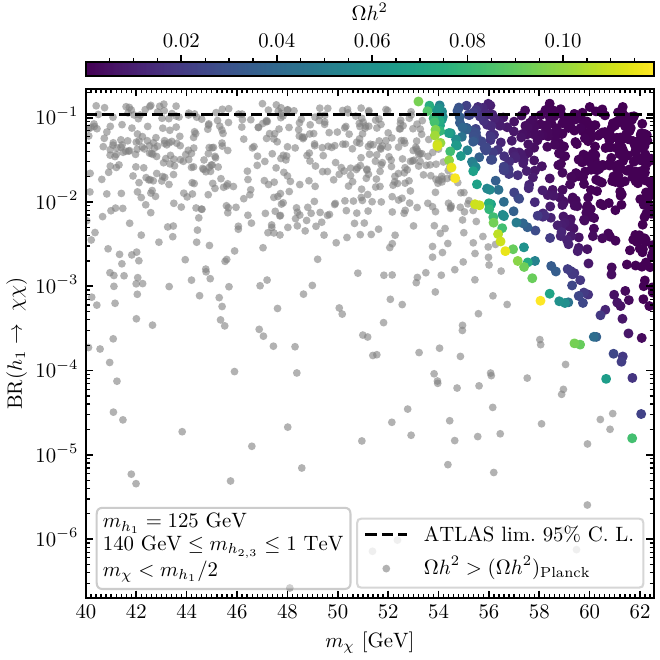}
\caption{\small
Left: $\tan\beta$ in dependence of $\sin ( \alpha_1
- \beta)$ for the parameter points that pass
all the constraints discussed
in \refse{secconsts}.
Right: Invisible branching ratio
$\mathrm{BR} ( h_1 \to \chi \chi)$ in
dependence of $m_\chi$ for the parameter points
with $m_\chi < m_{h_1} / 2$
that pass the constraints, not taking
into account the
experimental upper limit on
$\mathrm{BR} ( h_1 \to \chi \chi)$ as
reported by
ATLAS~\cite{ATLAS:2020kdi}.
The colour coding
indicates $\Sigma_{h_1}$.}
\label{figshsm}
\end{figure}

Complementary to the direct searches
for the BSM particles, the Higgs
sector of the S2HDM can also be probed
indirectly via the properties of the
Higgs boson $h_1 = h_{125}$ resembling
the Higgs boson that was
discovered at the LHC. In order to illustrate
the impact of such constraints, we show
in the left plot of
\reffi{figshsm} the allowed parameter points,
which all fulfill the criterion
$\chi^2_{125} \leq \chi^2_{\mathrm{SM},125}
+ 5.99$ (see \refse{secexpconstr} for details),
with $\sin(\alpha - \beta)$ on the horizontal
and $\tan\beta$ on the vertical axis.
In the case in which one of the heavier
states $h_2$ or $h_3$ has a singlet component
of almost $100\%$, the S2HDM features an alignment
limit similar to the 2HDM. In this limit
the couplings of $h_1$ reduce to the ones
of a SM Higgs boson, and the limit is
determined by the condition
$\sin(\alpha - \beta) = 0$ (see also
\citere{Muhlleitner:2016mzt}).
Consequently, departures from this condition
are associated with deviations of the
predictions for the signal rates of
$h_{125}$ with respect to the SM.
As can be seen in the
left plot of \reffi{figshsm}, our analysis
indicates that in order to be in agreement
with the measured signal rates, one has
to fulfill roughly $|\sin(\alpha - \beta)|
\lesssim 0.1$. The largest departures
from zero are found for the lower end of
the $\tan\beta$ range, whereas for larger
values of $\tan\beta$ the allowed range
of $|\sin(\alpha - \beta)|$ shrinks substantially.
The colour coding of the points indicates
the singlet component of the SM-like
Higgs boson $\Sigma_{h_1}$. 
Notably, we find that the current uncertainties
of the signal-rate measurements still
allow for a singlet-component of more than $14\%$.

Precise measurements of the properties of
$h_{125}$, for instance correlated
deviations from the SM prediction of the various different
couplings coefficients of $h_{125}$ to the
up- and down-type fermions $C_{h_{125} u \bar
u}$ and $C_{h_{125} d \bar d}$
and the gauge bosons
$C_{h_{125} VV}$, could help to
distinguish the type~II S2HDM from the
usual 2HDM.
Here the coupling coefficients
$C_{h_{125} u \bar u , d \bar d , VV}$ are defined
to be the couplings
normalized to the ones of a SM Higgs boson.
A sizable singlet-component of $h_{125}$, as found
in parts of our parameter points,
gives rise to a suppression of $C_{h_{125} VV}$.
In the usual 2HDM, a deviation from
$|C_{h_{125} VV}|\ =1$
is possible via departures from the
alignment limit,
and thus tightly constrained to
values of $C_{h_{125} VV}^2 \gtrsim
0.9$~\cite{Haller:2018nnx}.
Since we find parameter points
with $\Sigma_{h_{125}} > 0.1$,
and since in the S2HDM one has
$C_{h_{125} VV}^2 \leq
1 - \Sigma_{h_{125}}$, a possible future
measurement indicating
$C_{h_{125} VV}^2 \lesssim 0.9$ at the (HL)-LHC
would favor an S2HDM interpretation
instead of the 2HDM.
It is also interesting to compare the
maximum values of $\Sigma_{h_{125}} \sim 14\%$
with the corresponding values found
in the pNG DM model with only one Higgs doublet.
In \citere{Arina:2019tib} it was shown that in this
case the mixing of the SM-like Higgs boson
with the singlet state is more constrained,
and, except when the singlet scalar and
the doublet scalar are degenerate in mass,
only values of up to $10\%$ were found
to be in agreement with the Higgs-boson
measurements.
As a result, and under the assumption
that a deviation of the properties
of $h_{125}$ w.r.t.\ the SM
will be observed, one could potentially
distinguish the S2HDM
from the simpler model with only one
Higgs doublet via the precise measurements
of $C_{h_{125} VV}$, $C_{h_{125} u \bar u}$
and $C_{h_{125} d \bar d}$.
Moreover, the model with one Higgs doublet
predicts $C_{h_{125} u \bar u} =
C_{h_{125} d \bar d}$, such that
experimental indications for
$C_{h_{125} u \bar u}  \neq
C_{h_{125} d \bar d}$ would clearly
favour an S2HDM interpretation.
Another obvious possibility to
distinguish both models arises
from the fact that the S2HDM can
predict values of $|C_{h_{125} d \bar d}|,
|C_{h_{125} u \bar u}| > 1$
due to enhancements by factors
of $1 / c_\beta$ or $1 / s_\beta$
(depending on the Yukawa type),
while the pNG DM
model with one Higgs doublet can only
accommodate values equal or smaller than one.

The mixing among the CP-even
scalar fields in the S2HDM is identical
to the one of the N2HDM, such that it is not
surprising that we find similar effects
on the allowed parameter ranges
of $\alpha_i$
in the S2HDM. However, a crucial difference
between both models is the presence
of the additional particle $\chi$
in the S2HDM. Since we are focusing
here on the Higgs funnel region of the
model, it is possible that
$m_\chi < 125 \gev / 2$, giving rise
to an additional decay mode of $h_{125}$
into an invisible final state.
To illustrate the impact of this
additional decay on the allowed
parameter regions, we show
in the right plot of \reffi{figshsm}
the branching ratio for the
invisible decay of $h_{125}$
in dependence of $m_\chi$ for
the parameter points with
$m_\chi < 125 \gev / 2$.
Here we show also the parameter points
that would be excluded by the observed
upper limit on $\mathrm{BR}(h_1 \to
\chi \chi)$~\cite{ATLAS:2020kdi}
(indicated by the horizontal
dashed line).
In this way we can demonstrate the
interplay between the global constraints
from the \texttt{HiggsSignals} analysis
and the direct limit on the invisible branching
ratio. One can see that only a very small
fraction of the otherwise
allowed parameter points, which
in particular have passed
the constraint
$\chi^2_{125} \leq \chi^2_{\mathrm{SM},125}
+ 5.99$, lie above the ATLAS limit
on the invisible branching ratio.
Nevertheless, for some points we find
values of $\mathrm{BR} ( h_1 \to \chi \chi)$
that are 
about $50\%$ larger
than the upper limit in the whole range
of $m_\chi$ in which allowed points were
found.

The grey points in the right plot
of \reffi{figshsm} have to be discarded because
they feature a too large thermal relic
abundance of DM. For the allowed points
the DM relic abundance is indicated by
the colour coding of the points.
One can see that we find a
limit of $m_\chi \sim 53.8\gev$ below which
no allowed points were found.\footnote{A
very similar limit was found in the
pNG DM model with one Higgs
doublet~\cite{Arina:2019tib}.}
This limit arises from a combination of
the upper limit on $\mathrm{BR} ( h_1 \to \chi \chi)$,
on the one hand, and the constraint
$\Omega h^2 \leq (\Omega h^2)_{\rm Planck}$,
on the other hand. Parameter points with
$m_\chi \lesssim 53.8 \gev$ feature either
a $\chi$ that is weakly coupled to
$h_{125}$, in which case
$\mathrm{BR}_{\rm inv}$ can be in agreement
with the ATLAS limit but
the DM relic abundance 
is too large
because the annihiliation
process with $h_{125}$ in the
$s$-channel is not efficient
(see also discussion below), or $\chi$ is
coupled more strongly to $h_{125}$, in which
case the DM relic abundance can
be below the upper limit but the
invisible branching ratio
of $h_{125}$ is unacceptably large.
Note here that in the plot almost all
parameter points with $m_\chi \lesssim
53.8\gev$ belong to the first option,
predicting too large values of
$\Omega h^2$, while $\mathrm{BR}_{\rm inv}$
is below the experimental upper limit.
On the other hand, there are only three 
parameter points with $m_\chi \lesssim
53.8\gev$ belonging to the second option,
featuring too large values
of $\mathrm{BR}_{\rm inv}$ but with
$\Omega h^2$ below the Planck limit.
The reason for this lies in our procedure
to generate the parameter points
using the genetic algorithm.
Parameter points with values of
$\mathrm{BR}_{\rm inv} = \mathcal{O}(0.1)$
feature overall larger values
of $\chi^2_{125}$ and constitute therefore
only a very small part
of the sample of parameter
points, because the genetic
algorithm tries to find parameter points
that minimize $\chi^2_{125}$
(see the definition of the loss function
defined in \refeq{eqlossfunc}).

\begin{figure}
\centering
\includegraphics[width=0.48\textwidth]{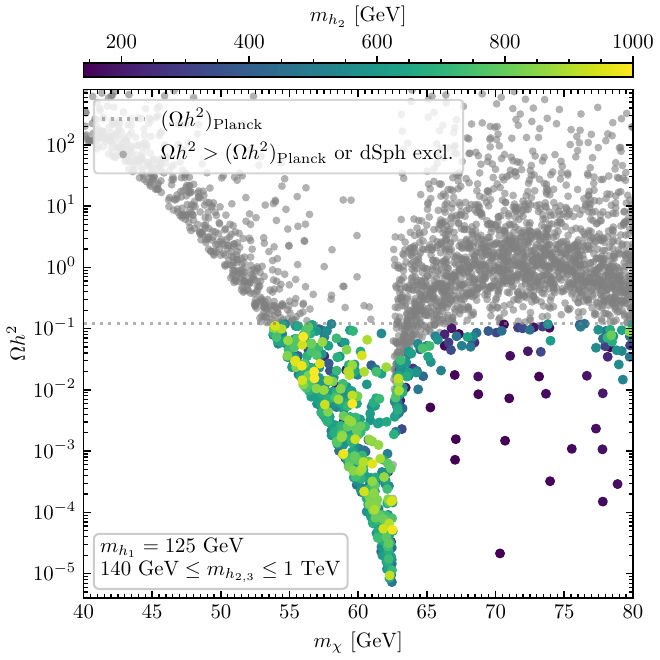}~
\includegraphics[width=0.48\textwidth]{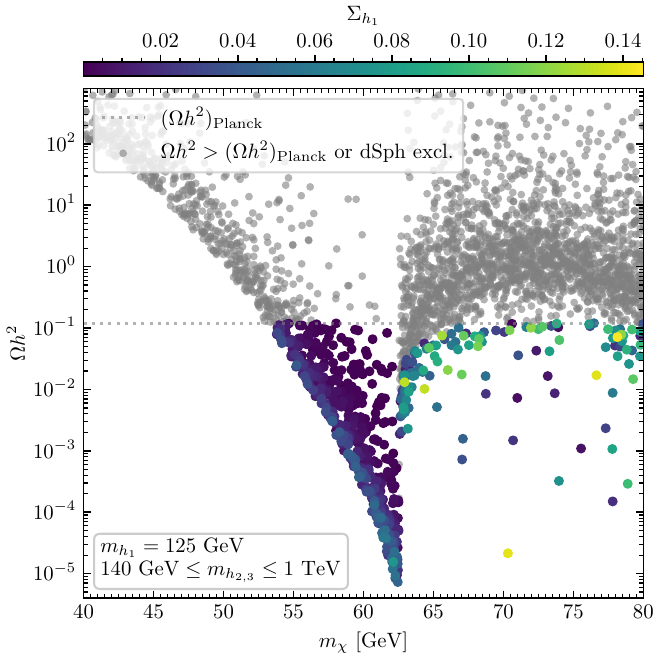}
\caption{\small Relic abundance
$\Omega h^2$ as
predicted by thermal freeze-out in
dependence of $m_\chi$.
The colour coding indicates the values
of $m_{h_2}$ (left) and the values
of $\Sigma_{h_2}$ (right).
Grey points are
excluded by $\Omega h^2 > (\Omega
h^2)_{\rm Planck}$
or Fermi dSph measurements.}
\label{figdms}
\end{figure}

The above discussed findings already
indicate the strong interplay between
the Higgs phenomenology and the DM
sector of the S2HDM, in particular
in the scenario discussed here that
fundamentally relies on the Higgs funnel
to predict a DM relic abundance in
agreement with experiments.
To shed more light
on this interplay, we show in \reffi{figdms}
the relic abundance as predicted
according to the freeze-out mechanism
in dependence of the DM mass $m_\chi$.
One can see the strong suppression of
$\Omega h^2$ for most parameter points
at $m_\chi \sim 125 / 2 \gev$, where
the DM annihilation cross sections
with $h_1$ in the $s$-channel are
resonantly enhanced. At this precise
resonance region, there are nevertheless
also a few parameter points featuring
values of $\Omega h^2$
within an order
of magnitude below
the experimentally
measured value $(\Omega h^2)_{\rm Planck}
= 0.119$ (indicated by the grey dashed
line in \reffi{figdms}).
For these parameter points
the resonant enhancement of the
annihilation cross sections is counteracted
by strongly suppressed couplings of
$\chi$ to~$h_{125}$.\footnote{
These parameter points
also have highly suppressed
DM-SM scattering processes
at finite temperatures, such that 
in some cases $\chi$ might be kinematically
decoupled already before the freeze-out
period. As a result, this effect of
\textit{early kinetic decoupling} of
DM~\cite{Binder:2017rgn,
Abe:2021jcz}
can give rise to an additional
source of uncertainty for
the prediction for $\Omega h^2$
for these points.}

For values of $m_\chi$ below
$m_{h_{125}}/2$,
there is a small band of values
$53 \gev \lesssim m_\chi \lesssim
m_{h_{125}}/2$ in which the measured value
of the relic abundance can be accommodated,
whereas for values below this range the
predicted amount of DM density is always
too large (grey points). As already mentioned before,
the reason for this lies in the constraints
on the properties of $h_{125}$. 
In order to predict
an allowed value for $\Omega h^2$ when
$m_\chi \lesssim 53\gev$ it is required that
the coupling of $\chi$ to $h_{125}$ is
large. However, this inevitably results in
values of the invisible branching ratio
for the decay $h_{125} \to \chi \chi$
above the experimental upper limit.
As a result, the lower limit on $m_\chi$
found here can be regarded as a robust
bound under the assumption that $h_{125}$
corresponds to the lightest scalar $h_1$.
One can compare also to
the right plot of \reffi{figdms}, where the colour
coding indicates the values of the singlet
component of the SM-like Higgs boson $h_1$.
A clear distinction is visible between the
points below and above the resonance
at $m_\chi = 125/2 \gev$. Points
with $m_\chi$ below the
resonance have substantially smaller
values of $\Sigma_{h_1}$, whereas points
with $m_\chi$ above the resonance
allow for values of
$\Sigma_{h_1} \gtrsim 0.1$.
Moreover, only points for which $m_\chi$
is relatively close to the kinematic
threshold of the decay $h_1 \to \chi \chi$,
i.e.\ $m_\chi \sim m_{h_{125}} / 2$,
feature sizable values of $\Sigma_{h_1}$
when $m_\chi < m_{h_{125}} / 2$.
The reason for this is that the couplings
$\lambda_7$ and $\lambda_8$ that couple
the singlet field to the doublet fields
(see \refeq{eqscalpot}) appear in the partial
decay width for the invisible decay
as shown in \refeq{eqgaminv}. In addition,
these couplings are responsible
for the possible singlet admixture of the
state $h_{1}$. Accordingly, parameter points with
sizable values of $\Sigma_{h_1}$
have sizable values of $\lambda_7$ and
$\lambda_8$, which in turn can
give rise to
too large values of $\mathrm{BR} ( h_1 \to \chi \chi)$
whenever this decay is kinematically
allowed.
In \refse{secnumanal2}
we will address the question
whether the bound $m_\chi \gtrsim 53\gev$
can be substantially
modified in a scenario featuring a scalar
$h_1$ with a mass smaller than $125\gev$,
and the second lightest scalar $h_2$ plays
the role of the discovered Higgs boson.
In this case $\chi$ has two possibilities
to annihilate resonantly, either with $h_1$
or with $h_2$ in the $s$-channel, and the
predictions for the relic abundance can
be substantially modified.

For values of $m_\chi > 125/2 \gev$ one
can see that the prediction for
$\Omega h^2$ rises quickly with increasing
value of $m_\chi$, because the resonant
enhancement of the annihilation cross
section is lost. As a result, most
parameter points predict a too large
DM relic abundance. Taking into account
the values of $m_{h_2}$ (indicated
by the colour coding of the points
in the left plot of \reffi{figdms}), one
can see that most parameter points with
$m_\chi \gtrsim 65\gev$ that are in
agreement with the upper limit on
$\Omega h^2$ feature a relatively
light scalar $h_2$ with masses
at the lower end of the scan range
of $m_{h_2}$.
As before, the reason for this
is that when $h_2$ is not much heavier
than twice the value of $m_\chi$, the
second $s$-channel contribution to
the annihilation cross section
becomes relevant. This gives rise
to a suppression of $\Omega h^2$
such that the prediction can be
below the experimental limit even
when $m_\chi$ is several GeV larger
than $125/2 \gev$. Again, this hints
to the fact that also in the
mass range $m_\chi > 125 / 2\gev$ the
prediction for $\Omega h^2$ 
could be substantially modified using
the \textit{inverted} mass hierarchy
in which $h_{125}$ is not the lightest
scalar, and we investigate this possibility
assuming a Higgs boson $h_1$ at $96\gev$
in \refse{secnumanal2}.

In both plots in \reffi{figdms} the grey
points are characterized by either being excluded
due to $\Omega h^2 > (\Omega h^2)_{\rm Planck}$,
as already mentioned before, or they are excluded
due to the constraints from DM indirect-detection
experiments. In most parts of the analyzed parameter
space, the more constraining experimental limit
results to be the upper limit on the predicted
relic abundance, as indicated by the fact that
most of the grey points lie above the
horizontal dashed line indicating the
Planck measurement. However, there is a small
region with $62.5\gev \lesssim m_\chi \lesssim
67\gev$ in which we find grey points below
the Planck limit. Consequently, in this
mass range of $\chi$ the indirect-detection
limits from the observation of dSph
by the Fermi satellite are more constraining.
Note that this is a region in which it appears
to be relatively easy to accommodate a value
of $\Omega h^2 \sim (\Omega h^2)_{\rm Planck}$
without being in tension with constraints
on $h_{125}$,
since it is just above the resonance of
the annihilation cross section, and
the decay $h_{125} \to \chi\chi$ is kinematically
forbidden. The fact that these parameter
points can be probed via indirect-detection
experiments is therefore crucial.
We remind the reader that the constraints
derived from the Fermi measurements are subject
to uncertainties, as was also discussed in
\refse{secexpconstr},
such that the respective limits
might change slightly in the future and
are currently possibly not as
robust as the Planck
limit on the relic abundance. Nevertheless,
our results indicate that when the DM
candidate of the S2HDM in this mass range
is responsible for a large fraction of
the measured relic abundance, the
observation of dSph and the resulting
constraints (or signals,
more optimistically speaking)
will be of great
importance for studies in the context of
pNG DM.

\begin{figure}
\centering
\includegraphics[width=0.48\textwidth]{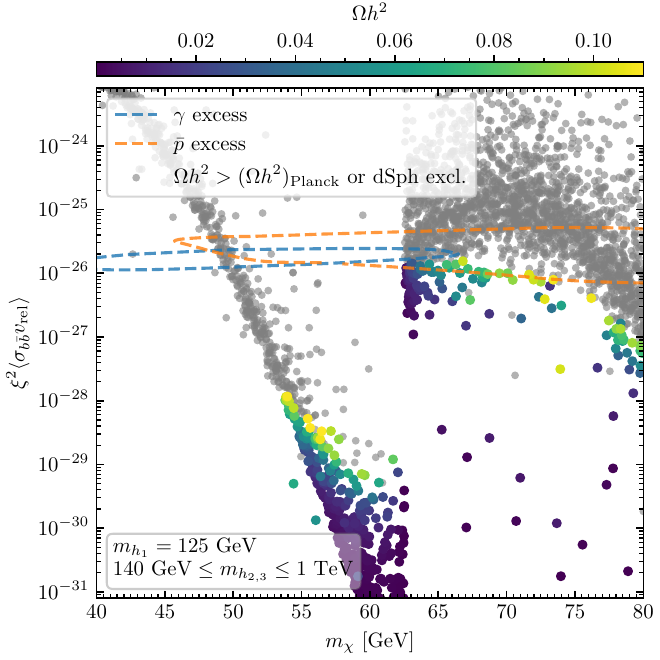}~
\includegraphics[width=0.48\textwidth]{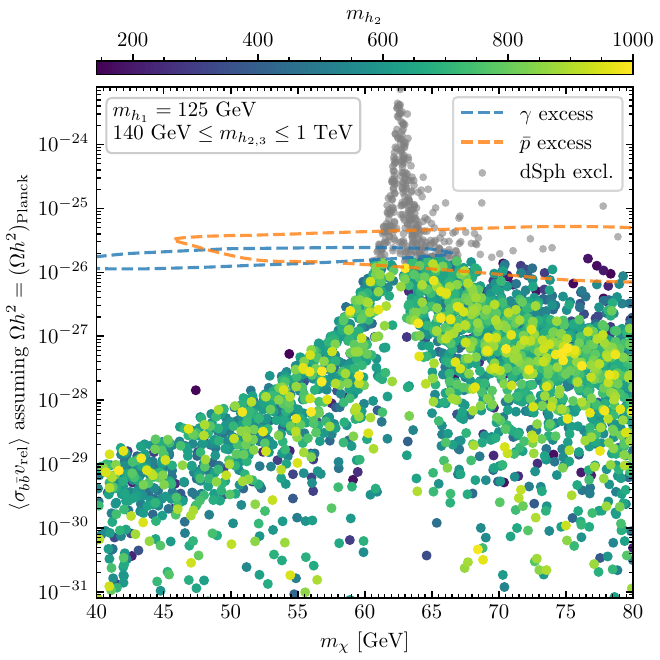}
\caption{\small Today's velocity averaged
annihilation cross section of $\chi$
into pairs of $b$ quarks taking into
account the number density as predicted
by thermal freeze-out (left) and
assuming $\Omega h^2 = 0.119$ (right).
The colour coding indicates the predicted
value of $\Omega h^2$ (left) and the value
of $m_{h_2}$ (right). Also indicated are
the regions in which the cosmic-rays
excesses could be explained within
the $2\sigma$ confidence level (blue and
orange dashed lines)~\cite{Cholis:2019ejx,
Cline:2019okt}.
Grey points are
excluded by
$\Omega h^2 > (\Omega
h^2_{\rm Planck})$
(left)
or Fermi dSph measurements (left and right).}
\label{figdmani}
\end{figure}

In this context it is interesting to
note that recent indirect-detection
experiments found anomalies in the
cosmic ray spectra. 
The first so-called 
galactic center
excess was found by the Fermi satellite,
which measured an intensity of gamma-rays
coming from the center of the galaxy
significantly above the predictions of
the standard model of cosmic rays
generation and propagation with
a peak in the spectrum around a
few GeV~\cite{Fermi-LAT:2017opo, Fermi-LAT:2015sau}.
Another anomalous cosmic-ray spectra
was measured by the Alpha
Magnetic Spectrometer (AMS)~\cite{AMS:2016oqu},
mounted on the international space station, which
reported an excess over the expected
flux of cosmic ray antiprotons
(see \refse{secintro} for
details).\footnote{The updated result
of the AMS collaboration could neither
definitively rule out nor confirm the
DM interpretation of the antiproton
excess~\cite{AMS:2021nhj}.}
While it is still under debate whether
the excesses arise from unresolved
astrophysical sources~\cite{Bartels:2015aea,
Lee:2015fea, Calore:2015bsx} or the treatment
of systematic uncertainties~\cite{Heisig:2020nse, Boudaud:2019efq}, or whether
their origin could be the annihilation
of DM, we will in the following assume
that the latter is the case.\footnote{See
also \citeres{DiMauro:2021raz,List:2021aer,
Kahlhoefer:2021sha,Abdughani:2021pdc,
Beck:2021xsv} 
for recent discussions of 
possible
explanations of the center-of-galaxy
excesses.} 
In \citere{Cholis:2019ejx} it
was shown that the excesses are compatible
with a DM interpretation, where the DM candidate
annihilates into pairs of $b$ quarks.
For the $\gamma$ excess
the allowed range of the mass of
the DM candidate at the $2\sigma$
confidence level was found to be
${37 \gev \leq m_{\rm DM} \leq 67 \gev}$.
For the $\bar p$ excess the allowed range
was found to be
${46 \gev \leq m_{\rm DM} \leq 94\gev}$,
which partially agrees with the mass
range preferred by the $\gamma$ excess.
Consequently, it is an interesting question
whether the S2HDM can explain both the excesses
simultaneously, while being in agreement with
all theoretical and experimental
constraints.\footnote{See
\citere{Cline:2019okt} for
an investigation of the
excesses in a singlet-extension of the
SM featuring pNG DM.}

In order to answer this question we show
in \reffi{figdmani} on the vertical axes
$\langle \sigma_{b \bar b}
v_{\rm rel} \rangle$,
being the predicted velocity-averaged
annihilation cross sections of
$\chi$ into pairs of $b$ quarks,
in dependence of the DM mass $m_\chi$ for the
parameter points of our scan.
The $2\sigma$ confidence level
regions of these two parameters required
to explain the $\gamma$ and the $\bar p$ excesses
are indicated with the blue and
orange dashed
lines, respectively~\cite{Cholis:2019ejx,
Cline:2019okt}.
We show the parameter points
in the two plots of \reffi{figdmani} under two
different assumptions. In the left plot we
assume that the usual thermal freeze-out scenario
can be applied, such that we have to take into
account the predicted values of the relic
abundace for each parameter point. Hence, the
values of 
$\langle \sigma_{b \bar b}
v_{\rm rel} \rangle$
on the vertical axis are scaled by the
factor $\xi^2$ as defined in
\refeq{eqxiani}.
On the other hand, in the right plot
we show the parameter
points under the assumption that the
relic abundance of DM is always accounted
for by $\chi$, independently of the
prediction from the thermal freeze-out.
As a result, they demand a non-standard
cosmological history giving rise to
the experimentally measured relic abundance,
which we will however not specify any further.
In the plots the grey points correspond to
parameter points that are excluded by a
too large predicted relic abundance (left)
or by constraints from dSph
observations (left and right).
In the right plot the dSph constraints are
consequently applied also assuming
$\Omega h^2 = (\Omega
h^2)_{\rm Planck}$.

Assuming the usual thermal freeze-out
scenario (left plot), one can see that
the resonant structure of the distribution
of the annihilation cross sections gives
rise to two distinct regions of $m_\chi$
in which points inside the blue and the
orange curves can be found. The first
region at lower DM masses of $m_\chi
\approx 50 \gev$ contains parameter points
that predict values of $\xi^2 \langle
\sigma_{b \bar b} v_{\rm rel} \rangle$
as required for an explanation of the
excesses, and where the values of $m_\chi$
lie roughly in the center of the values
preferred by the $\gamma$ excesses and
at the lower end of the range preferred
by the $\bar p$ excess. However, these
points are excluded because the predicted
values of $\Omega h^2$ are about an order
of magnitude larger than the experimentally
measured value, as can also be seen in
the left plot of \reffi{figdms}.
Accordingly, the parameter points in this
region of $m_\chi$ are excluded and
the cosmic-ray excesses cannot be
realized there.
The second region of DM masses in which
points within both the blue and the
orange curves are found is given
by $63 \gev \lesssim m_\chi \lesssim 67
\gev$. However, as before, the
corresponding points are shown in grey
and are consequently excluded. Interestingly,
here the responsible experimental constraint
do not arise from the Planck measurement of the
relic abundance, but from the Fermi-LAT
observations of dSph, as was already
discussed before. In fact, the predictions
for $\Omega h^2$ in this range of
$m_\chi$ are close or effectively
identical to the Planck measurement.
Hence, the points in this second region
of DM masses possibly predict the
correct DM relic abundance and could give rise
to both the cosmic $\gamma$- and the
$\bar p$-excesses, but they are in tension
with the null-results from the
observations of dSph. Here we remind the
reader, as was discussed already in
\refse{secexpconstr}, that the Fermi-LAT
dSph constraints
are subject to uncertainties in regards
to the astrophysical modelling of the
spectral curves, and as a result might
be slightly weaker as compared to
applied here. Nevertheless, with
future improvements of the dSph observations,
for instance, due to the inclusion of
more dSph and the increasing time periods
of data taking,
a firm exclusion (or confirmation if
a DM signal will actually be found)
of the parameter space region of
the second DM mass range discussed here
should be possible~\cite{Fermi-LAT:2016afa}.

Under the assumption of a non-standard
cosmological history that somehow gives
rise to a relic abundance of $\chi$
in agreement with the Planck measurement
(right plot), one can see that this
time only one DM mass region with parameter
points suitable for a realization of
the excesses is present.
Naturally, this region lies where
the resonant enhancement of
$\langle \sigma_{b \bar b} v_{\rm rel}
\rangle$ is present, i.e.\ at
$61 \gev \lesssim m_\chi \lesssim
67 \gev$, which consequently partially
coincides with the second region of
DM mass found in the left plot
of \reffi{figdmani}.
As before the points that
lie within both the blue and the
orange curves are in tension with
the dSph observations from the Fermi
satellite.

We end the discussion of the DM
properties in this scan by noting
that many of the above mentioned
findings crucially depend on the
assumed mass ordering of the
CP-even Higgs bosons. In particular,
the presence of a Higgs boson below
$125\gev$ can potentially impact
the predictions for the relic
abundance, as discussed in relation
to \reffi{figdms}. Moreover, the question
whether the cosmic-ray excesses can
be accommodated more easily when
a second $s$-channel resonance for
the annihilation cross section is
available can be addressed.
In \refse{secnumanal2} we will
investigate these questions following
the approach of \citere{Cline:2019okt}, in which
the presence of a Higgs boson at
around $96\gev$ was assumed in order to
simultaneously explain also
two collider excesses found at
LEP in the $b \bar b$ final state
and at the LHC in the diphoton
final state.

\subsection{pNG DM and a Higgs boson at 
\texorpdfstring{$96\gev$}{96GeV}}
\label{secnumanal2}
In \citere{Cline:2019okt} it was used that
the hypothetical
particle state $h_{96}$ at $96\gev$
can be coupled to new relatively
light charged states that can give rise
to additional contributions to the
loop induced coupling of $h_{96}$
to photons
in order to account for the
diphoton excess found by CMS.
In \citere{Biekotter:2019kde}
it was shown that in the N2HDM the
presence of the additional doublet Higgs
field and the real singlet field
are sufficient
to accurately describe
the collider excesses.
Here, the
diphoton rate was enhanced not
via an enhancement of the coupling
coefficient $|C_{h_{96} \gamma \gamma}|$,
where the coupling coefficients
$C_{h_{96} \dots}$ are defined as the
coupplings normalized to the one of a SM Higgs
boson of the same mass.
Instead, the branching ratio for
the diphoton decay of $h_{96}$ was
enhanced
via a
suppression of the couplings
of $h_{96}$ to $b$ quarks,
which then also gives rise to a suppression
of the total width of
$h_{96}$.\footnote{The presence of
a second doublet field gives rise to
the presence of the states $H^\pm$, such
that also in the S2HDM (compared
to the SM) new charged states
are present. However, the loop contributions
of $H^\pm$
to $|C_{h_{96} \gamma \gamma}|$ are not
relevant for the explanation of the
CMS excess, such that one can
have $m_{H^\pm} \gg 96\gev$.}
The required suppression of the coupling
coefficient $|C_{h_{96} b \bar b}|$
(without suppressing $|C_{h_{96} t \bar t}|$
in order to maintain sizable couplings
to photons via the $t$-quark loop)
can also be realized in the S2HDM
due to the possible
mixing patterns in the CP-even
sector and the presence of the
three mixing angles $\alpha_{1,2,3}$
in total analogy to the N2HDM.
In this regard, the only difference
in the S2HDM compared to the N2HDM is
the possible presence of the additional
decay modes $h_{96}/h_{125} \to \chi \chi$,
potentially
giving rise to a small suppression of
the decay modes $h_{96} \to \gamma \gamma$
relevant for the CMS excess and
$h_{96} \to b \bar b$ relevant for the
LEP excess, or to stronger constrains
on the properties of $h_{125}$.
In the following we will discuss
a scan to illustrate the impact of the
presence of $h_{96}$ on the phenomenology
of the DM candidate $\chi$, and whether
the collider excesses can be realized
in combination with the cosmic-ray excesses.

Before going into the description of
the parameter scan that we performed,
we briefly introduce the relevant details
of the collider excesses.
At LEP searches for Higgs bosons were
performed utilizing the $b \bar b$ final
state~\cite{LEPWorkingGroupforHiggsbosonsearches:2003ing},
which can be exploited
at a lepton collider in contrast to the
LHC due to the much smaller SM background.
Theoretically, the Higgs boson that is
searched for is assumed to be produced via
the Higgstrahlung pocess and subsequently
decays into a pair of $b$ quarks.
A local excess of about $2\sigma$ confidence
level was observed at a mass of roughly
$96\gev$, where the mass resolution is rather
poor due to the hadronic final
state. In \citere{Cao:2016uwt} it was shown
that the excess is consistent with a signal
interpretation corresponding to
a signal strength of
\begin{equation}
\mu_{\rm LEP}^{\rm exp} = 0.117 \pm 0.057 \ .
\end{equation}
Low-mass Higgs-boson searches have
also been performed at the LHC in
various final states. CMS searched
for light Higgs bosons in the diphoton
final state utilizing the 
$8\tev$ and parts of the $13\tev$
datasets~\cite{CMS:2018cyk}.
A local excess of roughly
$3\sigma$ confidence level was observed
at a mass of $96\gev$, hence
in agreement with the mass range compatible
with the LEP excess.
In this case the excess is consistent
with a signal interpretation corresponding
to a signal strength of
\begin{equation}
\mu_{\rm CMS}^{\rm exp} = 0.6 \pm 0.2 \ .
\end{equation}
In our scan, in which $h_1$ will play
the role of the state $h_{96}$, we
compare the theoretical predictions
for the signal strengths to the
experimental values given above.
The predictions were calculated by
\begin{equation}
\mu_{\rm LEP} \approx
\frac{C_{h_{96} V V}^2 \cdot\mathrm{BR}
    \left( h_{96} \to b \bar b \right)}
{\mathrm{BR}^{\rm SM} \left(
    H \to b \bar b \right)} \ , \quad
\mu_{\rm CMS} \approx
\frac{C_{h_{96} t \bar t}^2 \cdot\mathrm{BR}
    \left( h_{96} \to \gamma \gamma \right)}
{\mathrm{BR}^{\rm SM} \left(
    H \to \gamma \gamma \right)} \ .
\end{equation}
Hence, in both cases the cross section ratios
that enter the definitions of
the signal strengths are expressed
to a very good approximation
in terms of
the effective coupling coefficients
${C_{h_{96} V V} = c_{\alpha_2}
c_{\beta - \alpha_1}}$ and
${C_{h_{96} t \bar t} = s_{\alpha_1}
c_{\alpha_2} / s_\beta}$, which,
as mentioned already, are
defined as the couplings
of $h_{96}$ normalized to the respective
couplings of a SM Higgs boson with
the same mass.
The values for the SM branching ratios
in the denominator, again assuming a
SM Higgs boson at $96\gev$, can be
found in the
literature~\cite{LHCHiggsCrossSectionWorkingGroup:2016ypw}.
From the theoretical predicted values
$\mu_{\rm LEP, CMS}$ and the experimentally
determined values $\mu_{\rm LEP, CMS}^{\rm exp}$
and their uncertainties we construct
a $\chi^2$ function
\begin{equation}
\chi^2_{96} =
\frac{\left( \mu_{\rm LEP} - 0.117 \right)^2}
    {0.057^2}
+
\frac{\left( \mu_{\rm CMS} - 0.6 \right)^2}
    {0.2^2}
\ ,
\label{eqchisq96def}
\end{equation}
in order to quantify the goodness of
the fits to the excesses.
In this definition we assumed that there
is no correlation between both measurements.
 
Technically, the details of the scan that
we discuss here are very similar to the
ones of the scan discussed in \refse{secnumanal1}.
The scan ranges were set
as given in \refeq{eqranges1}, except for
the masses of the scalars, which were
chosen to be
\begin{equation}
m_{h_1} = 96 \gev \ , \quad
m_{h_2} = 125.09 \gev \ , \quad
m_{h_3} = m_H \leq 1 \tev \ ,
\end{equation}
such that $m_{h_3} = m_H$ is further constrained
by the condition $\Delta M_{\rm max}
< 200 \gev$, as defined
in \refeq{eqranges1} and substantially
heavier than $h_{1}$ and $h_2$ due to
the lower limit on $m_{H^\pm}$.
We again followed the two-step procedure.
In the first step, we used the genetic algorithm
to obtain parameter points in agreement
with the theoretical constraints and
the experimental constraints from the
Higgs phenomenology. To the loss function
defined in \refeq{eqlossfunc} we added
a term $10 \chi^2_{96}$ in order
to obtain parameter points
that potentially feature both a
good fit to the signal rates of
the SM-like Higgs boson $h_2 = h_{125}$
and to the signal rates $\mu_{\rm LEP}$
and $\mu_{\rm CMS}$. 
All parameter points obtained
by the help of the genetic algorithm
were subject to the constraint
\begin{equation}
\chi^2_{125} + \chi^2_{96} \leq
\chi^2_{\mathrm{SM}, 125} +
    \chi^2_{\mathrm{SM}, 96} \ , \quad
\chi^2_{\mathrm{SM}, 125} = 84.41 \ , \quad
\chi^2_{\mathrm{SM}, 96} = 13.99 \ ,
\label{eqhscond96}
\end{equation}
where the value of $\chi^2_{\mathrm{SM}, 96}$
is obtained from \refeq{eqchisq96def} assuming zero
values for both $\mu_{\rm LEP}$ and
$\mu_{\rm CMS}$ as predicted
by the SM, in which no particle is present
at a mass of $96\gev$.
As a result, in comparison to the analysis
discussed in \refse{secnumanal1}
in which the requirement
$\chi^2_{125} \leq \chi^2_{\mathrm{SM}, 125}$
was used, the requirement shown in
\refeq{eqhscond96} allows for larger values of
$\chi^2_{125}$ as long as the S2HDM parameter point
provides a good fit to the collider excesses, i.e.\
it features values of $\chi^2_{96} \ll
\chi^2_{\mathrm{SM}, 96}$. Here it should
be noted that even in the most extreme case
with $\chi^2_{96} = 0$ the allowed maximum value
of $\chi^2_{125}$ still does not indicate severe
modifications of the signal rates of $h_{125}$,
taking into account that the \texttt{HiggsSignals}
fit result applies a total amount of 107
observables, such that the reduced $\chi^2$
value remains substantially smaller than one
even in this case.
The second step is totally analogue
to the scan discussed in \refse{secnumanal1}.
All parameter points that
pass the constraint shown in \refeq{eqhscond96}
were confronted with the theoretical constraints
including now the RGE evolution of
the parameters. As before, we required the
scalar potential to be well behaved up
to energy scale of at least $\mu_v = 1 \tev$,
such that in particular the values of
the quartic couplings $\lambda_i$ allow for
a perturbative treatment at the range of
energy at which there are also particle
masses in our scan. Finally,
the remaining experimental
constraints regarding the DM phenomenology
were applied.

\begin{figure}
\centering
\includegraphics[width=0.48\textwidth]{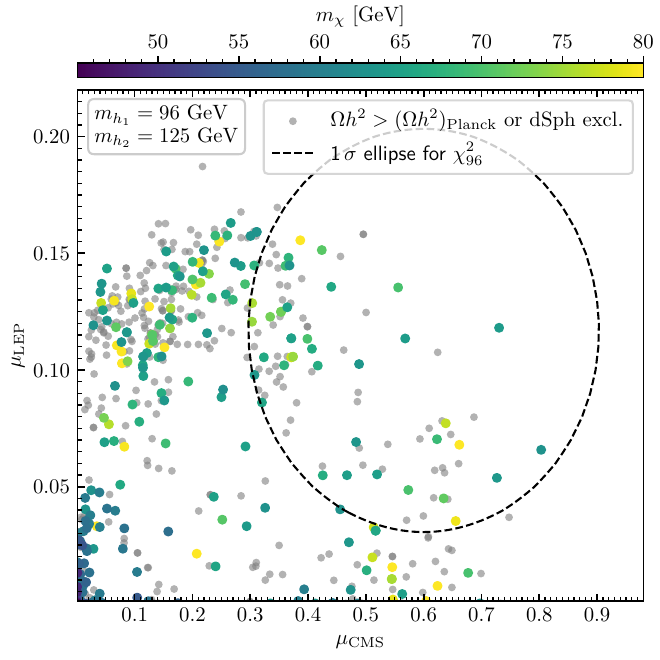}~
\includegraphics[width=0.48\textwidth]{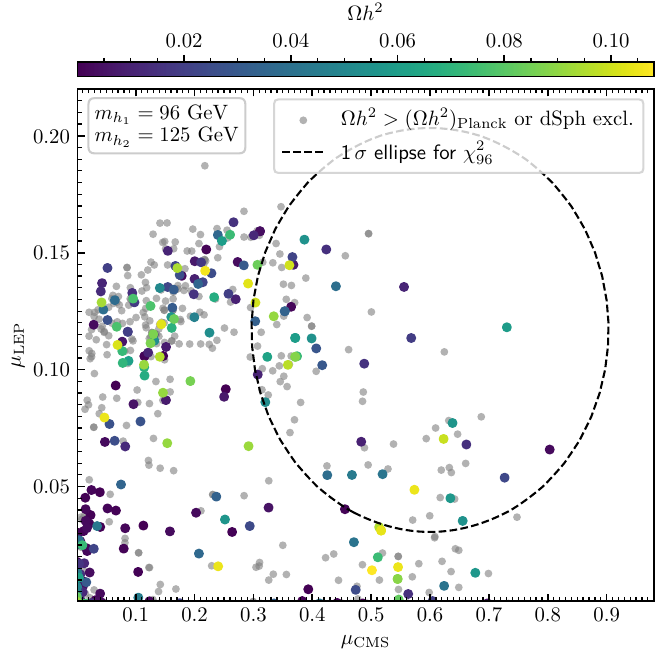}
\caption{\small $\mu_{\rm LEP}$ in dependence of
$\mu_{\rm CMS}$, with the colour coding
indicating the values of $m_\chi$
(left) and $\Omega h^2$ (right).
Grey points are excluded by $\Omega h^2 >
(\Omega h^2)_{\rm Planck}$ or Fermi
dSph measurements. The dashed ellipse
indicates the experimentally
preferred region of the collider excesses
at the $1\sigma$ confidence level.}
\label{fig96cmslep}
\end{figure}

We show the resulting parameter points
in \reffi{fig96cmslep}, where 
we display the signal rate
$\mu_{\rm LEP}$ in dependence of
$\mu_{\rm CMS}$. We indicate
with the colour coding of the points
the value of the DM
mass $m_{\chi}$ (left) 
and the DM relic abundance $\Omega h^2$
as predicted by the usual
thermal freeze-out scenario (right).
Also shown as grey
points are parameter points
that are excluded by a too
large prediction of the relic
abundance or by limits coming 
from observations of dSph.
The ellipse in both plots
indicates the region
in agreement with the
collider excesses
at the $1\sigma$
confidence level, i.e.\ $\chi^2_{96}
= 2.3$. One can see that we find parameter
points within the ellipses. Consequently,
both excesses 
can be explained simultaneously while
taking into account the constraints
described in \refse{secconsts}.
In the left plot, we observe that parameter points
with sizable values of $\mu_{\rm LEP}$
and $\mu_{\rm CMS}$ feature DM mass values
close to or larger than
$m_{h_{125}}/2$.
On the other hand, parameter points
with $m_\chi < m_{h_{125}}/2$ only predict
substantially smaller signal strengths, and
the collider excesses cannot be accounted for.
The reason for this is, as was also
discussed in \refse{secnumanal1}, that in
this case the decay $h_{125} \to \chi \chi$
is kinematically open. As a result, the
possible mixing of the singlet field
$h_1 = h_{96}$ with the SM-like Higgs boson
$h_2 = h_{125}$ is much more constrained.
However, a sizable mixing of $h_{96}$ and
$h_{125}$ is necessary to obtain values of
$\mu_{\rm LEP}$ and $\mu_{\rm CMS}$ of the
order of the experimentally measured values.
We therefore can conclude that a realization
of the collider excesses demands DM masses of
$m_\chi > m_{h_{125}}/2$.
In the right 
plot of \reffi{fig96cmslep}
we find
that several of the parameter
points that are able to explain both 
excesses 
also predict sizable values for
the relic abundance, with some
parameter points saturating the value
measured by the Planck
collaboration. Accordingly, we come to
the conclusion that the S2HDM can accommodate
the collider excesses at $96\gev$ while at
the same time accommodating a large fraction
or all
of the measured DM relic abundance.

\begin{figure}
\centering
\includegraphics[width=0.48\textwidth]{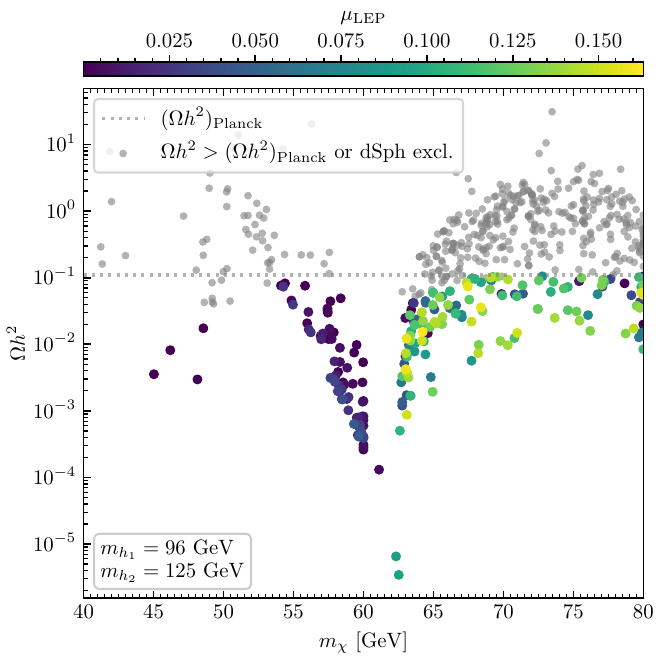}~
\includegraphics[width=0.48\textwidth]{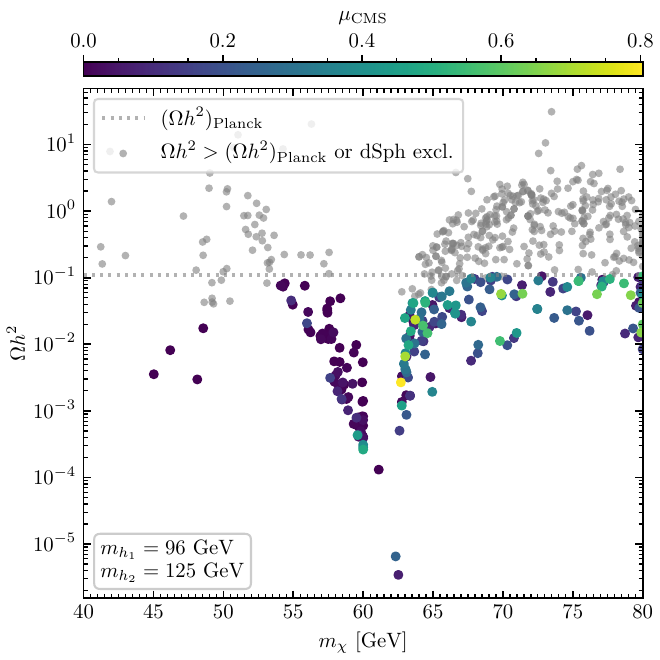}
\caption{\small $\Omega h^2$ in dependence
of $m_\chi$, with the colour coding indicating
the values of $\mu_{\rm LEP}$ (left) and
$\mu_{\rm CMS}$ (right).
Grey points are excluded by $\Omega h^2 >
(\Omega h^2)_{\rm Planck}$ or Fermi
dSph measurements.}
\label{fig96relic}
\end{figure}

In \reffi{fig96relic} the predicted 
relic abundance is shown in dependence of the
DM mass. The values of the signal rates
measured by LEP (left) and CMS (right) are
also indicated by the colour coding
of the points. We note a new prominent
feature in the distribution of the parameter points
with respect to \reffi{figdms}. Due to the 
opening of a new resonant $s$-channel mediated
by the $h_{96}$, parameter points featuring
DM masses smaller than
about $53 \gev$
can now be in agreement
with the upper limit imposed by the
observed
DM relic abundance.
Moreover, the presence of $h_{96}$ also
gives rise to the fact that a large fraction of
parameter points with $m_\chi > m_{h_{125}}
/ 2$ lie below the Planck limit, whereas
we found in \refse{secnumanal1} (compare to
\reffi{figdms}) that in this DM mass region
most points predict $\Omega h^2 >
(\Omega h^2)_{\rm Planck}$.
Grey points that
lay below the experimental
upper limit are excluded 
by dSph observations.
Here it is interesting to note
that we find, in addition to the
region around $m_\chi \sim 63 \gev$
already present in \reffi{figdms},
a second region at $48 \gev \lesssim
m_\chi \lesssim 58 \gev$ in which
the dSph constraints discard points
that would be in agreement with the
Planck measurement of the
DM relic abundance.
In the left plot of \reffi{fig96relic} we
find that for the points at the right side
of the resonance the predicted values
of $\mu_{\rm LEP}$ can be close to the
measured central value $\mu_{\rm LEP}^{\rm exp}
= 0.117$ independently of the precise
value of $m_\chi$. On the contrary, as
can be seen in the right plot of
\reffi{fig96relic}, values of $\mu_{\rm CMS}
\sim \mu_{\rm CMS}^{\rm exp} = 0.6$
that are in agreement with the constraints are mostly
found in the interval $62 \gev \lesssim m_\chi
\lesssim 65 \gev$. For larger values of
$m_\chi$ one can still find parameter points
that fit the CMS excess at the level
of $1\sigma$. However, they often predict
too large values of $\Omega h^2 > (\Omega
h^2)_{\rm Planck}$ and are therefore
shown mostly as grey points. The reason for this is
that, as discussed before, fitting the diphoton
excess requires a suppression of the couplings
of $h_{96}$ to $b$ quarks. However, this then
yields also a suppression of the annihilation
cross section via the process $\chi \chi \to
h_{96} \to b \bar b$.

\begin{figure}
\centering
\includegraphics[width=0.48\textwidth]{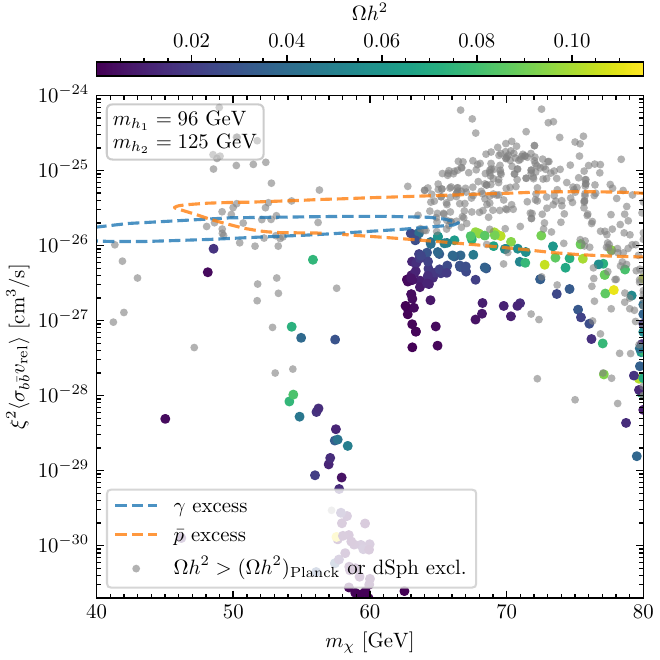}~
\includegraphics[width=0.48\textwidth]{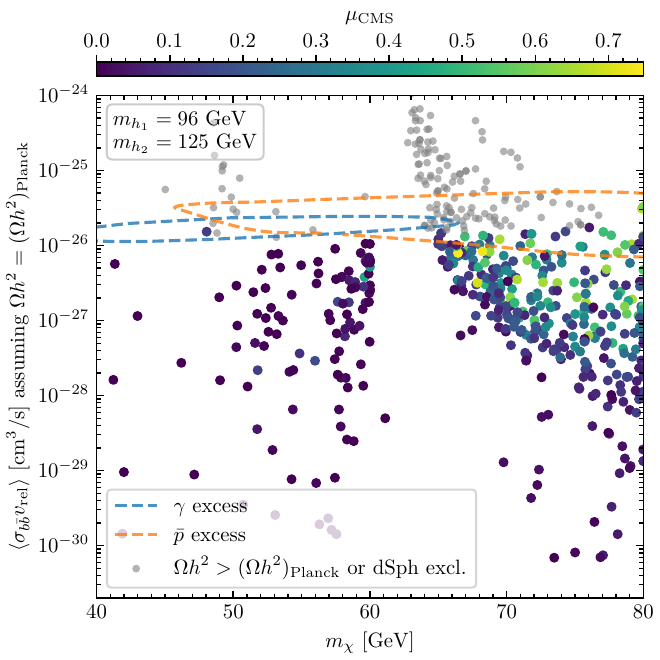}
\caption{\small Today's velocity averaged
annihilation cross section of $\chi$
into pairs of $b$ quarks taking into
account the number density as predicted
by thermal freeze-out (left) and
assuming $\Omega h^2 = 0.119$ (right).
The colour coding indicates the predicted
value of $\Omega h^2$ (left) and the value
of $m_{h_2}$ (right). Also indicated are
the regions in which the cosmic-rays
excesses could be explained (blue and
orange dashed
lines)~\cite{Calore:2014xka,Cholis:2019ejx}.
Grey points are
excluded by
$\Omega h^2 > (\Omega
h^2_{\rm Planck})$
(left)
or Fermi dSph measurements (left and right).}
\label{fig96anni}
\end{figure}

In order to discuss the gamma-ray and the 
antiproton excesses, we show in \reffi{fig96anni}
today's velocity-averaged annihilation
cross section of $\chi$ into pairs of
$b$ quarks taking into account the
number density as predicted by
thermal freeze-out (left) and
assuming $\Omega h^2 = 0.12$ (right), as explained
in \refse{secnumanal1}. In comparison
to \reffi{figdmani}, here we observe that
there are more regions of $m_\chi$ in which
points are found inside the preferred region
to explain both cosmic-ray excesses simultaneously.
These points
remain in tension with
present limits imposed by
the observation of dSph.
We remind the reader
about the 
uncertainties in determining those limits
(see \refse{secexpconstr} for more details). 
Regarding the agreement with the
signal rate $\mu_{\rm CMS}$,
only the parameter points situated towards the 
right end of the blue curve could 
simultaneously explain the two cosmic ray and the CMS
excesses. 
These points are again
in tension
with indirect-detection limits
from dSph observations.
Regardless of whether the collider
excesses are accommodated or not, we see
that the presence of $h_{96}$ gives rise
to more points at the lower end of $m_\chi$
that lie within the blue and the orange
curves. Thus, the new light scalar state
gives rise to new interesting regions
of parameter space with $m_\chi < 60\gev$
in the context of
the cosmic-ray anomalies. However,
as was already mentioned, the collider
excesses, which were the main motivation
to investigate a scenario with $m_{h_1}
= 96 \gev$ in the first place, cannot
be realized here.

\section{Conclusions}
\label{conclu}
In this paper we analyzed a
singlet-extended 2HDM, called S2HDM,
which is a model with a rich Higgs
phenomenology and that incorporates
a pseudo-Nambu-Goldstone boson
dark matter candidate.
We focused
on the parameter space 
of the S2HDM featuring 
DM masses in the Higgs funnel region,
i.e.\ $40 \gev \leq m_\chi \leq 80 \gev$.
One of the main purposes of this analysis
was to illustrate the combined impact
of various theoretical and experimental
constraints on the model parameters,
where in particular the strong interplay
between the Higgs-sector phenomenology
and the DM sector of the S2HDM
was demonstrated.
We required the scalar potential
to be well-behaved up to energy scales of $1\tev$,
i.e.\ to be bounded-from-below, to feature a stable
electroweak vacuum and to fulfill conditions 
derived from perturbative unitarity.
We also ensured that the parameter points
were in agreement with measurements
of electroweak precision observables,
flavour physics, properties of the discovered
Higgs boson at $125\gev$, searches for additional 
scalar states, and the DM observables.
The model exploration of the multi-dimensional
parameter space of the S2HDM was performed with
the help of a genetic algorithm, by which,
compared to random scans of the parameters,
a significant improvement of the computing
time required to find viable parameter
points was achieved.

In our numerical analysis,
we focused on
two benchmark scenarios.
Firstly, we performed a broad
parameter scan assuming that 
the SM-like Higgs boson $h_{125}$ was the lightest 
of the three CP-even Higgs bosons,
such that the predictions for
the DM relic abundance assuming
the usual thermal freeze-out mechanism
are mainly determined
by the resonant $s$-channel annihilation 
mediated by~$h_{125}$.
Secondly, we studied a scenario featuring
a singlet-like CP-even state
$h_{96}$ at $96 \gev$,
where the presence of $h_{96}$ also gives
rise to a second $s$-channel contribution
to the thermal freeze-out cross section
and today’s annihilation cross
section relevant for DM
indirect-detection experiments.

In the first
scenario in which $h_{125}$ is assumed 
to be the lightest CP-even scalar, DM masses
$62.5\gev \lesssim m_\chi \lesssim
67\gev$ were found to be able to 
explain the $\gamma$-ray and antiproton
cosmic rays excesses, while
simultaneously also predicting
values of the DM relic abundance in agreement
with the observations by the Planck
collaboration. However, these parameter
points are in tension with indirect-detection
limits derived from observations of 
dwarf spheroidal galaxies, where
it should be taken into account that
these constraints are
still subject to uncertainties with regards
to the astrophysical modelling of the spectral 
curves.
Concerning the Higgs phenomenology,
we found that
demanding that the theory can
be treated perturbatively
up to energy scales of at least
$1\tev$ has a strong impact on
the Higgs spectra that can be realized.
Namely,
the mass splittings among the 
heavy scalar states
$H^{\pm}$, $A$ and $h_{2,3}$ were
found to be smaller
than roughly $100 \gev$,
driving the model towards the
decoupling limit,
with the only exception of
a very
singlet-like CP-even state $h_{i}$,
which can be substantially
lighter (or heavier)
than the other BSM state
without giving rise to issues with
unperturbative effects at energy
scales below $1\tev$.

In the second scenario,
we studied whether
the S2HDM could offer an explanation
for the collider excesses
observed at about $96 \gev$
at LEP and CMS in
the $b \bar b$ and the diphoton final
state, respectively.
Here we found that a singlet-like
CP-even Higgs boson at $96\gev$
can reproduce both
collider excesses
under the constraint that
$m_\chi > m_{h_{125}}/2$ in order to
allow for a sizable mixing between
$h_{96}$ and $h_{125}$.
Furthermore, it is possible to accommodate 
at the same time a large fraction or all of the 
measured DM relic abundance.
Finally, we found that the
simultaneous explanation of the 
cosmic-ray excesses and the collider
excess at $96\gev$ is in principle
possible, but, as in the first scenario,
the parameter points are also in tension 
with limits arising from observations
of dwarf spheroidal galaxies.

To summarize, we demonstrated that the
S2HDM is an attractive model that
can accommodate a rich phenomenology
and an interesting interplay between
the DM sector and the Higgs sector.
We also showed that it is crucial to
take into account the various theoretical
and experimental constraints on the
model parameters. For future studies,
we make our implementation of the model
predictions and the application of the
constraints available to the public in
the form of a \texttt{python} package called
\texttt{s2hdmTools}, which is briefly
described in \refap{seccode}.

\section*{Acknowledgements}
We thank V.~Mart\'in-Lozano for interesting
discussions and for bringing the
investigated model to our attention.
The work of T.B.\ is supported by the Deutsche
Forschungsgemeinschaft under Germany’s Excellence
Strategy EXC2121 ``Quantum Universe'' - 390833306.

\appendix

\section{Parameter transformations}
\label{secparas}
In the following we give the transformation
formulae between the basis of the
Lagrangian parameters and the physical basis chosen
to scan the parameter space of the
S2HDM defined in \refeq{eqparas}.
The quartic couplings $\lambda_{i}$ can be
written in terms of the physical basis as
\begin{align}
    \lambda_{1} &= \frac{1}{v^2c_{\beta}^2}\left(-M^2s_{\beta}^2+  \sum_{n=1}^{3}m_{h_{i}}^2R_{i1}^2\right) \ , \\
    \lambda_{2} &= \frac{1}{v^2s_{\beta}^2}\left(-M^2c_{\beta}^2+  \sum_{n=1}^{3}m_{h_{i}}^2R_{i2}^2\right) \ , \\
    \lambda_{3} &= \frac{1}{v^2}\left(-M^2 + \frac{1}{c_{\beta}s_{\beta}}\left( \sum_{n=1}^{3}m_{h_{i}}^2R_{i1}^2R_{i2}^2\right) + 2m_{H^{\pm}}^2\right) \ , \\
    \lambda_{4} &= \frac{1}{v^2}\left(M^2 + m_{A}^{2} - 2 m_{H^{\pm}}^2 \right) \ , \\
    \lambda_{5} &= \frac{1}{v^2}\left(M^2 - m_{A}^{2} \right) \ , \\
    \lambda_{6} &= \frac{1}{v_{S}^2}\left(\sum_{n=1}^{3}m_{h_{i}}^2R_{i3}^2 \right) \ , \\
    \lambda_{7} &= \frac{1}{v v_{S}c_{\beta}}\left(\sum_{n=1}^{3}m_{h_{i}}^2R_{i1}R_{i3} \right) \ , \\
    \lambda_{8} &= \frac{1}{v v_{S}s_{\beta}}\left(\sum_{n=1}^{3}m_{h_{i}}^2R_{i2}R_{i3} \right) \ , 
\end{align}
where the matrix elements $R_{ij}$ have been
defined in terms of the mixing
angles $\alpha_{1,2,3}$ in \refeq{mixingmatrix}.
With the previous transformations, one
can also compute the mass parameters in
the scalar potential using the tadpole equations as follows,
\begin{align}
    \mu_{11}^2 &= m_{12}^{2}\tan\beta -\frac{1}{2}\left(\lambda_{1}v^{2}c_{\beta}^{2}+(\lambda_{3}+\lambda_{4}+\lambda_{5})v^2s_{\beta}^2 + \lambda_{7}v_{S}^{2}\right) \ , \\
    \mu_{22}^2 &= \frac{m_{12}^{2}}{\tan\beta} -\frac{1}{2}\left(\lambda_{2} v^2 s_{\beta}^2 + (\lambda_{3}+\lambda_{4}+\lambda_{5})v^2c_{\beta}^2 +  \lambda_{8}v_{S}^{2}\right) \ , \\
    \mu_\chi^2 &= m_{\chi}^{2} \ , \\
    \mu_{S}^2 &= m_{\chi}^{2} -  \left(\lambda_{7} v^2 c_{\beta}^2 + \lambda_{8} v^2 s_{\beta}^2 + \lambda_{6}v_{S}^2 \right) \ .
\end{align}

\section{Tree-level perturbative unitarity constraints}
\label{apppert}
Here we list the tree-level perturbative unitarity
conditions that were applied in our analysis
in order to exclude parameter points which cannot
be treated perturbatively:
\begin{align}
\left| \lambda _3+\lambda _4 \right| &\leq 8 \pi \\
\frac{1}{2} \left| \lambda _1+\lambda _2-
\sqrt{\lambda _1^2-2 \lambda _2 \lambda _1+
\lambda _2^2+4 \lambda _4^2}\right| &\leq 8 \pi \\
\frac{1}{2} \left| \lambda _1+\lambda _2+
\sqrt{\lambda _1^2-2 \lambda _2 \lambda _1+
\lambda _2^2+4 \lambda _4^2}\right| &\leq 8 \pi \\
\left| \lambda _3+2 \lambda _4-3 \lambda _5\right| &\leq 8 \pi
\label{eqkiller} \\
\left| \lambda _3-\lambda _5\right| &\leq 8 \pi \\
\left| \lambda _3+\lambda _5\right| &\leq 8 \pi \\
\left| \lambda _3+2 \lambda _4+3 \lambda _5\right| &\leq 8 \pi \\
\frac{1}{2} \left| \lambda _1+\lambda _2-
\sqrt{\lambda _1^2-2 \lambda _2 \lambda _1+
\lambda _2^2+4 \lambda _5^2}\right| &\leq 8 \pi \\
\frac{1}{2} \left| \lambda _1+\lambda _2+
\sqrt{\lambda _1^2-2 \lambda _2 \lambda _1+
\lambda _2^2+4 \lambda _5^2}\right| &\leq 8 \pi
 \\
\left| \lambda _6\right| &\leq 8 \pi \\
\left| \lambda _7\right| &\leq 8 \pi \\
\left| \lambda _8\right| &\leq 8 \pi \\
\left| \lambda _3-\lambda _4\right| &\leq 8 \pi \\
\frac{1}{2} \Bigg| \text{Roots}\Big[x^3+\left(-6 \lambda _1-
6 \lambda _2-4 \lambda _6\right) x^2+\big(-16 \lambda _3^2-16
\lambda _4 \lambda _3-4 \lambda _4^2-&8 \lambda _7^2-
8 \lambda _8^2+ \notag \\
36 \lambda _1 \lambda _2+24 \lambda _1
\lambda _6+24 \lambda _2 \lambda _6\big) x+
48 \lambda _2 \lambda _7^2+48 \lambda _1 \lambda _8^2+
64 \lambda _3^2 \lambda _6&+16 \lambda _4^2 \lambda _6- \notag \\
144 \lambda _1 \lambda _2 \lambda _6+
64 \lambda _3 \lambda _4 \lambda _6-
64 \lambda _3 \lambda _7 \lambda _8-
&32 \lambda _4 \lambda _7 \lambda _8, x\Big]\Bigg| \leq 8 \pi
\label{eqroots}
\end{align}

\section{The python package \texttt{s2hdmTools}}
\label{seccode}
In order to make the analysis of the S2HDM
as performed here publicly available, we
developed the python package \texttt{s2hdmTools}.
The code can be downloaded at
\url{https://gitlab.com/thomas.biekoetter/s2hdmtools}.
The installation requires a \texttt{python3}
environment and compilers for \texttt{Fortran},
\texttt{C} and \texttt{C++}. In addition,
a \texttt{python2} installation is required for
the external code \texttt{MadDM}.
During the installation process, the following
external codes are downloaded and installed:
\begin{itemize}
\item[-] \texttt{AnyHdecay}~\cite{Djouadi:1997yw,
Butterworth:2010ym,Djouadi:2018xqq,
Muhlleitner:2016mzt,Engeln:2018mbg}:
Calculates partial decay widths
of the Higgs bosons
\item[-] \texttt{HiggsBounds}~\cite{Bechtle:2008jh,
Bechtle:2011sb,Bechtle:2013gu,Bechtle:2013wla,
Bechtle:2015pma,Bechtle:2020pkv}: Tests against constraints from
Higgs-boson searches at colliders
\item[-] \texttt{HiggsSignals}~\cite{Bechtle:2013xfa,
Stal:2013hwa,Bechtle:2014ewa,Bechtle:2020uwn}:
Test against constraints from
measurements of $h_{125}$
\item[-] \texttt{Hom4PS2}~\cite{Lee:hom4ps2}:
Solver of system of polynomial equations
\item[-] \texttt{MicrOmegas}~\cite{Belanger:2018ccd}:
Calculation of relic
abundance of dark matter
\item[-] \texttt{MadGraph}~\cite{Alwall:2014hca}: Monte-Carlo event
generator
\item[-] \texttt{MadDM}~\cite{Backovic:2013dpa,
Ambrogi:2018jqj}: Test against constraints
from indirect-detecion experiments for dark matter
\end{itemize}
The user interface of
\texttt{s2hdmTools} is defined in the class
\texttt{ParamPoint} and briefly summarized
in the \texttt{README}. The most important
features can be accessed via the following
functions defined in \texttt{ParamPoint}:
\begin{itemize}
\item[-] \texttt{pt = ParamPoint(dc)}: Initializes
a parameter point given a dictionary \texttt{dc}
containing the values of the input parameters
\item[-] \texttt{pt.check_theory_constraints()}:
Verifies whether the theoretical constraints
are fulfilled, optionally up to an energy scale
as provided by the user
\item[-] \texttt{hbhs.check_point(pt)}: Applies
the \texttt{HiggsBounds}- and \texttt{HiggsSignals}
test
\item[-] \texttt{pt.check_ewpo()}: Checks against
constraints from electroweak precision observables
\item[-] \texttt{pt.check_darkmatter()}: Computes
dark matter observables and checks against constraints
on the relic abundance and on
today's annihilation cross sections
from dSph observations
\end{itemize}
More instructions regarding the
installation and the usage of
the package can
be found
in the documentation under
the link:
\begin{center}
\url{https://www.desy.de/~biek/s2hdmtoolsdocu/site/}
\end{center}
In addition, the
application folder of the \texttt{git}
repository contains some of the
scripts that were used in order to produce
the results discussed here.

\bibliographystyle{JHEP}
\bibliography{lit}

\providecommand{\href}[2]{#2}\begingroup\raggedright\begin{thebibliography}{100}

\bibitem{ATLAS:2012yve}
{\scshape ATLAS} collaboration, \emph{{Observation of a new particle in the
  search for the Standard Model Higgs boson with the ATLAS detector at the
  LHC}}, \href{https://doi.org/10.1016/j.physletb.2012.08.020}{\emph{Phys.
  Lett. B} {\bfseries 716} (2012) 1}
  [\href{https://arxiv.org/abs/1207.7214}{{\ttfamily 1207.7214}}].

\bibitem{CMS:2012qbp}
{\scshape CMS} collaboration, \emph{{Observation of a New Boson at a Mass of
  125 GeV with the CMS Experiment at the LHC}},
  \href{https://doi.org/10.1016/j.physletb.2012.08.021}{\emph{Phys. Lett. B}
  {\bfseries 716} (2012) 30} [\href{https://arxiv.org/abs/1207.7235}{{\ttfamily
  1207.7235}}].

\bibitem{Khachatryan:2016vau}
{\scshape ATLAS, CMS} collaboration, \emph{{Measurements of the Higgs boson
  production and decay rates and constraints on its couplings from a combined
  ATLAS and CMS analysis of the LHC pp collision data at $ \sqrt{s}=7 $ and 8
  TeV}}, \href{https://doi.org/10.1007/JHEP08(2016)045}{\emph{JHEP} {\bfseries
  08} (2016) 045} [\href{https://arxiv.org/abs/1606.02266}{{\ttfamily
  1606.02266}}].

\bibitem{Aad:2019mbh}
{\scshape ATLAS} collaboration, \emph{{Combined measurements of Higgs boson
  production and decay using up to $80$ fb$^{-1}$ of proton-proton collision
  data at $\sqrt{s}=$ 13 TeV collected with the ATLAS experiment}},
  \href{https://doi.org/10.1103/PhysRevD.101.012002}{\emph{Phys. Rev. D}
  {\bfseries 101} (2020) 012002}
  [\href{https://arxiv.org/abs/1909.02845}{{\ttfamily 1909.02845}}].

\bibitem{Sirunyan:2018koj}
{\scshape CMS} collaboration, \emph{{Combined measurements of Higgs boson
  couplings in proton\textendash{}proton collisions at $\sqrt{s}=13\,\text
  {Te}\text {V} $}},
  \href{https://doi.org/10.1140/epjc/s10052-019-6909-y}{\emph{Eur. Phys. J. C}
  {\bfseries 79} (2019) 421}
  [\href{https://arxiv.org/abs/1809.10733}{{\ttfamily 1809.10733}}].

\bibitem{Zwicky:1933gu}
F.~Zwicky, \emph{{Die Rotverschiebung von extragalaktischen Nebeln}},
  \href{https://doi.org/10.1007/s10714-008-0707-4}{\emph{Helv. Phys. Acta}
  {\bfseries 6} (1933) 110}.

\bibitem{Rubin:1970zza}
V.C.~Rubin and W.K.~Ford, Jr., \emph{{Rotation of the Andromeda Nebula from a
  Spectroscopic Survey of Emission Regions}},
  \href{https://doi.org/10.1086/150317}{\emph{Astrophys. J.} {\bfseries 159}
  (1970) 379}.

\bibitem{Massey:2010hh}
R.~Massey, T.~Kitching and J.~Richard, \emph{{The dark matter of gravitational
  lensing}}, \href{https://doi.org/10.1088/0034-4885/73/8/086901}{\emph{Rept.
  Prog. Phys.} {\bfseries 73} (2010) 086901}
  [\href{https://arxiv.org/abs/1001.1739}{{\ttfamily 1001.1739}}].

\bibitem{Clowe:2006eq}
D.~Clowe, M.~Bradac, A.H.~Gonzalez, M.~Markevitch, S.W.~Randall, C.~Jones
  et~al., \emph{{A direct empirical proof of the existence of dark matter}},
  \href{https://doi.org/10.1086/508162}{\emph{Astrophys. J. Lett.} {\bfseries
  648} (2006) L109} [\href{https://arxiv.org/abs/astro-ph/0608407}{{\ttfamily
  astro-ph/0608407}}].

\bibitem{Planck:2018vyg}
{\scshape Planck} collaboration, \emph{{Planck 2018 results. VI. Cosmological
  parameters}},
  \href{https://doi.org/10.1051/0004-6361/201833910}{\emph{Astron. Astrophys.}
  {\bfseries 641} (2020) A6}
  [\href{https://arxiv.org/abs/1807.06209}{{\ttfamily 1807.06209}}].

\bibitem{Patt:2006fw}
B.~Patt and F.~Wilczek, \emph{{Higgs-field portal into hidden sectors}},
  \href{https://arxiv.org/abs/hep-ph/0605188}{{\ttfamily hep-ph/0605188}}.

\bibitem{Barbieri:2006dq}
R.~Barbieri, L.J.~Hall and V.S.~Rychkov, \emph{{Improved naturalness with a
  heavy Higgs: An Alternative road to LHC physics}},
  \href{https://doi.org/10.1103/PhysRevD.74.015007}{\emph{Phys. Rev. D}
  {\bfseries 74} (2006) 015007}
  [\href{https://arxiv.org/abs/hep-ph/0603188}{{\ttfamily hep-ph/0603188}}].

\bibitem{Schumann:2019eaa}
M.~Schumann, \emph{{Direct Detection of WIMP Dark Matter: Concepts and
  Status}}, \href{https://doi.org/10.1088/1361-6471/ab2ea5}{\emph{J. Phys. G}
  {\bfseries 46} (2019) 103003}
  [\href{https://arxiv.org/abs/1903.03026}{{\ttfamily 1903.03026}}].

\bibitem{Barger:2008jx}
V.~Barger, P.~Langacker, M.~McCaskey, M.~Ramsey-Musolf and G.~Shaughnessy,
  \emph{{Complex Singlet Extension of the Standard Model}},
  \href{https://doi.org/10.1103/PhysRevD.79.015018}{\emph{Phys. Rev. D}
  {\bfseries 79} (2009) 015018}
  [\href{https://arxiv.org/abs/0811.0393}{{\ttfamily 0811.0393}}].

\bibitem{Barger:2010yn}
V.~Barger, M.~McCaskey and G.~Shaughnessy, \emph{{Complex Scalar Dark Matter
  vis-\textbackslash{}`{a}-vis CoGeNT, DAMA/LIBRA and XENON100}},
  \href{https://doi.org/10.1103/PhysRevD.82.035019}{\emph{Phys. Rev. D}
  {\bfseries 82} (2010) 035019}
  [\href{https://arxiv.org/abs/1005.3328}{{\ttfamily 1005.3328}}].

\bibitem{Barducci:2016fue}
D.~Barducci, A.~Bharucha, N.~Desai, M.~Frigerio, B.~Fuks, A.~Goudelis et~al.,
  \emph{{Monojet searches for momentum-dependent dark matter interactions}},
  \href{https://doi.org/10.1007/JHEP01(2017)078}{\emph{JHEP} {\bfseries 01}
  (2017) 078} [\href{https://arxiv.org/abs/1609.07490}{{\ttfamily
  1609.07490}}].

\bibitem{Gross:2017dan}
C.~Gross, O.~Lebedev and T.~Toma, \emph{{Cancellation Mechanism for
  Dark-Matter\textendash{}Nucleon Interaction}},
  \href{https://doi.org/10.1103/PhysRevLett.119.191801}{\emph{Phys. Rev. Lett.}
  {\bfseries 119} (2017) 191801}
  [\href{https://arxiv.org/abs/1708.02253}{{\ttfamily 1708.02253}}].

\bibitem{Balkin:2018tma}
R.~Balkin, M.~Ruhdorfer, E.~Salvioni and A.~Weiler, \emph{{Dark matter shifts
  away from direct detection}},
  \href{https://doi.org/10.1088/1475-7516/2018/11/050}{\emph{JCAP} {\bfseries
  11} (2018) 050} [\href{https://arxiv.org/abs/1809.09106}{{\ttfamily
  1809.09106}}].

\bibitem{Huitu:2018gbc}
K.~Huitu, N.~Koivunen, O.~Lebedev, S.~Mondal and T.~Toma, \emph{{Probing
  pseudo-Goldstone dark matter at the LHC}},
  \href{https://doi.org/10.1103/PhysRevD.100.015009}{\emph{Phys. Rev. D}
  {\bfseries 100} (2019) 015009}
  [\href{https://arxiv.org/abs/1812.05952}{{\ttfamily 1812.05952}}].

\bibitem{Karamitros:2019ewv}
D.~Karamitros, \emph{{Pseudo Nambu-Goldstone Dark Matter: Examples of Vanishing
  Direct Detection Cross Section}},
  \href{https://doi.org/10.1103/PhysRevD.99.095036}{\emph{Phys. Rev. D}
  {\bfseries 99} (2019) 095036}
  [\href{https://arxiv.org/abs/1901.09751}{{\ttfamily 1901.09751}}].

\bibitem{Azevedo:2018exj}
D.~Azevedo, M.~Duch, B.~Grzadkowski, D.~Huang, M.~Iglicki and R.~Santos,
  \emph{{One-loop contribution to dark-matter-nucleon scattering in the
  pseudo-scalar dark matter model}},
  \href{https://doi.org/10.1007/JHEP01(2019)138}{\emph{JHEP} {\bfseries 01}
  (2019) 138} [\href{https://arxiv.org/abs/1810.06105}{{\ttfamily
  1810.06105}}].

\bibitem{Ishiwata:2018sdi}
K.~Ishiwata and T.~Toma, \emph{{Probing pseudo Nambu-Goldstone boson dark
  matter at loop level}},
  \href{https://doi.org/10.1007/JHEP12(2018)089}{\emph{JHEP} {\bfseries 12}
  (2018) 089} [\href{https://arxiv.org/abs/1810.08139}{{\ttfamily
  1810.08139}}].

\bibitem{Arina:2019tib}
C.~Arina, A.~Beniwal, C.~Degrande, J.~Heisig and A.~Scaffidi, \emph{{Global fit
  of pseudo-Nambu-Goldstone Dark Matter}},
  \href{https://doi.org/10.1007/JHEP04(2020)015}{\emph{JHEP} {\bfseries 04}
  (2020) 015} [\href{https://arxiv.org/abs/1912.04008}{{\ttfamily
  1912.04008}}].

\bibitem{Chiang:2019oms}
C.-W.~Chiang and B.-Q.~Lu, \emph{{First-order electroweak phase transition in a
  complex singlet model with $\mathbb{Z}_3$ symmetry}},
  \href{https://doi.org/10.1007/JHEP07(2020)082}{\emph{JHEP} {\bfseries 07}
  (2020) 082} [\href{https://arxiv.org/abs/1912.12634}{{\ttfamily
  1912.12634}}].

\bibitem{Cline:2019okt}
J.M.~Cline and T.~Toma, \emph{{Pseudo-Goldstone dark matter confronts cosmic
  ray and collider anomalies}},
  \href{https://doi.org/10.1103/PhysRevD.100.035023}{\emph{Phys. Rev. D}
  {\bfseries 100} (2019) 035023}
  [\href{https://arxiv.org/abs/1906.02175}{{\ttfamily 1906.02175}}].

\bibitem{Ahmed:2020hiw}
A.~Ahmed, S.~Najjari and C.B.~Verhaaren, \emph{{A Minimal Model for Neutral
  Naturalness and pseudo-Nambu-Goldstone Dark Matter}},
  \href{https://doi.org/10.1007/JHEP06(2020)007}{\emph{JHEP} {\bfseries 06}
  (2020) 007} [\href{https://arxiv.org/abs/2003.08947}{{\ttfamily
  2003.08947}}].

\bibitem{Abe:2020dut}
Y.~Abe, Y.~Hamada, T.~Ohata, K.~Suzuki and K.~Yoshioka, \emph{{TeV-scale
  Majorogenesis}}, \href{https://doi.org/10.1007/JHEP07(2020)105}{\emph{JHEP}
  {\bfseries 07} (2020) 105}
  [\href{https://arxiv.org/abs/2004.00599}{{\ttfamily 2004.00599}}].

\bibitem{Glaus:2020ihj}
S.~Glaus, M.~M\"uhlleitner, J.~M\"uller, S.~Patel, T.~R\"omer and R.~Santos,
  \emph{{Electroweak Corrections in a Pseudo-Nambu Goldstone Dark Matter Model
  Revisited}}, \href{https://doi.org/10.1007/JHEP12(2020)034}{\emph{JHEP}
  {\bfseries 12} (2020) 034}
  [\href{https://arxiv.org/abs/2008.12985}{{\ttfamily 2008.12985}}].

\bibitem{Zhang:2021alu}
Z.~Zhang, C.~Cai, X.-M.~Jiang, Y.-L.~Tang, Z.-H.~Yu and H.-H.~Zhang,
  \emph{{Phase transition gravitational waves from pseudo-Nambu-Goldstone dark
  matter and two Higgs doublets}},
  \href{https://doi.org/10.1007/JHEP05(2021)160}{\emph{JHEP} {\bfseries 05}
  (2021) 160} [\href{https://arxiv.org/abs/2102.01588}{{\ttfamily
  2102.01588}}].

\bibitem{Haisch:2021ugv}
U.~Haisch, G.~Polesello and S.~Schulte, \emph{{Searching for pseudo
  Nambu-Goldstone boson dark matter production in association with top
  quarks}},  \href{https://arxiv.org/abs/2107.12389}{{\ttfamily 2107.12389}}.

\bibitem{Branco:2011iw}
G.C.~Branco, P.M.~Ferreira, L.~Lavoura, M.N.~Rebelo, M.~Sher and J.P.~Silva,
  \emph{{Theory and phenomenology of two-Higgs-doublet models}},
  \href{https://doi.org/10.1016/j.physrep.2012.02.002}{\emph{Phys. Rept.}
  {\bfseries 516} (2012) 1} [\href{https://arxiv.org/abs/1106.0034}{{\ttfamily
  1106.0034}}].

\bibitem{Kannike:2019wsn}
K.~Kannike and M.~Raidal, \emph{{Phase Transitions and Gravitational Wave Tests
  of Pseudo-Goldstone Dark Matter in the Softly Broken U(1) Scalar Singlet
  Model}}, \href{https://doi.org/10.1103/PhysRevD.99.115010}{\emph{Phys. Rev.
  D} {\bfseries 99} (2019) 115010}
  [\href{https://arxiv.org/abs/1901.03333}{{\ttfamily 1901.03333}}].

\bibitem{Biekotter:2021ysx}
T.~Biek\"otter, S.~Heinemeyer, J.M.~No, M.O.~Olea and G.~Weiglein, \emph{{Fate
  of electroweak symmetry in the early Universe: Non-restoration and trapped
  vacua in the N2HDM}},
  \href{https://doi.org/10.1088/1475-7516/2021/06/018}{\emph{JCAP} {\bfseries
  06} (2021) 018} [\href{https://arxiv.org/abs/2103.12707}{{\ttfamily
  2103.12707}}].

\bibitem{Kuzmin:1985mm}
V.A.~Kuzmin, V.A.~Rubakov and M.E.~Shaposhnikov, \emph{{On the Anomalous
  Electroweak Baryon Number Nonconservation in the Early Universe}},
  \href{https://doi.org/10.1016/0370-2693(85)91028-7}{\emph{Phys. Lett. B}
  {\bfseries 155} (1985) 36}.

\bibitem{Cline:1996mga}
J.M.~Cline and P.-A.~Lemieux, \emph{{Electroweak phase transition in two Higgs
  doublet models}}, \href{https://doi.org/10.1103/PhysRevD.55.3873}{\emph{Phys.
  Rev. D} {\bfseries 55} (1997) 3873}
  [\href{https://arxiv.org/abs/hep-ph/9609240}{{\ttfamily hep-ph/9609240}}].

\bibitem{Jiang:2019soj}
X.-M.~Jiang, C.~Cai, Z.-H.~Yu, Y.-P.~Zeng and H.-H.~Zhang,
  \emph{{Pseudo-Nambu-Goldstone dark matter and two-Higgs-doublet models}},
  \href{https://doi.org/10.1103/PhysRevD.100.075011}{\emph{Phys. Rev. D}
  {\bfseries 100} (2019) 075011}
  [\href{https://arxiv.org/abs/1907.09684}{{\ttfamily 1907.09684}}].

\bibitem{Veltman:1980mj}
M.J.G.~Veltman, \emph{{The Infrared - Ultraviolet Connection}}, {\emph{Acta
  Phys. Polon. B} {\bfseries 12} (1981) 437}.

\bibitem{Dimopoulos:1981zb}
S.~Dimopoulos and H.~Georgi, \emph{{Softly Broken Supersymmetry and SU(5)}},
  \href{https://doi.org/10.1016/0550-3213(81)90522-8}{\emph{Nucl. Phys. B}
  {\bfseries 193} (1981) 150}.

\bibitem{Witten:1981nf}
E.~Witten, \emph{{Dynamical Breaking of Supersymmetry}},
  \href{https://doi.org/10.1016/0550-3213(81)90006-7}{\emph{Nucl. Phys. B}
  {\bfseries 188} (1981) 513}.

\bibitem{Kim:1986ax}
J.E.~Kim, \emph{{Light Pseudoscalars, Particle Physics and Cosmology}},
  \href{https://doi.org/10.1016/0370-1573(87)90017-2}{\emph{Phys. Rept.}
  {\bfseries 150} (1987) 1}.

\bibitem{Fayet:2020bmb}
P.~Fayet, \emph{{$U$ boson interpolating between a generalized dark photon or
  dark $Z$ , an axial boson, and an axionlike particle}},
  \href{https://doi.org/10.1103/PhysRevD.103.035034}{\emph{Phys. Rev. D}
  {\bfseries 103} (2021) 035034}
  [\href{https://arxiv.org/abs/2010.04673}{{\ttfamily 2010.04673}}].

\bibitem{Frigerio:2012uc}
M.~Frigerio, A.~Pomarol, F.~Riva and A.~Urbano, \emph{{Composite Scalar Dark
  Matter}}, \href{https://doi.org/10.1007/JHEP07(2012)015}{\emph{JHEP}
  {\bfseries 07} (2012) 015} [\href{https://arxiv.org/abs/1204.2808}{{\ttfamily
  1204.2808}}].

\bibitem{Mrazek:2011iu}
J.~Mrazek, A.~Pomarol, R.~Rattazzi, M.~Redi, J.~Serra and A.~Wulzer, \emph{{The
  Other Natural Two Higgs Doublet Model}},
  \href{https://doi.org/10.1016/j.nuclphysb.2011.07.008}{\emph{Nucl. Phys. B}
  {\bfseries 853} (2011) 1} [\href{https://arxiv.org/abs/1105.5403}{{\ttfamily
  1105.5403}}].

\bibitem{DeCurtis:2016tsm}
S.~De~Curtis, S.~Moretti, K.~Yagyu and E.~Yildirim, \emph{{LHC Phenomenology of
  Composite 2-Higgs Doublet Models}},
  \href{https://doi.org/10.1140/epjc/s10052-017-5082-4}{\emph{Eur. Phys. J. C}
  {\bfseries 77} (2017) 513}
  [\href{https://arxiv.org/abs/1610.02687}{{\ttfamily 1610.02687}}].

\bibitem{DeCurtis:2018zvh}
S.~De~Curtis, L.~Delle~Rose, S.~Moretti and K.~Yagyu, \emph{{A Concrete
  Composite 2-Higgs Doublet Model}},
  \href{https://doi.org/10.1007/JHEP12(2018)051}{\emph{JHEP} {\bfseries 12}
  (2018) 051} [\href{https://arxiv.org/abs/1810.06465}{{\ttfamily
  1810.06465}}].

\bibitem{Vieu:2018nfq}
T.~Vieu, A.P.~Morais and R.~Pasechnik, \emph{{Electroweak phase transitions in
  multi-Higgs models: the case of Trinification-inspired THDSM}},
  \href{https://doi.org/10.1088/1475-7516/2018/07/014}{\emph{JCAP} {\bfseries
  07} (2018) 014} [\href{https://arxiv.org/abs/1801.02670}{{\ttfamily
  1801.02670}}].

\bibitem{Fermi-LAT:2016uux}
{\scshape Fermi-LAT, DES} collaboration, \emph{{Searching for Dark Matter
  Annihilation in Recently Discovered Milky Way Satellites with Fermi-LAT}},
  \href{https://doi.org/10.3847/1538-4357/834/2/110}{\emph{Astrophys. J.}
  {\bfseries 834} (2017) 110}
  [\href{https://arxiv.org/abs/1611.03184}{{\ttfamily 1611.03184}}].

\bibitem{Fermi-LAT:2017opo}
{\scshape Fermi-LAT} collaboration, \emph{{The Fermi Galactic Center GeV Excess
  and Implications for Dark Matter}},
  \href{https://doi.org/10.3847/1538-4357/aa6cab}{\emph{Astrophys. J.}
  {\bfseries 840} (2017) 43}
  [\href{https://arxiv.org/abs/1704.03910}{{\ttfamily 1704.03910}}].

\bibitem{Fermi-LAT:2015sau}
{\scshape Fermi-LAT} collaboration, \emph{{Fermi-LAT Observations of
  High-Energy $\gamma$-Ray Emission Toward the Galactic Center}},
  \href{https://doi.org/10.3847/0004-637X/819/1/44}{\emph{Astrophys. J.}
  {\bfseries 819} (2016) 44}
  [\href{https://arxiv.org/abs/1511.02938}{{\ttfamily 1511.02938}}].

\bibitem{Hooper:2010mq}
D.~Hooper and L.~Goodenough, \emph{{Dark Matter Annihilation in The Galactic
  Center As Seen by the Fermi Gamma Ray Space Telescope}},
  \href{https://doi.org/10.1016/j.physletb.2011.02.029}{\emph{Phys. Lett. B}
  {\bfseries 697} (2011) 412}
  [\href{https://arxiv.org/abs/1010.2752}{{\ttfamily 1010.2752}}].

\bibitem{Hooper:2011ti}
D.~Hooper and T.~Linden, \emph{{On The Origin Of The Gamma Rays From The
  Galactic Center}},
  \href{https://doi.org/10.1103/PhysRevD.84.123005}{\emph{Phys. Rev. D}
  {\bfseries 84} (2011) 123005}
  [\href{https://arxiv.org/abs/1110.0006}{{\ttfamily 1110.0006}}].

\bibitem{Hooper:2013rwa}
D.~Hooper and T.R.~Slatyer, \emph{{Two Emission Mechanisms in the Fermi
  Bubbles: A Possible Signal of Annihilating Dark Matter}},
  \href{https://doi.org/10.1016/j.dark.2013.06.003}{\emph{Phys. Dark Univ.}
  {\bfseries 2} (2013) 118} [\href{https://arxiv.org/abs/1302.6589}{{\ttfamily
  1302.6589}}].

\bibitem{Daylan:2014rsa}
T.~Daylan, D.P.~Finkbeiner, D.~Hooper, T.~Linden, S.K.N.~Portillo, N.L.~Rodd
  et~al., \emph{{The characterization of the gamma-ray signal from the central
  Milky Way: A case for annihilating dark matter}},
  \href{https://doi.org/10.1016/j.dark.2015.12.005}{\emph{Phys. Dark Univ.}
  {\bfseries 12} (2016) 1} [\href{https://arxiv.org/abs/1402.6703}{{\ttfamily
  1402.6703}}].

\bibitem{Calore:2014xka}
F.~Calore, I.~Cholis and C.~Weniger, \emph{{Background Model Systematics for
  the Fermi GeV Excess}},
  \href{https://doi.org/10.1088/1475-7516/2015/03/038}{\emph{JCAP} {\bfseries
  03} (2015) 038} [\href{https://arxiv.org/abs/1409.0042}{{\ttfamily
  1409.0042}}].

\bibitem{Zhou:2014lva}
B.~Zhou, Y.-F.~Liang, X.~Huang, X.~Li, Y.-Z.~Fan, L.~Feng et~al., \emph{{GeV
  excess in the Milky Way: The role of diffuse galactic gamma-ray emission
  templates}}, \href{https://doi.org/10.1103/PhysRevD.91.123010}{\emph{Phys.
  Rev. D} {\bfseries 91} (2015) 123010}
  [\href{https://arxiv.org/abs/1406.6948}{{\ttfamily 1406.6948}}].

\bibitem{Abazajian:2014fta}
K.N.~Abazajian, N.~Canac, S.~Horiuchi and M.~Kaplinghat, \emph{{Astrophysical
  and Dark Matter Interpretations of Extended Gamma-Ray Emission from the
  Galactic Center}},
  \href{https://doi.org/10.1103/PhysRevD.90.023526}{\emph{Phys. Rev. D}
  {\bfseries 90} (2014) 023526}
  [\href{https://arxiv.org/abs/1402.4090}{{\ttfamily 1402.4090}}].

\bibitem{Navarro:1996gj}
J.F.~Navarro, C.S.~Frenk and S.D.M.~White, \emph{{A Universal density profile
  from hierarchical clustering}},
  \href{https://doi.org/10.1086/304888}{\emph{Astrophys. J.} {\bfseries 490}
  (1997) 493} [\href{https://arxiv.org/abs/astro-ph/9611107}{{\ttfamily
  astro-ph/9611107}}].

\bibitem{Kaplinghat:2013xca}
M.~Kaplinghat, R.E.~Keeley, T.~Linden and H.-B.~Yu, \emph{{Tying Dark Matter to
  Baryons with Self-interactions}},
  \href{https://doi.org/10.1103/PhysRevLett.113.021302}{\emph{Phys. Rev. Lett.}
  {\bfseries 113} (2014) 021302}
  [\href{https://arxiv.org/abs/1311.6524}{{\ttfamily 1311.6524}}].

\bibitem{Bartels:2015aea}
R.~Bartels, S.~Krishnamurthy and C.~Weniger, \emph{{Strong support for the
  millisecond pulsar origin of the Galactic center GeV excess}},
  \href{https://doi.org/10.1103/PhysRevLett.116.051102}{\emph{Phys. Rev. Lett.}
  {\bfseries 116} (2016) 051102}
  [\href{https://arxiv.org/abs/1506.05104}{{\ttfamily 1506.05104}}].

\bibitem{Lee:2015fea}
S.K.~Lee, M.~Lisanti, B.R.~Safdi, T.R.~Slatyer and W.~Xue, \emph{{Evidence for
  Unresolved $\gamma$-Ray Point Sources in the Inner Galaxy}},
  \href{https://doi.org/10.1103/PhysRevLett.116.051103}{\emph{Phys. Rev. Lett.}
  {\bfseries 116} (2016) 051103}
  [\href{https://arxiv.org/abs/1506.05124}{{\ttfamily 1506.05124}}].

\bibitem{Calore:2015bsx}
F.~Calore, M.~Di~Mauro, F.~Donato, J.W.T.~Hessels and C.~Weniger, \emph{{Radio
  detection prospects for a bulge population of millisecond pulsars as
  suggested by Fermi LAT observations of the inner Galaxy}},
  \href{https://doi.org/10.3847/0004-637X/827/2/143}{\emph{Astrophys. J.}
  {\bfseries 827} (2016) 143}
  [\href{https://arxiv.org/abs/1512.06825}{{\ttfamily 1512.06825}}].

\bibitem{AMS:2016oqu}
{\scshape AMS} collaboration, \emph{{Antiproton Flux, Antiproton-to-Proton Flux
  Ratio, and Properties of Elementary Particle Fluxes in Primary Cosmic Rays
  Measured with the Alpha Magnetic Spectrometer on the International Space
  Station}}, \href{https://doi.org/10.1103/PhysRevLett.117.091103}{\emph{Phys.
  Rev. Lett.} {\bfseries 117} (2016) 091103}.

\bibitem{Cirelli:2014lwa}
M.~Cirelli, D.~Gaggero, G.~Giesen, M.~Taoso and A.~Urbano, \emph{{Antiproton
  constraints on the GeV gamma-ray excess: a comprehensive analysis}},
  \href{https://doi.org/10.1088/1475-7516/2014/12/045}{\emph{JCAP} {\bfseries
  12} (2014) 045} [\href{https://arxiv.org/abs/1407.2173}{{\ttfamily
  1407.2173}}].

\bibitem{Cuoco:2016eej}
A.~Cuoco, M.~Kr\"amer and M.~Korsmeier, \emph{{Novel Dark Matter Constraints
  from Antiprotons in Light of AMS-02}},
  \href{https://doi.org/10.1103/PhysRevLett.118.191102}{\emph{Phys. Rev. Lett.}
  {\bfseries 118} (2017) 191102}
  [\href{https://arxiv.org/abs/1610.03071}{{\ttfamily 1610.03071}}].

\bibitem{Cui:2016ppb}
M.-Y.~Cui, Q.~Yuan, Y.-L.S.~Tsai and Y.-Z.~Fan, \emph{{Possible dark matter
  annihilation signal in the AMS-02 antiproton data}},
  \href{https://doi.org/10.1103/PhysRevLett.118.191101}{\emph{Phys. Rev. Lett.}
  {\bfseries 118} (2017) 191101}
  [\href{https://arxiv.org/abs/1610.03840}{{\ttfamily 1610.03840}}].

\bibitem{Cholis:2019ejx}
I.~Cholis, T.~Linden and D.~Hooper, \emph{{A Robust Excess in the Cosmic-Ray
  Antiproton Spectrum: Implications for Annihilating Dark Matter}},
  \href{https://doi.org/10.1103/PhysRevD.99.103026}{\emph{Phys. Rev. D}
  {\bfseries 99} (2019) 103026}
  [\href{https://arxiv.org/abs/1903.02549}{{\ttfamily 1903.02549}}].

\bibitem{Lin:2019ljc}
S.-J.~Lin, X.-J.~Bi and P.-F.~Yin, \emph{{Investigating the dark matter signal
  in the cosmic ray antiproton flux with the machine learning method}},
  \href{https://doi.org/10.1103/PhysRevD.100.103014}{\emph{Phys. Rev. D}
  {\bfseries 100} (2019) 103014}
  [\href{https://arxiv.org/abs/1903.09545}{{\ttfamily 1903.09545}}].

\bibitem{Carena:2019pwq}
M.~Carena, J.~Osborne, N.R.~Shah and C.E.M.~Wagner, \emph{{Return of the WIMP:
  Missing energy signals and the Galactic Center excess}},
  \href{https://doi.org/10.1103/PhysRevD.100.055002}{\emph{Phys. Rev. D}
  {\bfseries 100} (2019) 055002}
  [\href{https://arxiv.org/abs/1905.03768}{{\ttfamily 1905.03768}}].

\bibitem{Biekotter:2019kde}
T.~Biek\"otter, M.~Chakraborti and S.~Heinemeyer, \emph{{A 96 GeV Higgs boson
  in the N2HDM}},
  \href{https://doi.org/10.1140/epjc/s10052-019-7561-2}{\emph{Eur. Phys. J. C}
  {\bfseries 80} (2020) 2} [\href{https://arxiv.org/abs/1903.11661}{{\ttfamily
  1903.11661}}].

\bibitem{Cao:2016uwt}
J.~Cao, X.~Guo, Y.~He, P.~Wu and Y.~Zhang, \emph{{Diphoton signal of the light
  Higgs boson in natural NMSSM}},
  \href{https://doi.org/10.1103/PhysRevD.95.116001}{\emph{Phys. Rev. D}
  {\bfseries 95} (2017) 116001}
  [\href{https://arxiv.org/abs/1612.08522}{{\ttfamily 1612.08522}}].

\bibitem{Biekotter:2017xmf}
T.~Biek\"otter, S.~Heinemeyer and C.~Mu\~noz, \emph{{Precise prediction for the
  Higgs-boson masses in the $\mu \nu $ SSM}},
  \href{https://doi.org/10.1140/epjc/s10052-018-5978-7}{\emph{Eur. Phys. J. C}
  {\bfseries 78} (2018) 504}
  [\href{https://arxiv.org/abs/1712.07475}{{\ttfamily 1712.07475}}].

\bibitem{Domingo:2018uim}
F.~Domingo, S.~Heinemeyer, S.~Pa\ss{}ehr and G.~Weiglein, \emph{{Decays of the
  neutral Higgs bosons into SM fermions and gauge bosons in the
  $\mathcal{CP}$-violating NMSSM}},
  \href{https://doi.org/10.1140/epjc/s10052-018-6400-1}{\emph{Eur. Phys. J. C}
  {\bfseries 78} (2018) 942}
  [\href{https://arxiv.org/abs/1807.06322}{{\ttfamily 1807.06322}}].

\bibitem{Hollik:2018yek}
W.G.~Hollik, S.~Liebler, G.~Moortgat-Pick, S.~Pa\ss{}ehr and G.~Weiglein,
  \emph{{Phenomenology of the inflation-inspired NMSSM at the electroweak
  scale}}, \href{https://doi.org/10.1140/epjc/s10052-019-6561-6}{\emph{Eur.
  Phys. J. C} {\bfseries 79} (2019) 75}
  [\href{https://arxiv.org/abs/1809.07371}{{\ttfamily 1809.07371}}].

\bibitem{Choi:2019yrv}
K.~Choi, S.H.~Im, K.S.~Jeong and C.B.~Park, \emph{{Light Higgs bosons in the
  general NMSSM}},
  \href{https://doi.org/10.1140/epjc/s10052-019-7473-1}{\emph{Eur. Phys. J. C}
  {\bfseries 79} (2019) 956}
  [\href{https://arxiv.org/abs/1906.03389}{{\ttfamily 1906.03389}}].

\bibitem{Biekotter:2019gtq}
T.~Biek\"otter, S.~Heinemeyer and C.~Mu\~noz, \emph{{Precise prediction for the
  Higgs-Boson masses in the $\mu\nu$SSM with three right-handed neutrino
  superfields}},
  \href{https://doi.org/10.1140/epjc/s10052-019-7175-8}{\emph{Eur. Phys. J. C}
  {\bfseries 79} (2019) 667}
  [\href{https://arxiv.org/abs/1906.06173}{{\ttfamily 1906.06173}}].

\bibitem{Muhlleitner:2016mzt}
M.~Muhlleitner, M.O.P.~Sampaio, R.~Santos and J.~Wittbrodt, \emph{{The N2HDM
  under Theoretical and Experimental Scrutiny}},
  \href{https://doi.org/10.1007/JHEP03(2017)094}{\emph{JHEP} {\bfseries 03}
  (2017) 094} [\href{https://arxiv.org/abs/1612.01309}{{\ttfamily
  1612.01309}}].

\bibitem{Klimenko:1984qx}
K.G.~Klimenko, \emph{{On Necessary and Sufficient Conditions for Some Higgs
  Potentials to Be Bounded From Below}},
  \href{https://doi.org/10.1007/BF01034825}{\emph{Theor. Math. Phys.}
  {\bfseries 62} (1985) 58}.

\bibitem{Staub:2017ktc}
F.~Staub, \emph{{Reopen parameter regions in Two-Higgs Doublet Models}},
  \href{https://doi.org/10.1016/j.physletb.2017.11.065}{\emph{Phys. Lett. B}
  {\bfseries 776} (2018) 407}
  [\href{https://arxiv.org/abs/1705.03677}{{\ttfamily 1705.03677}}].

\bibitem{Lee:hom4ps2}
T.L.~Lee, T.Y.~Li and C.H.~Tsai, \emph{{HOM4PS-2.0: a software package for
  solving polynomial systems by the ployhedral homotopy continuation method}},
  \href{https://doi.org/10.1007/s00607-008-0015-6}{\emph{Computing} {\bfseries
  83} (2008) 109}.

\bibitem{Ferreira:2019iqb}
P.M.~Ferreira, M.~M\"uhlleitner, R.~Santos, G.~Weiglein and J.~Wittbrodt,
  \emph{{Vacuum Instabilities in the N2HDM}},
  \href{https://doi.org/10.1007/JHEP09(2019)006}{\emph{JHEP} {\bfseries 09}
  (2019) 006} [\href{https://arxiv.org/abs/1905.10234}{{\ttfamily
  1905.10234}}].

\bibitem{Staub:2013tta}
F.~Staub, \emph{{SARAH 4 : A tool for (not only SUSY) model builders}},
  \href{https://doi.org/10.1016/j.cpc.2014.02.018}{\emph{Comput. Phys. Commun.}
  {\bfseries 185} (2014) 1773}
  [\href{https://arxiv.org/abs/1309.7223}{{\ttfamily 1309.7223}}].

\bibitem{Schienbein:2018fsw}
I.~Schienbein, F.~Staub, T.~Steudtner and K.~Svirina, \emph{{Revisiting RGEs
  for general gauge theories}},
  \href{https://doi.org/10.1016/j.nuclphysb.2018.12.001}{\emph{Nucl. Phys. B}
  {\bfseries 939} (2019) 1} [\href{https://arxiv.org/abs/1809.06797}{{\ttfamily
  1809.06797}}].

\bibitem{Machacek:1983fi}
M.E.~Machacek and M.T.~Vaughn, \emph{{Two Loop Renormalization Group Equations
  in a General Quantum Field Theory. 2. Yukawa Couplings}},
  \href{https://doi.org/10.1016/0550-3213(84)90533-9}{\emph{Nucl. Phys. B}
  {\bfseries 236} (1984) 221}.

\bibitem{Machacek:1984zw}
M.E.~Machacek and M.T.~Vaughn, \emph{{Two Loop Renormalization Group Equations
  in a General Quantum Field Theory. 3. Scalar Quartic Couplings}},
  \href{https://doi.org/10.1016/0550-3213(85)90040-9}{\emph{Nucl. Phys. B}
  {\bfseries 249} (1985) 70}.

\bibitem{Machacek:1983tz}
M.E.~Machacek and M.T.~Vaughn, \emph{{Two Loop Renormalization Group Equations
  in a General Quantum Field Theory. 1. Wave Function Renormalization}},
  \href{https://doi.org/10.1016/0550-3213(83)90610-7}{\emph{Nucl. Phys. B}
  {\bfseries 222} (1983) 83}.

\bibitem{Sartore:2020gou}
L.~Sartore and I.~Schienbein, \emph{{PyR@TE 3}},
  \href{https://doi.org/10.1016/j.cpc.2020.107819}{\emph{Comput. Phys. Commun.}
  {\bfseries 261} (2021) 107819}
  [\href{https://arxiv.org/abs/2007.12700}{{\ttfamily 2007.12700}}].

\bibitem{Bechtle:2008jh}
P.~Bechtle, O.~Brein, S.~Heinemeyer, G.~Weiglein and K.E.~Williams,
  \emph{{HiggsBounds: Confronting Arbitrary Higgs Sectors with Exclusion Bounds
  from LEP and the Tevatron}},
  \href{https://doi.org/10.1016/j.cpc.2009.09.003}{\emph{Comput. Phys. Commun.}
  {\bfseries 181} (2010) 138}
  [\href{https://arxiv.org/abs/0811.4169}{{\ttfamily 0811.4169}}].

\bibitem{Bechtle:2011sb}
P.~Bechtle, O.~Brein, S.~Heinemeyer, G.~Weiglein and K.E.~Williams,
  \emph{{HiggsBounds 2.0.0: Confronting Neutral and Charged Higgs Sector
  Predictions with Exclusion Bounds from LEP and the Tevatron}},
  \href{https://doi.org/10.1016/j.cpc.2011.07.015}{\emph{Comput. Phys. Commun.}
  {\bfseries 182} (2011) 2605}
  [\href{https://arxiv.org/abs/1102.1898}{{\ttfamily 1102.1898}}].

\bibitem{Bechtle:2013gu}
P.~Bechtle, O.~Brein, S.~Heinemeyer, O.~St{\aa}l, T.~Stefaniak, G.~Weiglein
  et~al., \emph{{Recent Developments in HiggsBounds and a Preview of
  HiggsSignals}}, \href{https://doi.org/10.22323/1.156.0024}{\emph{PoS}
  {\bfseries CHARGED2012} (2012) 024}
  [\href{https://arxiv.org/abs/1301.2345}{{\ttfamily 1301.2345}}].

\bibitem{Bechtle:2013wla}
P.~Bechtle, O.~Brein, S.~Heinemeyer, O.~St{\aa}l, T.~Stefaniak, G.~Weiglein
  et~al., \emph{{$\mathsf{HiggsBounds}-4$: Improved Tests of Extended Higgs
  Sectors against Exclusion Bounds from LEP, the Tevatron and the LHC}},
  \href{https://doi.org/10.1140/epjc/s10052-013-2693-2}{\emph{Eur. Phys. J.}
  {\bfseries C74} (2014) 2693}
  [\href{https://arxiv.org/abs/1311.0055}{{\ttfamily 1311.0055}}].

\bibitem{Bechtle:2015pma}
P.~Bechtle, S.~Heinemeyer, O.~Stal, T.~Stefaniak and G.~Weiglein,
  \emph{{Applying Exclusion Likelihoods from LHC Searches to Extended Higgs
  Sectors}}, \href{https://doi.org/10.1140/epjc/s10052-015-3650-z}{\emph{Eur.
  Phys. J. C} {\bfseries 75} (2015) 421}
  [\href{https://arxiv.org/abs/1507.06706}{{\ttfamily 1507.06706}}].

\bibitem{Bechtle:2020pkv}
P.~Bechtle, D.~Dercks, S.~Heinemeyer, T.~Klingl, T.~Stefaniak, G.~Weiglein
  et~al., \emph{{HiggsBounds-5: Testing Higgs Sectors in the LHC 13 TeV Era}},
  \href{https://doi.org/10.1140/epjc/s10052-020-08557-9}{\emph{Eur. Phys. J. C}
  {\bfseries 80} (2020) 1211}
  [\href{https://arxiv.org/abs/2006.06007}{{\ttfamily 2006.06007}}].

\bibitem{Bechtle:2013xfa}
P.~Bechtle, S.~Heinemeyer, O.~St{\aa}l, T.~Stefaniak and G.~Weiglein,
  \emph{{$HiggsSignals$: Confronting arbitrary Higgs sectors with measurements
  at the Tevatron and the LHC}},
  \href{https://doi.org/10.1140/epjc/s10052-013-2711-4}{\emph{Eur. Phys. J.}
  {\bfseries C74} (2014) 2711}
  [\href{https://arxiv.org/abs/1305.1933}{{\ttfamily 1305.1933}}].

\bibitem{Stal:2013hwa}
O.~St{\aa}l and T.~Stefaniak, \emph{{Constraining extended Higgs sectors with
  HiggsSignals}}, \href{https://doi.org/10.22323/1.180.0314}{\emph{PoS}
  {\bfseries EPS-HEP2013} (2013) 314}
  [\href{https://arxiv.org/abs/1310.4039}{{\ttfamily 1310.4039}}].

\bibitem{Bechtle:2014ewa}
P.~Bechtle, S.~Heinemeyer, O.~St{\aa}l, T.~Stefaniak and G.~Weiglein,
  \emph{{Probing the Standard Model with Higgs signal rates from the Tevatron,
  the LHC and a future ILC}},
  \href{https://doi.org/10.1007/JHEP11(2014)039}{\emph{JHEP} {\bfseries 11}
  (2014) 039} [\href{https://arxiv.org/abs/1403.1582}{{\ttfamily 1403.1582}}].

\bibitem{Bechtle:2020uwn}
P.~Bechtle, S.~Heinemeyer, T.~Klingl, T.~Stefaniak, G.~Weiglein and
  J.~Wittbrodt, \emph{{HiggsSignals-2: Probing new physics with precision Higgs
  measurements in the LHC 13 TeV era}},
  \href{https://doi.org/10.1140/epjc/s10052-021-08942-y}{\emph{Eur. Phys. J. C}
  {\bfseries 81} (2021) 145}
  [\href{https://arxiv.org/abs/2012.09197}{{\ttfamily 2012.09197}}].

\bibitem{Djouadi:1997yw}
A.~Djouadi, J.~Kalinowski and M.~Spira, \emph{{HDECAY: A Program for Higgs
  boson decays in the standard model and its supersymmetric extension}},
  \href{https://doi.org/10.1016/S0010-4655(97)00123-9}{\emph{Comput. Phys.
  Commun.} {\bfseries 108} (1998) 56}
  [\href{https://arxiv.org/abs/hep-ph/9704448}{{\ttfamily hep-ph/9704448}}].

\bibitem{Butterworth:2010ym}
J.M.~Butterworth et~al., \emph{{THE TOOLS AND MONTE CARLO WORKING GROUP Summary
  Report from the Les Houches 2009 Workshop on TeV Colliders}},  in \emph{{6th
  Les Houches Workshop on Physics at TeV Colliders}}, 3, 2010
  [\href{https://arxiv.org/abs/1003.1643}{{\ttfamily 1003.1643}}].

\bibitem{Djouadi:2018xqq}
A.~Djouadi, J.~Kalinowski, M.~Muehlleitner and M.~Spira, \emph{{HDECAY:
  Twenty$_{++}$ years after}},
  \href{https://doi.org/10.1016/j.cpc.2018.12.010}{\emph{Comput. Phys. Commun.}
  {\bfseries 238} (2019) 214}
  [\href{https://arxiv.org/abs/1801.09506}{{\ttfamily 1801.09506}}].

\bibitem{Engeln:2018mbg}
I.~Engeln, M.~M\"uhlleitner and J.~Wittbrodt, \emph{{N2HDECAY: Higgs Boson
  Decays in the Different Phases of the N2HDM}},
  \href{https://doi.org/10.1016/j.cpc.2018.07.020}{\emph{Comput. Phys. Commun.}
  {\bfseries 234} (2019) 256}
  [\href{https://arxiv.org/abs/1805.00966}{{\ttfamily 1805.00966}}].

\bibitem{ATLAS:2020kdi}
{\scshape ATLAS} collaboration, \emph{{Combination of searches for invisible
  Higgs boson decays with the ATLAS experiment}}, {\emph{ATLAS-CONF-2020-052}
  (2020) }.

\bibitem{Peskin:1990zt}
M.E.~Peskin and T.~Takeuchi, \emph{{A New constraint on a strongly interacting
  Higgs sector}}, \href{https://doi.org/10.1103/PhysRevLett.65.964}{\emph{Phys.
  Rev. Lett.} {\bfseries 65} (1990) 964}.

\bibitem{Peskin:1991sw}
M.E.~Peskin and T.~Takeuchi, \emph{{Estimation of oblique electroweak
  corrections}}, \href{https://doi.org/10.1103/PhysRevD.46.381}{\emph{Phys.
  Rev. D} {\bfseries 46} (1992) 381}.

\bibitem{Grimus:2007if}
W.~Grimus, L.~Lavoura, O.M.~Ogreid and P.~Osland, \emph{{A Precision constraint
  on multi-Higgs-doublet models}},
  \href{https://doi.org/10.1088/0954-3899/35/7/075001}{\emph{J. Phys. G}
  {\bfseries 35} (2008) 075001}
  [\href{https://arxiv.org/abs/0711.4022}{{\ttfamily 0711.4022}}].

\bibitem{Grimus:2008nb}
W.~Grimus, L.~Lavoura, O.M.~Ogreid and P.~Osland, \emph{{The Oblique parameters
  in multi-Higgs-doublet models}},
  \href{https://doi.org/10.1016/j.nuclphysb.2008.04.019}{\emph{Nucl. Phys. B}
  {\bfseries 801} (2008) 81} [\href{https://arxiv.org/abs/0802.4353}{{\ttfamily
  0802.4353}}].

\bibitem{Haller:2018nnx}
J.~Haller, A.~Hoecker, R.~Kogler, K.~M\"onig, T.~Peiffer and J.~Stelzer,
  \emph{{Update of the global electroweak fit and constraints on
  two-Higgs-doublet models}},
  \href{https://doi.org/10.1140/epjc/s10052-018-6131-3}{\emph{Eur. Phys. J. C}
  {\bfseries 78} (2018) 675}
  [\href{https://arxiv.org/abs/1803.01853}{{\ttfamily 1803.01853}}].

\bibitem{Misiak:2020vlo}
M.~Misiak, A.~Rehman and M.~Steinhauser, \emph{{Towards $ \overline{B}\to
  {X}_s\gamma $ at the NNLO in QCD without interpolation in m$_{c}$}},
  \href{https://doi.org/10.1007/JHEP06(2020)175}{\emph{JHEP} {\bfseries 06}
  (2020) 175} [\href{https://arxiv.org/abs/2002.01548}{{\ttfamily
  2002.01548}}].

\bibitem{Bernlochner:2020jlt}
{\scshape SIMBA} collaboration, \emph{{Precision Global Determination of the
  $B\to X_s\gamma$ Decay Rate}},
  \href{https://arxiv.org/abs/2007.04320}{{\ttfamily 2007.04320}}.

\bibitem{Christensen:2008py}
N.D.~Christensen and C.~Duhr, \emph{{FeynRules - Feynman rules made easy}},
  \href{https://doi.org/10.1016/j.cpc.2009.02.018}{\emph{Comput. Phys. Commun.}
  {\bfseries 180} (2009) 1614}
  [\href{https://arxiv.org/abs/0806.4194}{{\ttfamily 0806.4194}}].

\bibitem{Christensen:2009jx}
N.D.~Christensen, P.~de~Aquino, C.~Degrande, C.~Duhr, B.~Fuks, M.~Herquet
  et~al., \emph{{A Comprehensive approach to new physics simulations}},
  \href{https://doi.org/10.1140/epjc/s10052-011-1541-5}{\emph{Eur. Phys. J. C}
  {\bfseries 71} (2011) 1541}
  [\href{https://arxiv.org/abs/0906.2474}{{\ttfamily 0906.2474}}].

\bibitem{Alloul:2013bka}
A.~Alloul, N.D.~Christensen, C.~Degrande, C.~Duhr and B.~Fuks, \emph{{FeynRules
  2.0 - A complete toolbox for tree-level phenomenology}},
  \href{https://doi.org/10.1016/j.cpc.2014.04.012}{\emph{Comput. Phys. Commun.}
  {\bfseries 185} (2014) 2250}
  [\href{https://arxiv.org/abs/1310.1921}{{\ttfamily 1310.1921}}].

\bibitem{Belyaev:2012qa}
A.~Belyaev, N.D.~Christensen and A.~Pukhov, \emph{{CalcHEP 3.4 for collider
  physics within and beyond the Standard Model}},
  \href{https://doi.org/10.1016/j.cpc.2013.01.014}{\emph{Comput. Phys. Commun.}
  {\bfseries 184} (2013) 1729}
  [\href{https://arxiv.org/abs/1207.6082}{{\ttfamily 1207.6082}}].

\bibitem{Belanger:2018ccd}
G.~B\'elanger, F.~Boudjema, A.~Goudelis, A.~Pukhov and B.~Zaldivar,
  \emph{{micrOMEGAs5.0 : Freeze-in}},
  \href{https://doi.org/10.1016/j.cpc.2018.04.027}{\emph{Comput. Phys. Commun.}
  {\bfseries 231} (2018) 173}
  [\href{https://arxiv.org/abs/1801.03509}{{\ttfamily 1801.03509}}].

\bibitem{Degrande:2011ua}
C.~Degrande, C.~Duhr, B.~Fuks, D.~Grellscheid, O.~Mattelaer and T.~Reiter,
  \emph{{UFO - The Universal FeynRules Output}},
  \href{https://doi.org/10.1016/j.cpc.2012.01.022}{\emph{Comput. Phys. Commun.}
  {\bfseries 183} (2012) 1201}
  [\href{https://arxiv.org/abs/1108.2040}{{\ttfamily 1108.2040}}].

\bibitem{Backovic:2013dpa}
M.~Backovic, K.~Kong and M.~McCaskey, \emph{{MadDM v.1.0: Computation of Dark
  Matter Relic Abundance Using MadGraph5}},
  \href{https://doi.org/10.1016/j.dark.2014.04.001}{\emph{Physics of the Dark
  Universe} {\bfseries 5-6} (2014) 18}
  [\href{https://arxiv.org/abs/1308.4955}{{\ttfamily 1308.4955}}].

\bibitem{Ambrogi:2018jqj}
F.~Ambrogi, C.~Arina, M.~Backovic, J.~Heisig, F.~Maltoni, L.~Mantani et~al.,
  \emph{{MadDM v.3.0: a Comprehensive Tool for Dark Matter Studies}},
  \href{https://doi.org/10.1016/j.dark.2018.11.009}{\emph{Phys. Dark Univ.}
  {\bfseries 24} (2019) 100249}
  [\href{https://arxiv.org/abs/1804.00044}{{\ttfamily 1804.00044}}].

\bibitem{Alwall:2014hca}
J.~Alwall, R.~Frederix, S.~Frixione, V.~Hirschi, F.~Maltoni, O.~Mattelaer
  et~al., \emph{{The automated computation of tree-level and next-to-leading
  order differential cross sections, and their matching to parton shower
  simulations}}, \href{https://doi.org/10.1007/JHEP07(2014)079}{\emph{JHEP}
  {\bfseries 07} (2014) 079} [\href{https://arxiv.org/abs/1405.0301}{{\ttfamily
  1405.0301}}].

\bibitem{Bauer:2017qwy}
M.~Bauer and T.~Plehn, \emph{{Yet Another Introduction to Dark Matter}: {The
  Particle Physics Approach}}, vol.~959 of \emph{Lecture Notes in Physics},
  Springer (2019),
  \href{https://doi.org/10.1007/978-3-030-16234-4}{10.1007/978-3-030-16234-4},
  [\href{https://arxiv.org/abs/1705.01987}{{\ttfamily 1705.01987}}].

\bibitem{DEAP_JMLR2012}
F.-A.~Fortin, F.-M.~{De Rainville}, M.-A.~Gardner, M.~Parizeau and C.~Gagn\'e,
  \emph{{DEAP}: Evolutionary algorithms made easy}, {\emph{Journal of Machine
  Learning Research} {\bfseries 13} (2012) 2171}.

\bibitem{vonBuddenbrock:2016rmr}
S.~von Buddenbrock, N.~Chakrabarty, A.S.~Cornell, D.~Kar, M.~Kumar, T.~Mandal
  et~al., \emph{{Phenomenological signatures of additional scalar bosons at the
  LHC}}, \href{https://doi.org/10.1140/epjc/s10052-016-4435-8}{\emph{Eur. Phys.
  J. C} {\bfseries 76} (2016) 580}
  [\href{https://arxiv.org/abs/1606.01674}{{\ttfamily 1606.01674}}].

\bibitem{Baum:2018zhf}
S.~Baum and N.R.~Shah, \emph{{Two Higgs Doublets and a Complex Singlet:
  Disentangling the Decay Topologies and Associated Phenomenology}},
  \href{https://doi.org/10.1007/JHEP12(2018)044}{\emph{JHEP} {\bfseries 12}
  (2018) 044} [\href{https://arxiv.org/abs/1808.02667}{{\ttfamily
  1808.02667}}].

\bibitem{ATLAS:2018oht}
{\scshape ATLAS} collaboration, \emph{{Search for a heavy Higgs boson decaying
  into a $Z$ boson and another heavy Higgs boson in the $\ell\ell bb$ final
  state in $pp$ collisions at $\sqrt{s}=13$ TeV with the ATLAS detector}},
  \href{https://doi.org/10.1016/j.physletb.2018.07.006}{\emph{Phys. Lett. B}
  {\bfseries 783} (2018) 392}
  [\href{https://arxiv.org/abs/1804.01126}{{\ttfamily 1804.01126}}].

\bibitem{ATLAS:2018sbw}
{\scshape ATLAS} collaboration, \emph{{Combination of searches for heavy
  resonances decaying into bosonic and leptonic final states using 36 fb$^{-1}$
  of proton-proton collision data at $\sqrt{s} = 13$ TeV with the ATLAS
  detector}}, \href{https://doi.org/10.1103/PhysRevD.98.052008}{\emph{Phys.
  Rev. D} {\bfseries 98} (2018) 052008}
  [\href{https://arxiv.org/abs/1808.02380}{{\ttfamily 1808.02380}}].

\bibitem{CMS:2015hra}
{\scshape CMS} collaboration, \emph{{Search for a Higgs boson in the mass range
  from 145 to 1000 GeV decaying to a pair of W or Z bosons}},
  \href{https://doi.org/10.1007/JHEP10(2015)144}{\emph{JHEP} {\bfseries 10}
  (2015) 144} [\href{https://arxiv.org/abs/1504.00936}{{\ttfamily
  1504.00936}}].

\bibitem{ATLAS:2017tlw}
{\scshape ATLAS} collaboration, \emph{{Search for heavy ZZ resonances in the
  $\ell ^+\ell ^-\ell ^+\ell ^-$ and $\ell ^+\ell ^-\nu \bar{\nu }$ final
  states using proton\textendash{}proton collisions at $\sqrt{s}= 13$ $\text
  {TeV}$ with the ATLAS detector}},
  \href{https://doi.org/10.1140/epjc/s10052-018-5686-3}{\emph{Eur. Phys. J. C}
  {\bfseries 78} (2018) 293}
  [\href{https://arxiv.org/abs/1712.06386}{{\ttfamily 1712.06386}}].

\bibitem{CMS:2018amk}
{\scshape CMS} collaboration, \emph{{Search for a new scalar resonance decaying
  to a pair of Z bosons in proton-proton collisions at $\sqrt{s}=13 $ TeV}},
  \href{https://doi.org/10.1007/JHEP06(2018)127}{\emph{JHEP} {\bfseries 06}
  (2018) 127} [\href{https://arxiv.org/abs/1804.01939}{{\ttfamily
  1804.01939}}].

\bibitem{ATLAS:2020zms}
{\scshape ATLAS} collaboration, \emph{{Search for heavy Higgs bosons decaying
  into two tau leptons with the ATLAS detector using $pp$ collisions at
  $\sqrt{s}=13$ TeV}},
  \href{https://doi.org/10.1103/PhysRevLett.125.051801}{\emph{Phys. Rev. Lett.}
  {\bfseries 125} (2020) 051801}
  [\href{https://arxiv.org/abs/2002.12223}{{\ttfamily 2002.12223}}].

\bibitem{ATLAS:2018rnh}
{\scshape ATLAS} collaboration, \emph{{Search for pair production of Higgs
  bosons in the $b\bar{b}b\bar{b}$ final state using proton-proton collisions
  at $\sqrt{s} = 13$ TeV with the ATLAS detector}},
  \href{https://doi.org/10.1007/JHEP01(2019)030}{\emph{JHEP} {\bfseries 01}
  (2019) 030} [\href{https://arxiv.org/abs/1804.06174}{{\ttfamily
  1804.06174}}].

\bibitem{CMS:2019qcx}
{\scshape CMS} collaboration, \emph{{Search for a heavy pseudoscalar boson
  decaying to a Z and a Higgs boson at $\sqrt{s} =$ 13 TeV}},
  \href{https://doi.org/10.1140/epjc/s10052-019-7058-z}{\emph{Eur. Phys. J. C}
  {\bfseries 79} (2019) 564}
  [\href{https://arxiv.org/abs/1903.00941}{{\ttfamily 1903.00941}}].

\bibitem{ATLAS:2021upq}
{\scshape ATLAS} collaboration, \emph{{Search for charged Higgs bosons decaying
  into a top quark and a bottom quark at $ \sqrt{\mathrm{s}} $ = 13 TeV with
  the ATLAS detector}},
  \href{https://doi.org/10.1007/JHEP06(2021)145}{\emph{JHEP} {\bfseries 06}
  (2021) 145} [\href{https://arxiv.org/abs/2102.10076}{{\ttfamily
  2102.10076}}].

\bibitem{ATLAS:2019qdc}
{\scshape ATLAS} collaboration, \emph{{Combination of searches for Higgs boson
  pairs in $pp$ collisions at $\sqrt{s} = $13 TeV with the ATLAS detector}},
  \href{https://doi.org/10.1016/j.physletb.2019.135103}{\emph{Phys. Lett. B}
  {\bfseries 800} (2020) 135103}
  [\href{https://arxiv.org/abs/1906.02025}{{\ttfamily 1906.02025}}].

\bibitem{CMS:2019pzc}
{\scshape CMS} collaboration, \emph{{Search for heavy Higgs bosons decaying to
  a top quark pair in proton-proton collisions at $\sqrt{s} =$ 13 TeV}},
  \href{https://doi.org/10.1007/JHEP04(2020)171}{\emph{JHEP} {\bfseries 04}
  (2020) 171} [\href{https://arxiv.org/abs/1908.01115}{{\ttfamily
  1908.01115}}].

\bibitem{Binder:2017rgn}
T.~Binder, T.~Bringmann, M.~Gustafsson and A.~Hryczuk, \emph{{Early kinetic
  decoupling of dark matter: when the standard way of calculating the thermal
  relic density fails}},
  \href{https://doi.org/10.1103/PhysRevD.96.115010}{\emph{Phys. Rev. D}
  {\bfseries 96} (2017) 115010}
  [\href{https://arxiv.org/abs/1706.07433}{{\ttfamily 1706.07433}}].

\bibitem{Abe:2021jcz}
T.~Abe, \emph{{Early kinetic decoupling and a pseudo-Nambu-Goldstone dark
  matter model}},  \href{https://arxiv.org/abs/2106.01956}{{\ttfamily
  2106.01956}}.

\bibitem{AMS:2021nhj}
{\scshape AMS} collaboration, \emph{{The Alpha Magnetic Spectrometer (AMS) on
  the international space station: Part II \textemdash{} Results from the first
  seven years}},
  \href{https://doi.org/10.1016/j.physrep.2020.09.003}{\emph{Phys. Rept.}
  {\bfseries 894} (2021) 1}.

\bibitem{Heisig:2020nse}
J.~Heisig, M.~Korsmeier and M.W.~Winkler, \emph{{Dark matter or correlated
  errors: Systematics of the AMS-02 antiproton excess}},
  \href{https://doi.org/10.1103/PhysRevResearch.2.043017}{\emph{Phys. Rev.
  Res.} {\bfseries 2} (2020) 043017}
  [\href{https://arxiv.org/abs/2005.04237}{{\ttfamily 2005.04237}}].

\bibitem{Boudaud:2019efq}
M.~Boudaud, Y.~G\'enolini, L.~Derome, J.~Lavalle, D.~Maurin, P.~Salati et~al.,
  \emph{{AMS-02 antiprotons' consistency with a secondary astrophysical
  origin}}, \href{https://doi.org/10.1103/PhysRevResearch.2.023022}{\emph{Phys.
  Rev. Res.} {\bfseries 2} (2020) 023022}
  [\href{https://arxiv.org/abs/1906.07119}{{\ttfamily 1906.07119}}].

\bibitem{DiMauro:2021raz}
M.~Di~Mauro, \emph{{Characteristics of the Galactic Center excess measured with
  11 years of $Fermi$-LAT data}},
  \href{https://doi.org/10.1103/PhysRevD.103.063029}{\emph{Phys. Rev. D}
  {\bfseries 103} (2021) 063029}
  [\href{https://arxiv.org/abs/2101.04694}{{\ttfamily 2101.04694}}].

\bibitem{List:2021aer}
F.~List, N.L.~Rodd and G.F.~Lewis, \emph{{Dim but not entirely dark: Extracting
  the Galactic Center Excess' source-count distribution with neural nets}},
  \href{https://arxiv.org/abs/2107.09070}{{\ttfamily 2107.09070}}.

\bibitem{Kahlhoefer:2021sha}
F.~Kahlhoefer, M.~Korsmeier, M.~Kr\"amer, S.~Manconi and K.~Nippel,
  \emph{{Constraining dark matter annihilation with cosmic ray antiprotons
  using neural networks}},  \href{https://arxiv.org/abs/2107.12395}{{\ttfamily
  2107.12395}}.

\bibitem{Abdughani:2021pdc}
M.~Abdughani, Y.-Z.~Fan, L.~Feng, Y.-L.S.~Tsai, L.~Wu and Q.~Yuan, \emph{{A
  common origin of muon g-2 anomaly, Galaxy Center GeV excess and AMS-02
  anti-proton excess in the NMSSM}},
  \href{https://doi.org/10.1016/j.scib.2021.07.029}{\emph{Sci. Bull.}
  {\bfseries 66} (2021) 2170}
  [\href{https://arxiv.org/abs/2104.03274}{{\ttfamily 2104.03274}}].

\bibitem{Beck:2021xsv}
G.~Beck, R.~Temo, E.~Malwa, M.~Kumar and B.~Mellado, \emph{{Connecting
  multi-lepton anomalies at the LHC and in Astrophysics with MeerKAT/SKA}},  2,
  2021 [\href{https://arxiv.org/abs/2102.10596}{{\ttfamily 2102.10596}}].

\bibitem{Fermi-LAT:2016afa}
{\scshape Fermi-LAT} collaboration, \emph{{Sensitivity Projections for Dark
  Matter Searches with the Fermi Large Area Telescope}},
  \href{https://doi.org/10.1016/j.physrep.2016.05.001}{\emph{Phys. Rept.}
  {\bfseries 636} (2016) 1} [\href{https://arxiv.org/abs/1605.02016}{{\ttfamily
  1605.02016}}].

\bibitem{LEPWorkingGroupforHiggsbosonsearches:2003ing}
{\scshape LEP Working Group for Higgs boson searches, ALEPH, DELPHI, L3, OPAL}
  collaboration, \emph{{Search for the standard model Higgs boson at LEP}},
  \href{https://doi.org/10.1016/S0370-2693(03)00614-2}{\emph{Phys. Lett. B}
  {\bfseries 565} (2003) 61}
  [\href{https://arxiv.org/abs/hep-ex/0306033}{{\ttfamily hep-ex/0306033}}].

\bibitem{CMS:2018cyk}
{\scshape CMS} collaboration, \emph{{Search for a standard model-like Higgs
  boson in the mass range between 70 and 110 GeV in the diphoton final state in
  proton-proton collisions at $\sqrt{s}=$ 8 and 13 TeV}},
  \href{https://doi.org/10.1016/j.physletb.2019.03.064}{\emph{Phys. Lett. B}
  {\bfseries 793} (2019) 320}
  [\href{https://arxiv.org/abs/1811.08459}{{\ttfamily 1811.08459}}].

\bibitem{LHCHiggsCrossSectionWorkingGroup:2016ypw}
{\scshape LHC Higgs Cross Section Working Group} collaboration, \emph{{Handbook
  of LHC Higgs Cross Sections: 4. Deciphering the Nature of the Higgs Sector}},
   \href{https://arxiv.org/abs/1610.07922}{{\ttfamily 1610.07922}}.

\end{thebibliography}\endgroup

\end{document}